\documentclass{lmcs}
\pdfoutput=1

\usepackage{lastpage}
\lmcsdoi{20}{4}{19}
\lmcsheading{}{\pageref{LastPage}}{}{}%
{Sep.~27,~2023}{Dec.~03,~2024}{}

\pdfoutput=1 
\usepackage[utf8]{inputenc}

\usepackage{amsmath}  
\usepackage{verbatim}
\usepackage{graphicx}
\usepackage{epstopdf}
\usepackage{bussproofs}
\usepackage{xfrac}
\usepackage{color,xcolor}
\usepackage[normalem]{ulem}
\definecolor{hlcolor}{rgb}{1, 0.9, 0.5}
\newcommand{\hl}{}
\usepackage{dsfont}
\usepackage{stmaryrd}
\usepackage{amsthm}
\usepackage{mathtools}
\usepackage{amssymb}
\usepackage{tikz}
\usepackage{booktabs,tabularx}
\usepackage{todonotes}

\usetikzlibrary{matrix}

\usepackage{tikz-cd}

\usepackage{quiver}

\usepackage{framed}
\usepackage{comment}
\usepackage{hyperref}
\definecolor{shadecolor}{gray}{0.95}

\newcommand{\tmonad}[1]{\terms{\sig, #1}}

\newcommand{\qtmonad}[1]{\qterms{\sig, #1}}

\newcommand{\sem}[1] {  \llbracket #1 \rrbracket  }

\newcommand{\Set}{\mathbf{Set}}
\newcommand{\GMet}{\mathbf{GMet}}
\newcommand{\Met}{\mathbf{Met}}

\newcommand{\MEq}{\textnormal{QEq}}

\newcommand{\Terms}[1]{\textnormal{Terms}_\Sigma(#1)}
\newcommand{\TermsA}{\textnormal{Terms}_\Sigma(A)}
\newcommand{\TermsB}{\textnormal{Terms}_\Sigma(B)}

\newcommand{\Freedist}[1]
{\Delta^{F(#1)}}

\newcommand{\FreedistA}{\Freedist{A,d_A}}
\newcommand{\FreedistB}{\Freedist{B,d_B}}

\newcommand{\algA}{\mathbb{A}}
\newcommand{\algB}{\mathbb{B}}
\newcommand{\algC}{\mathbb{C}}
\newcommand{\algD}{\mathbb{D}}

\newcommand{\catC}{\mathcal{C}}
\newcommand{\catD}{\mathcal{D}}
\newcommand{\catE}{\mathcal{E}}

\newcommand{\mon}{M}
\newrobustcmd{\lift}[1]{\widehat{#1}}
\newrobustcmd{\sig}{\Sigma}
\newrobustcmd{\qsig}{\widehat{\sig}}
\renewrobustcmd{\th}{\Phi}
\newrobustcmd{\qth}{\widehat{\th}}
\newrobustcmd{\terms}[1]{T^\Set_{#1}}
\newrobustcmd{\qterms}[1]{{T}^\FRel_{#1}}
\newrobustcmd{\qtermsmet}[1]{{T}_{#1}^{\Met}}
\newrobustcmd{\qtermsgmet}[1]{{T}^{\GMet}_{#1}}
\newrobustcmd{\op}{op}

\newrobustcmd{\unit}{\eta}
\newrobustcmd{\mult}{\mu}

\newcommand{\id}{\mathrm{id}}

\newcommand{\Theory}{\mathbf{Th}_{\Sigma}}

\newcommand{\QTheory}{\mathbf{QTh}_{\Sigma}}

\newrobustcmd{\FRel}{\mathbf{FRel}}

\newcommand{\Models}{{\bf Mod}_\Sigma}

\newcommand{\QModels}{{\bf QMod}_\Sigma}

\newcommand{\ModelsCat}{{\bf Mod}_{\Sigma}}

\newcommand{\QModelsCat}{{\bf QMod}_{\Sigma}}
\newcommand{\QModelsCatNoOp}{{\bf QMod}_{\emptyset}}

\newcommand{\Eq}
{\textnormal{Eq}(\Sigma)}

\newcommand{\EM}{{\bf EM}}

\newcommand{\Alg}{{\bf Alg}(\Sigma)}

\newcommand{\Catalg}{\ModelsCat(\Phi)}

\newcommand{\Catqalg}{\QModelsCat({\Phi})}
\newcommand{\Catqalgempty}{\Qalg}

\newcommand{\Qalg}{{\bf QAlg}{(\Sigma)}}

\newcommand{\QalgFRel}{{\bf QAlg}^{\FRel}{(\Sigma})}
\newcommand{\QalgGMet}{{\bf QAlg}^{\GMet}{(\Sigma})}
\newcommand{\QalgMet}{{\bf QAlg}^{\Met}{(\Sigma})}

\newcommand{\imap}{\tau}
\newcommand{\freemap}{\alpha}

\newcommand{\derivefrel}{\mathrel{\vdash}}
\newcommand{\derivefrelfrel}{\mathrel{\vdash_{\FRel}}}
\newcommand{\implyfrel}{\mathrel{\Vvdash}}
\newcommand{\implyfrelfrel}{\mathrel{\Vvdash_{\FRel}}}

\newcommand{\derive}{\mathrel{\vdash_{\Set}}}
\newcommand{\imply}{\mathrel{\Vvdash_{\Set}}}

\newcommand{\derivegmet}{\mathrel{\vdash_\GMet}}
\newcommand{\implygmet}{\Vvdash_\GMet}

\newcommand{\Hmap}{\foi}
\newcommand{\Hphi}{\phi_H}
\newcommand{\HPhi}{\Phi_{\Hset}}
\newcommand{\HGMet}{\GMet_{\Hset}}
\newcommand{\Hset}{\mathcal{H}}
\newrobustcmd{\Hqterms}[1]{\widehat{T}^{\Hset}_{#1}}

\newcommand{\QmodFRel}{{\bf QMod}^{\FRel}_\Sigma}
\newcommand{\QmodGMet}{{\bf QMod}^{\GMet}_\Sigma}
\newcommand{\QmodMet}{{\bf QMod}^{\Met}_\Sigma}

\newrobustcmd{\Cat}{\catC}

\makeatletter
\newcommand{\doublewidehat}[1]{{%
  \mathpalette\double@widehat{#1}%
}}
\newcommand{\double@widehat}[2]{%
  \sbox\z@{$\m@th#1\widehat{#2}$}%
  \ht\z@=.83\ht\z@
  \widehat{\box\z@}%
}
\makeatother

\newcommand{\ruleorange}[1]
{{#1}}

\usepackage{mathrsfs}

\newcommand{\foi}{\iota}

\newcommand{\funE}{E}

\newcommand{\fomodels}{\models^{\folang}}
\newcommand{\folang}{\mathscr{L}}

\newcommand{\mP}{\mathscr P}
\newcommand{\mD}{\mathscr D}
\newcommand{\LK}[1]{{#1}_{\textnormal{\L K}}}
\newcommand{\lmP}{\lift{\mathscr P}}


\title[Universal Quantitative Algebra]{Universal Quantitative Algebra\\ for Fuzzy Relations and Generalised Metric Spaces}
\author[M.~Mio]{Matteo Mio \lmcsorcid{0000-0003-4050-3617}}[a]
\author[R.~Sarkis]{Ralph Sarkis \lmcsorcid{0000-0002-9037-2435}}[b]
\author[V.~Vignudelli]{Valeria Vignudelli \lmcsorcid{0000-0002-7864-9892}}[a]
\address{CNRS \& ENS de Lyon, LIP, UMR 5668, France}
\address{ENS de Lyon, LIP, UMR 5668, France}

\usepackage{array}
\begin{document}
\begin{abstract}
     We present a generalisation of the theory of quantitative algebras of Mardare, Panangaden and Plotkin where (i) the carriers of quantitative algebras are not restricted to be metric spaces and can be arbitrary fuzzy relations or generalised metric spaces, and (ii) the interpretations of the algebraic operations are not required to be nonexpansive. Our main results include: a novel sound and complete proof system, the proof that free quantitative algebras always exist, the proof of strict monadicity of the induced Free-Forgetful adjunction, the result that all monads (on fuzzy relations) that lift finitary monads (on sets) admit a quantitative equational presentation.
\end{abstract}
\maketitle
\section{Introduction}\label{section_intro}

Equational reasoning and algebraic methods are widespread in all areas of computer science, and in particular in program semantics. Indeed, initial algebra semantics and monads are cornerstones of the modern theory of functional programming and are used to reason about inductive definitions, computational effects and specifications in a formal way (see, e.g., \cite{DBLP:journals/acs/PlotkinP03,DBLP:conf/rex/RuttenT93,DBLP:journals/iandc/Moggi91, DBLP:journals/entcs/HylandP07}).

In the last few decades, with the growth of quantitative methods in computing (e.g., from artificial intelligence, probabilistic programming, cyber-physical systems, \emph{etc.}) it has become evident that traditional program equations
\begin{center}
$P = Q$  \ \ \ \ $\Leftrightarrow$ \ \ \ \ the programs $P$ and $Q$ have the same behavior
\end{center}
are not always adequate to reason about the behaviour of programs that are similar, in a certain quantitative sense, but that, strictly speaking, have different behavior. Examples include programs  that differ only by small perturbations in some of their numeric constants such as probabilities, values measured from noisy sensors, scalars in a neural network, \emph{etc.} The intuitive notion of ``similar in a quantitative sense'', has been formally captured in many works by means of program distances $d(P,Q)\in [0,\infty]$ expressing numerically the divergence in behavior. See \cite{DBLP:conf/ifip2/GiacaloneJS90, DBLP:conf/concur/DesharnaisGJP99, DBLP:conf/lics/DesharnaisJGP02, DBLP:journals/iandc/DesharnaisEP02,DBLP:conf/icalp/BreugelW01, DBLP:conf/concur/BreugelW01, 
DBLP:journals/tcs/BreugelW05} for a selection of \hl{influential} papers.  

In a recent work \cite{DBLP:conf/lics/MardarePP16}, Mardare, Panangaden and Plotkin introduced a novel abstract mathematical ``framework'', called \emph{Quantitative Algebra}, which extends ordinary Universal Algebra (see, e.g., \cite{DBLP:books/daglib/0067494, Wechler1992UniversalAF}) and is designed to reason about distances that are metrics.\footnote{\label{footnote_intro_1}Precisely, they consider \emph{extended metrics}, which are distances $d:X^2 \rightarrow [0,\infty]$ satisfying the following constraints for all $x,y,z\in X$: $d(x,y)=0 \Leftrightarrow x=y$, $d(x,y)=d(y,x)$ and $d(x,z)\leq d(x,y) + d(y,z) $.} The standard equality judgment ($s=t$) of Universal Algebra is replaced by \emph{quantitative equations} ($s=_\epsilon t$), intuitively expressing that $d(s,t)\leq\epsilon$. In the program semantics context, we thus have
\begin{center}
$P =_\epsilon Q$  \ \ \ \ $\Leftrightarrow$ \ \ \ \ the difference in behavior between $P$ and $Q$ is at most $\epsilon$.
\end{center}

The usual notion from Universal Algebra of \emph{algebra} $(A,\{op^{A}\}_{op\in\Sigma})$ for a signature $\Sigma$, that is a carrier set $A$ together with interpretations $op^A$ for all function symbols in $\Sigma$, is replaced by that of \emph{quantitative algebra}: 
\begin{center}
$\big((A,d_A)\ , \ \{op^{A}\}_{op\in\Sigma}\big)$,
\end{center}
where the carrier is a metric space $(A,d_A)$ and the interpretations $op^A$, for all $op\in\Sigma$, are nonexpansive maps.\footnote{\label{footcategoricalproduct}More precisely, the interpretation $op^A:(A^n,d^n_A)\rightarrow (A,d_A)$ of $op\in\Sigma$ is nonexpansive, where $d^n_A$ is the (categorical) product metric on $A^n$. See \cite{DBLP:conf/lics/MardarePP16} and Section \ref{comparison_section} for a detailed discussion.}

A number of recent works have built on top of the results of the seminal \cite{DBLP:conf/lics/MardarePP16}. For a non-exhaustive list, see, e.g., \cite{DBLP:conf/lics/MardarePP17, DBLP:conf/lics/BacciMPP18, DBLP:conf/concur/MioV20, DBLP:conf/lics/MioSV22, DBLP:conf/lics/MioSV21,DBLP:conf/lics/MardarePP21, DBLP:conf/calco/FordMS21, DBLP:conf/calco/BacciMPP21, DBLP:conf/lics/MioSV22, DBLP:conf/lics/Adamek22, dallago_et_al:LIPIcs.FSCD.2022.4, Adamek2023, DBLP:journals/pacmpl/GavazzoF23, Jurka2024, Rozowski2024}. Key theoretical results in quantitative algebra include: sound and complete deductive systems, existence of free quantitative algebras generated by metric spaces, monads and composition techniques for \hl{monads on} the category $\Met$ of metric spaces and nonexpansive maps, completion results, variety ``HSP-type'' theorems, \emph{etc}. Applications of this framework can be found in the identification of useful \hl{monads on} $\Met$ as ``free quantitative algebra'' monads (see, e.g., \cite{DBLP:conf/lics/MardarePP16, DBLP:conf/concur/MioV20, DBLP:conf/lics/MioSV21, DBLP:conf/lics/MioSV22}) and in the quantitative axiomatisation of behavioral metrics \cite{DBLP:conf/lics/BacciMPP18,DBLP:journals/lmcs/BacciBLM18, DBLP:journals/entcs/Bacci0LM18, DBLP:conf/lics/MioSV21, Rozowski2024}.

Furthermore, some works have proposed extensions or modifications of the framework of \cite{DBLP:conf/lics/MardarePP16}. For instance, \cite{DBLP:conf/lics/MioSV22} has considered quantitative algebras $\big((A,d_A)\ , \ \{op^{A}\}_{op\in\Sigma}\big)$ where $(A,d_A)$ is not necessarily a metric space but, more generally, a \emph{generalised metric space}\footnote{The terminology ``generalised metric space'' has already appeared in the literature (see, e.g., \cite{DBLP:journals/tcs/BonsangueBR98}, \cite{Branciari2000}) with slightly different meanings.} %
(e.g., 
pseudometrics, quasimetrics \cite{Wilson1931quasi}, ultrametrics \cite{DBLP:journals/tcs/BonsangueBR98}, semimetrics \cite{Wilson1931semi}, diffuse metrics \cite{Hitzler2000,Castro2021}). %
In \cite{DBLP:conf/calco/FordMS21} this type of generalisation is pushed even further, allowing the carrier to be an arbitrary relational structure. In a different direction, in \hl{the already mentioned} \cite{DBLP:conf/lics/MioSV22} (see also \cite{DBLP:journals/lmcs/BacciBLM18} and, in the different context of ordered algebras, \hl{\cite{DBLP:journals/mscs/AdamekFMS21}}) the authors have considered quantitative algebras where the interpretations $op^A$, of all $op\in\Sigma$, are not
\hl{required to be} nonexpansive maps.  This extends considerably the applicability of the theory, as witnessed by interesting examples (e.g., from concurrency theory in \cite{DBLP:journals/lmcs/BacciBLM18}  and artificial intelligence in \cite{DBLP:conf/lics/MioSV22}).

\subsection{Contributions}

The main contribution of this paper is to present a generalisation of the framework of \cite{DBLP:conf/lics/MardarePP16} in a self-contained and coherent way and to prove in full details some fundamental results. \hl{This is partially based on the previous conference paper \cite{DBLP:conf/lics/MioSV22} by the authors, which is here significantly simplified and generalised. A precise comparison between the present work and \cite{DBLP:conf/lics/MioSV22} is discussed in Section \ref{subsec:msvcomp}}.

We extend \cite{DBLP:conf/lics/MardarePP16} along two orthogonal lines, by considering quantitative algebras  $\big((A,d_A)\ , \ \{op^{A}\}_{op\in\Sigma}\big)$ where
\begin{enumerate}
\item[] (1st line): the carrier $(A,d_A)$ is an arbitrary fuzzy relation space \cite{Zadeh71}, that is, a set $A$ together with an arbitrary map $d_A:A^2\rightarrow [0,1]$, \hl{and}
\item[] (2nd line): the interpretations $op^A$, for $op\in\Sigma$, are arbitrary \hl{functions}, and not required to be nonexpansive. 
\end{enumerate}

\hl{Regarding the first line of extension, considering fuzzy relation spaces $d_A:A^2\rightarrow [0,1]$  is a somewhat arbitrary choice between the full generality of arbitrary relational structures (as in \cite{DBLP:conf/calco/FordMS21} and Section \ref{subsec:fmscomp}) and the simplicity of working with a concrete notion of numerical distance, in the style of the original \cite{DBLP:conf/lics/MardarePP16}, which is still general enough to include, e.g., all the generalised metric spaces considered in \cite{DBLP:conf/lics/MioSV22}. We expect that our results can be generalised to arbitrary relational structures.}

Regarding the second line of extension, allowing $op^A$ to be an arbitrary \hl{function} results in greater generality in the definition. \hl{This is a novelty with respect to both \cite{DBLP:conf/lics/MardarePP16}  and \cite{DBLP:conf/calco/FordMS21}}, and \hl{it also constitutes a simplification of \cite{DBLP:conf/lics/MioSV22},  as we detail in Section \ref{subsec:msvcomp}.}

From a logical point of view, since we work with arbitrary fuzzy relations (for which the property $x=y \Leftrightarrow d(x,y)=0$ might not hold), we have to decouple the notion of equality from that of distance. As a result, our theory of quantitative algebras deals with two types of formal judgments: equations and quantitative equations, respectively of the form
\[
\forall (X,d_X). s=t \ \ \ \  \ \ \ \ \ \ \ \forall (X,d_X). s=_\epsilon t,
\]
where $\epsilon\in[0,1]$ and $s,t\in \Terms{X}$ are terms built from a set of variables $X$ which is endowed with a fuzzy relation $d_X:X^2\rightarrow [0,1]$. We have followed the Universal Algebra textbook \cite{Wechler1992UniversalAF} in using the ``$\forall$'' symbol, in the formal judgments, to explicitly remind that the stated equality (equation) or bound on distance (quantitative equation) is universally quantified over all interpretations of the variables in $X$. 

Crucially, as the set of variables is endowed with a fuzzy relation $d_X$, an interpretation is required to be nonexpansive. The idea of restricting attention to interpretations that ``preserve structure'' has already appeared in the literature in, e.g., \cite{DBLP:journals/mscs/AdamekFMS21} in the study of ordered algebras, \hl{and \cite{DBLP:conf/calco/FordMS21} in the context of algebras over arbitrary relational structures.}  This has important consequences.  Consider for example the following ``gap'' property:
\[
\textnormal{For all $x,y$, if the distance between $x$ and $y$ is $\leq \tfrac{1}{2}$, then the distance is in fact $\leq \tfrac{1}{4}$.}
\]
\hl{In \cite{DBLP:conf/lics/MardarePP16}} and most subsequent works, \hl{including our conference paper \cite{DBLP:conf/lics/MioSV22}}, to express the above property one needs to consider \hl{implications}\footnote{In  \cite{DBLP:conf/lics/MardarePP16} such implications are referred to as ``quantitative inferences'' and denoted by $x=_\frac{1}{2} y \vdash x=_\frac{1}{4} y$.} between quantitative equations:
\[
x=_\frac{1}{2} y \Rightarrow x=_\frac{1}{4} y.
\]
The syntactic deductive apparatus presented in \cite{DBLP:conf/lics/MardarePP16} is designed to manipulate \hl{this type of judgments},
and not just quantitative equations. In our setting, instead, the ``gap'' property is directly expressed by the quantitative equation,
\[
\forall (\{x,y\}, d_X). x=_\frac{1}{4}y,
\]
where the fuzzy relation $d_X$ on the set $\{x,y\}$ of variables assigns value $\frac{1}{2}$ to $(x,y)$:
\begin{center}$d_X(x,x) =1$,  $d_X(x,y) =\frac{1}{2}$,   $d_X(y,x) =1$, \hl{$d_X(y,y) =1$}.
\end{center}
Requiring interpretations to be nonexpansive then corresponds precisely to requiring that the premise of the implication is satisfied. As a consequence, we are able to work just with equations and quantitative equations, thus avoiding higher-level logical concepts such as implications. This is a novelty with respect to both \cite{DBLP:conf/lics/MardarePP16} and \cite{DBLP:conf/lics/MioSV22}.

In this new setting, we can recover the framework of Mardare, Panangaden and Plotkin as a specific ``quantitative equationally'' defined subclass of quantitative algebras.
For instance, we can now define, by means of equations and quantitative equations, the subclass of quantitative algebras whose $d_A$ satisfy the constraints of a chosen generalised metric (including, e.g.,  the standard conditions of metrics, as in Footnote \ref{footnote_intro_1}), in the same way that, in Universal Algebra, the subclass of abelian groups can be defined equationally from groups. 
At the same time, it is possible to  define by means of equations and quantitative equations the subclass of quantitative algebras whose $op^A$ is nonexpansive or, more generally, Lipschitz with constant $\alpha>1$, or other useful notions.

The following is a list of our main results \ref{item:introDeductivesys}--\ref{item:introFReltoGMet} concerning the generalised quantitative algebra theory presented in this work:
\begin{enumerate}[(I)]
\item\label{item:introDeductivesys} We present a sound and complete ``Birkhoff-style" deductive system to derive valid equations and quantitative equations. The novelty of such proof system is that it only manipulates equations and quantitative equations, rather than implications.\footnote{We expand the discussion on the novelties of our system with full technical details in Section \ref{comparison_section}.}  
 \item We show that, for any class of quantitative algebras defined by equations and quantitative equations, the free quantitative algebra generated by a fuzzy relation space $(A,d_A)$ always exists in the class, and we give an explicit construction.
\item We prove that the adjunction induced by the free construction above is \emph{strictly monadic}. Strict monadicity is a key property in the context of Universal Algebra, and the fact that it holds in our theory of quantitative algebras suggests that we have indeed identified an ``equational'' (in a categorical sense) quantitative setting.
\item We show that all monads on $\FRel$ which are liftings of finitary $\Set$ monads, i.e., $\Set$ monads with an equational presentation, 
can be presented by a given set of equations and quantitative equations. This includes most examples from the literature on quantitative algebras, e.g, the finite powerset monad with the Hausdorff metric and the probability distributions monad with the Kantorovich metric \cite{DBLP:conf/lics/MardarePP16}, among others (see, e.g., \cite{DBLP:conf/concur/MioV20, DBLP:conf/lics/MioSV21, DBLP:conf/lics/MioSV22}). 
\item\label{item:introFReltoGMet} We prove that all the results above, stated for the category $\FRel$ of fuzzy relation spaces, can be restricted to any chosen category $\GMet$ of generalised metric spaces (including, for example, the familiar category $\Met$ of metric spaces). 
\end{enumerate}

\subsection{Organisation of the Paper}

The rest of the paper is organised as follows.

In Section \ref{sec:background} we provide the necessary technical background regarding Universal Algebra, Category Theory and some basic notions regarding fuzzy relations. This section is relatively lengthy as we have included all notions
required to make this article self-contained. 
In Section \ref{framework:presentation:section} we formally introduce our theory of quantitative algebras with all its key definitions: quantitative algebras, equations and quantitative equations, their semantics, quantitative theories, and so on. In Subsection \ref{results_announcement_section} we enumerate our key results, which will be proved in the subsequent sections.
Our main results \ref{item:introDeductivesys}--\ref{item:introFReltoGMet} are formally stated and proved in Sections \ref{section:result:1}, \ref{section:result:2}, \ref{section:result:3}, \ref{section:lifting} and \ref{section:fromFRel_to_GMET}, respectively.

After having developed all our technical results, in Section \ref{comparison_section}
 we give a formal comparison \hl{of our theory of quantitative algebras, in three separate subsections, with} 

 \begin{itemize}
 \item \hl{the original framework introduced in  \cite{DBLP:conf/lics/MardarePP16} by  Mardare, Panangaden and Plotkin,}%
 
 \item \hl{the framework introduced in \cite{DBLP:conf/calco/FordMS21} by  Ford, Milius and Schröder, and }%
  \item \hl{the framework introduced in our earlier conference paper  \cite{DBLP:conf/lics/MioSV22}.}%
 \end{itemize}

Finally, we conclude in Section \ref{conclusion_section} suggesting possible \hl{future lines of work}.

\section{Technical Background}
\label{sec:background}
In this section we provide the mathematical background needed to formally state our results and to verify the proofs. 
In Section  \ref{background_universal_algebra} we cover material from Universal Algebra. In Section \ref{sec_intro_category} from category theory. And in Section \ref{sec:gen:metric:spaces} we give the necessary definitions regarding fuzzy relations and generalised metric spaces.

\subsection{Universal Algebra}
\label{background_universal_algebra}
We recall in this section some basic definitions of Universal Algebra. We refer the reader to \cite{DBLP:books/daglib/0067494} and \cite{Wechler1992UniversalAF}  as standard references, the latter is specifically intended for computer scientists.

A \emph{signature} $\Sigma$ is a (possibly infinite) set of function symbols $op
\in \Sigma$ each having a finite arity $ar(op)\in\mathbb{N}$. Operations of arity $0$ are referred to as constants.

\begin{defi}[$\Sigma$-algebra]\label{defn:sigalg}
Given a signature \hl{$\Sigma$}, a \emph{$\Sigma$-algebra} $\mathbb{A}$ is a pair of the form $\mathbb{A}=(A, \{ op^A\}_{op\in \Sigma})$, where
\begin{enumerate}
\item $A$ is a (possibly empty) set, and
\item $\{ op^A\}_{op\in \Sigma}$ is a collection of {interpretations of the operations} containing, for each function symbol in $\Sigma$, a function of type
$
op^A: A^{ar(op)}\rightarrow A
$.
\end{enumerate}
A \emph{homomorphism} between $\Sigma$-algebras $(A,\{ op^A\}_{op\in \Sigma})$ and $(B,\{ op^B\}_{op\in \Sigma})$ is a function $f:A\rightarrow B$ satisfying, for all  $a_1,\dots, a_n\in A$ and $n$-ary $op\in\Sigma$,
\[ f(op^A(a_1,\dots, a_n)) = op^B(f(a_1),\dots, f(a_n)).\]
We denote with $\Alg$ both the collection (a proper class) of all $\Sigma$-algebras and the category of $\Sigma$-algebras and their homomorphisms. 
\end{defi}

\begin{defi}[Terms over $\Sigma$]
Given a signature $\Sigma$ and a set $A$, we denote with $\TermsA$ the collection of all $\Sigma$-\emph{terms} built from $A$, i.e., the set inductively defined as follows:
$$
a\in \TermsA \ \ \ \ \ t_1,\dots, t_n \in \TermsA \Longrightarrow op(t_1,\dots, t_n)\in  \TermsA 
$$
for all $a\in A$ and $n$-ary $op\in \Sigma$. 
\end{defi}

The following definition follows the notational approach of \cite{Wechler1992UniversalAF}, denoting equations by ``$\forall A.s=t$'', where ``$\forall A.$'' explicitly indicates the set $A$ of variables involved.

\begin{defi}[Equations]\label{standard_equations_def}
Given a signature $\Sigma$, a \emph{$\Sigma$-equation} is a triple $(A,s,t)$ where $A$ is a set, \hl{such that $A\cap \Sigma=\emptyset$}, and $s,t\in\TermsA$. We write such triple as
$$
\forall A. s = t,
$$
and we denote with $\Eq$ the class of all $\Sigma$-equations.
\end{defi}

\begin{rem}\label{remark:infinitevariables1}
    \hl{In \cite{Wechler1992UniversalAF} and many other accounts of Universal Algebra, the set $A$ is assumed to be a finite subset of some fixed infinite set of variables. Our definition, instead, allows for a potentially infinite set $A$. This choice has no effect on the technical development and on expressiveness, in the context of Universal Algebra. However, as we discuss in Remark \ref{remark_inf_variables_2} later on, infinite sets of variables are used to develop our theory of quantitative algebras. Thus, our choice here is motivated by uniformity reasons.}
\end{rem}

In the rest of the paper, the signature $\Sigma$ will often be clear from the context and we will just talk about ``equations'' rather than $\Sigma$-equations. We use the letters $\phi,\psi$ to range over equations, and $\Phi,\Psi$ to range over classes of equations.

\begin{defi}[Interpretation]\label{interpretation_set}
Given a $\Sigma$-algebra $\mathbb{A}=(A, \{ op^A\}_{op\in \Sigma})$ and a set $B$, an \emph{interpretation of $B$ in $\mathbb{A}$} is a function
$
\imap: B\rightarrow A 
$. The interpretation $\imap$ extends to a function of type $\sem{\_}_\imap^A:\Terms{B}\rightarrow A$ defined inductively by
$$
\sem{b}_\imap^A = \imap(b)  \ \ \ \ \ \ \sem{op(t_1,\dots,t_n)}_\imap^A = op^A(\sem{t_1}_\imap^A,\dots, \sem{t_n}_\imap^A). 
$$
\end{defi}

The following definition gives semantics to equations and motivates the \hl{notation} ``$\forall (\_ ).$'' we adopted, which hints at the universal quantification over interpretations of the variables.

\begin{defi}[Semantics of equations]
Given a $\Sigma$-algebra $\mathbb{A}$ and an equation $\phi$ of the form $\forall B. s = t$, we say that $\mathbb{A}$ \emph{satisfies} $\phi$ (written
$
\mathbb{A}\models\phi$)
if, for all interpretations $\imap:B\rightarrow A$, it holds that
$
\sem{s}_\imap^A = \sem{t}_\imap^A 
$.
\end{defi}

\begin{defi}[Equational theory of a class of models]
Let $\mathcal{K}\subseteq \Alg$ be a class of $\Sigma$-algebras. The \emph{equational theory of $\mathcal{K}$} is defined as the class of equations satisfied by all $\Sigma$-algebras in $\mathcal{K}$, formally
$$
\Theory (\mathcal{K}) = \{ \phi \in \Eq \mid \forall \mathbb{A}\in \mathcal{K}.  \mathbb{A}\models \phi \}.
$$
\end{defi}

\begin{defi}[Models and equationally defined classes]
Let $\Phi\subseteq \Eq$ be a class of $\Sigma$-equations. The \emph{models} of $\Phi$ are the $\Sigma$-algebras that satisfy all equations in $\Phi$, formally
\[
\Models (\Phi) = \{ \mathbb{A}\in \Alg \mid \forall \phi \in \Phi.  \mathbb{A}\models \phi \}.
\]
If $\mathcal{K}\subseteq \Alg$ is such that $\mathcal{K}=\Models (\Phi)$ for some $\Phi\subseteq \Eq$, we say that $\mathcal{K}$ is an \emph{equationally defined} (by $\Phi$) class.
\end{defi}

\begin{defi}[Model theoretic entailment relation]\label{defn:entailment}
We define a binary relation\footnote{$\mathcal{P}(\Eq)$ denotes the collection of all classes of $\sig$-equations, i.e. all subclasses of $\Eq$. Therefore, it is a conglomerate in the sense of \cite[2.3]{AdamekACC}, and so is $\imply$.} ${\imply} \subseteq \mathcal{P}(\Eq)\times \Eq$ that describes how an equation $\phi$ can be a consequence of a class of equations $\Phi \subseteq \Eq$. It is defined by 
$$
\Phi \imply \phi \ \ \ \ \ \Longleftrightarrow   \ \ \ \ \  \phi \in \Theory(\Models(\Phi)).
$$
Therefore, the meaning of $\Phi \imply \phi$ is that any $\Sigma$-algebra that satisfies $\Phi$ (i.e., all the equations in $\Phi$) necessarily also satisfies the equation $\phi$.
\end{defi}

A fundamental result of Birkhoff establishes that $\imply$ coincides with the derivability relation $\Phi \derive \phi$ of the deductive system of ``equational logic'' (the relation $\derive$ is inductively defined, see, e.g., \cite[\S 3.2.4, Definition 8]{Wechler1992UniversalAF}). Thus, this celebrated result is a logical axiomatisation of the entailment relation $\imply$.

\subsection{Category Theory}
\label{sec_intro_category}

We assume basic knowledge on category theory and we recall here only some crucial definitions and results used in the rest of the paper. We refer to \cite{MacLane71} and \cite{Awodey} as standard references.

Let $\ModelsCat(\Phi)$ denote both the class of $\Sigma$-algebras satisfying $\Phi$ and the full subcategory of $\Alg$ whose objects are in $\Models(\Phi)$. In other words, $\ModelsCat(\Phi)$ is the category having as objects $\Sigma$-algebras $\mathbb{A}$ such that $\mathbb{A}\models \phi$, for all $\phi\in\Phi$, and as morphisms all their homomorphisms of $\Sigma$-algebras.
There is a forgetful functor
$
U_{\Catalg \rightarrow \Set}: \Catalg\rightarrow \Set 
$
mapping an algebra in $\ModelsCat(\Phi)$ to its carrier.
We often write $U$ when no confusion arises. 
\subsubsection{Monads and Adjunctions}

\begin{defi}[Monad]
\label{monad:main_definition}
Given a category $\catC$, a \emph{monad} on $\catC$ is a triple $(\mon, \eta, \mu)$ composed of a functor $\mon\colon\catC \rightarrow \catC$ together with two
natural transformations: a \emph{unit} $\eta\colon \id_{\catC}
\Rightarrow \mon$, where $\id_{\catC}$ is the identity functor on $\catC$, and a \emph{multiplication} $\mu \colon \mon^{2} \Rightarrow
\mon$, satisfying  %
$\mu \circ \eta_\mon = \mu \circ \mon\eta = \id_{\mon} $ and 
$  \mu\circ \mon\mu = \mu \circ\mu_\mon$.
\end{defi}
We often refer to a monad $(M,\eta,\mu)$ simply with its underlying functor $M$.

\begin{exa}
\label{ex_monad_termseq}
For any class $\Phi\subseteq \Eq$ of $\Sigma$-equations, we have an associated monad $(\terms{\Sigma,\Phi}, \eta,\mu)$ on $\Set$, defined as follows:
\begin{itemize}
    \item The functor $\terms{\Sigma,\Phi}$ maps a set $A$ to the set $\TermsA/_{\equiv}$ of terms over $A$ quotiented by the relation $\equiv$ defined as follows, for all $s,t \in \TermsA/_{\equiv}$,
    $$
    s\equiv t \ \ \ \Longleftrightarrow \ \ \ \Phi \Vvdash_{\Set }\forall A. s=t, 
    $$
    and maps a function $f:A \to B$ to $\terms{\Sigma,\Phi}(f)$ defined by induction on terms:
    $$\terms{\Sigma,\Phi}(f)([a]_\equiv)= [f(a)]_\equiv
    $$
    $$\terms{\Sigma,\Phi}(f)([op(t_1,...,t_n)]_\equiv)=op^{F(A)}(\terms{\Sigma,\Phi}(f)([t_1]_\equiv),...,\terms{\Sigma,\Phi}(f)([t_n]_\equiv)),
$$ 
where $op^{F(A)}$ is defined as $op^{F(A)}([t_1]_\equiv,...,[t_n]_\equiv)=[op(t_1,...,t_n)]_\equiv$ (the reason why we denote the interpretation of operations by $op^{F(A)}$ will become clear in \autoref{ex_freealgebra}).
    \item For each set $A$, the unit $\eta_{A}: A \to \TermsA/_{\equiv}$ sends $a$ to $[a]_{\equiv}$.

    \item For each set $A$, the multiplication $$\mu_{A}:\Terms{\TermsA/_{\equiv}}/_{\equiv}\rightarrow \TermsA/_{\equiv}$$
    is defined by the following ``flattening" operation:
$$
    [s([t_1]_{\equiv}, \dots, [t_n]_{\equiv})]_{\equiv} \mapsto [s\{t_1/[t_1]_{\equiv},\dots, t_n/[t_n]_{\equiv}\}]_{\equiv},
$$
where $s([t_1]_{\equiv}, \dots, [t_n]_{\equiv})$ denotes that $[t_1]_{\equiv}, \dots, [t_n]_{\equiv}$ are all and only the elements of $\TermsA/_{\equiv}$ appearing in the term $s$, and $s\{t_1/[t_1]_{\equiv},\dots, t_n/[t_n]_{\equiv}\}$ denotes the simultaneous substitution in $s$ of each of these equivalence classes with one representative.

\end{itemize}
It can be shown that, indeed, the above  definitions do not depend on specific choices of representatives of the $\equiv$-equivalence classes.
\end{exa}
A monad $\mon$ has an associated category of $\mon$-algebras.
\begin{defi}[Eilenberg--Moore algebra for a monad]\label{def:algebra-of-a-monad}
Let $(\mon,\unit,\mult)$ be a monad on $\Cat$. An \emph{algebra} for $\mon$ (or \emph{$\mon$-algebra}) is a pair $(A,\alpha)$ where $A\in\Cat$ is an object and $\alpha:\mon (A)\rightarrow A$ is a morphism such that (1) $ \alpha \circ  \unit_A = \id_A$ and (2) $\alpha\circ \mon \alpha= \alpha \circ \mult_A $ hold. An \emph{$\mon$-algebra morphism} between two $\mon$-algebras $(A,\alpha)$ and $(A^\prime,\alpha^\prime)$ is a morphism $f:A\rightarrow A^\prime$ in $\Cat$ such that
$f\circ \alpha = \alpha^\prime \circ \mon(f)$. The category of $\mon$-algebras and their morphisms, denoted by $\EM(\mon)$, is called the Eilenberg--Moore category for $\mon$. There is a forgetful functor $\EM(\mon) \rightarrow \Cat$ that forgets the algebra structures.
\end{defi}
\begin{defi}[Monad morphism]\label{defn:monadmorph}
    Let $(\mon,\unit,\mult)$ and $(\mon',\unit',\mult')$ be two monads on $\Cat$. A \emph{monad morphism} from $\mon$ to $\mon'$ is a natural transformation $\lambda: \mon \Rightarrow \mon'$ such that (1) $\lambda \circ \eta^{\mon} = \eta^{\mon'}$ and (2) $\lambda \circ \mu^{\mon} = \mu^{\mon'} \circ \lambda \mon' \circ \mon \lambda$. It is a \emph{monad isomorphism} whenever each component $\lambda_X: \mon X \to \mon'X$ is an isomorphism in $\Cat$.
\end{defi}

\begin{propC}[{\cite[Corollary of Theorem 6.3 in Chapter 3]{TTT}}]\label{prop:monadisocatiso}
    \hl{Let $(\mon,\unit,\mult)$ and $(\mon',\unit',\mult')$ be two monads on $\Cat$. There is a monad isomorphism $\mon \cong \mon'$ if and only if there is an isomorphism of categories $\EM(\mon') \cong \EM(\mon)$ that commutes with the forgetful functors to $\Cat$.}
\end{propC}

Monads can be defined as arising from adjunctions.\footnote{See, e.g., \cite[Chapter 9]{Awodey} for several equivalent definitions.}

\begin{propC}[{\cite[Proposition 10.3]{Awodey}}]
\label{prop_adjunction_monad}
    Every adjunction $F: \catC \to \catD \dashv U: \catD \to \catC$ defines a monad $(M,\eta, \mu)$ where
\begin{itemize}
    \item $M$ is the functor $U\circ F$,
    \item the unit $\eta:\id_\catC \Rightarrow M$ of the monad is the unit of the adjunction, and
    \item the multiplication $\mu:M^2\Rightarrow M$ is given by $\mu_X = U(\varepsilon_{F(X)})$, where $\varepsilon: F\circ U \to \id_\catD$ is the counit of the adjunction.
\end{itemize}
\end{propC}
As we discuss in the following section, the monad of quotiented terms from \autoref{ex_monad_termseq} arises from the adjunction between the forgetful functor $U: \Catalg \to \Set$ and the functor mapping sets to free objects in $\Catalg$.

\subsubsection{Free Objects}
\label{sec_free_background}

\begin{defi}[Free object]\label{free:object:def}
Let $U: \catD \to \catC$ be a functor, $X\in\catC$, $Y\in\catD$ and $\freemap: X\rightarrow U(Y)$. We say that $Y$ is a $U$-free object generated by $X$ with respect to $\freemap$ if the following \textbf{UMP} (Universal Mapping Property) holds: for every $B\in\catD$ and every $\catC$-morphism $f:X\rightarrow U(B)$, there exists a unique $\catD$-morphism $g:Y\rightarrow B$ such that $f=U(g)\circ \freemap$, as indicated in the following diagram. %
\begin{equation*}
\begin{tikzcd}[column sep=0.9em, row sep=0.9em]
	Y && X & {U(Y)} \\
	B &&& {U(B)}
	\arrow["{U(g)}", from=1-4, to=2-4]
	\arrow["g"', dashed, from=1-1, to=2-1]
	\arrow["\alpha", from=1-3, to=1-4]
	\arrow["f"', from=1-3, to=2-4]
\end{tikzcd}
\end{equation*}
We say that the category $\catD$ has $U$-free objects if for every $X\in\catC$ there exist an object $D_X\in\catD$ and a function $\freemap_X: X\rightarrow U(D_X)$ such that $D_X$ is a $U$-free object generated by $X$ with respect to $\freemap_X$. 
When they exist, $U$-free objects are unique up to isomorphism.
\end{defi}

If the functor $U$ and the map $\alpha_X$  are clear from the context, we just refer to ``the free object in $\catD$ generated by $X$'' instead of ``$U$-free object generated by $X$ with respect to $\alpha_X$''.

\begin{exa}
\label{ex_freealgebra}

It is a standard result in Universal Algebra that the forgetful functor $U: \Catalg\rightarrow \Set$  has $U$-free objects. Concretely, for any set $A$, take the algebra $F(A)=(\TermsA/_{\equiv}, \{op^{F(A)}\}_{op\in \Sigma})$, where $\TermsA/_{\equiv}$ and $op^{F(A)}$ are defined as in \autoref{ex_monad_termseq}.
Then $F(A)$ is the $U$-free object generated by  $A$ with respect to the function $\freemap: A \to F(A)$ that sends every $a \in A$ to $\alpha(a) = [a]_{\equiv}$.

\end{exa}

The following proposition states that if $U: \catD \to \catC$ is a functor such that  $\catD$ has 
$U$-free objects, then there is a functor $F$, called the free functor, which assigns to objects of $\catC$ the $U$-free object they generate,
and which gives an adjunction $F \dashv U$.

\begin{propC}[{\cite[\S IV.1, Theorem 2.(ii)]{MacLane71}}]
\label{prop_freeadjunction}
Let $U: \catD \to \catC$ be a functor such
that free $U$-objects exist in $\catD$, i.e., such that for every $X\in\catC$ there exist an object $D_X\in\catD$ and a function $\freemap_X: X\rightarrow U(D_X)$ such that $D_X$ is the $U$-free object generated by $X$ with respect to $\freemap_X$.
Then $U:\catD\rightarrow\catC$ has a left adjoint $F:\catC\rightarrow\catD$ ($F \dashv U$) with $F$ the functor mapping an object $X$ to the $U$-free object $D_X$ and mapping a morphism $f:X \to Y$ to the unique $\catD$-morphism $F (f): F(X) \to F (Y)$ that makes the 
following diagram commute:%
\begin{equation*}
\begin{tikzcd}[row sep=1.3em]
	{F(X)} && X & {U(F(X))} \\
	{F(Y)} &&& {U(F(Y))}
	\arrow["{U(F(f))}", from=1-4, to=2-4]
	\arrow["{F(f)}"', dashed, from=1-1, to=2-1]
	\arrow["{\alpha_X}", from=1-3, to=1-4]
	\arrow["{\alpha_Y \circ f}"', from=1-3, to=2-4]
\end{tikzcd}
\end{equation*}
\end{propC}

From this adjunction, we can build the monad of $U$-free objects using \autoref{prop_adjunction_monad}. 

\begin{exa}
\label{ex_monad_termseq_adjunction}
    In \autoref{ex_freealgebra} we have identified $U$-free objects, for $U: \Catalg \to \Set$. As explained in \autoref{prop_freeadjunction}, we can then define a functor $F$  such that we obtain an adjunction $F \dashv U$. By \autoref{prop_adjunction_monad}, we have a monad with underlying functor
    $U\circ F$, which is exactly the monad of quotiented terms $\terms{\sig,\Phi}$ from \autoref{ex_monad_termseq}. 
\end{exa}

\subsubsection{Strict Monadicity}

\begin{prop}[Comparison functor]
    Let $F: \catC \to \catD \dashv U: \catD \to \catC$ be an adjunction, and let $UF$ be the induced monad. Then there exists a functor 
    $K: \catD \to \mathbf{EM}(UF)$
    called the (canonical) comparison functor (see, e.g., \cite[\S VI.3, Theorem 1]{MacLane71} for its construction).
\end{prop}

There are interesting cases in which this comparison functor is an isomorphism. In such cases, we say that the functor is strictly monadic.

\begin{defi}[Strict monadicity]%
\label{monadic:functor:def}
Let $F: \catC \to \catD \dashv U: \catD \to \catC$ be an adjunction.
We say that the adjunction is \emph{strictly monadic} if the comparison functor is an isomorphism. 
Given a functor $U: \catD \to \catC$, we say that $U$ is strictly monadic if it has a left adjoint $F$ such that the adjunction is strictly monadic.
\end{defi}

The next theorem, due to Beck, gives useful equivalent characterisations of strict monadicity relying on coequalizers.

\begin{prop}[Beck's monadicity theorem]
\label{thm_beck_absolute}
Let $F: \catC \to \catD \dashv U: \catD \to \catC$ be an adjunction. The following are equivalent:
\begin{enumerate}
\item  $U$ is strictly monadic.
\item $U: \catD \to \catC$  strictly creates coequalizers for all $\catD$-arrows $f,g$ such that $U(f), U(g)$ has an absolute coequalizer (in $\catC$).
\item $U: \catD \to \catC$  strictly creates coequalizers for all $\catD$-arrows $f,g$ such that $U(f), U(g)$ has a split coequalizer (in $\catC$).
\end{enumerate}
where
\begin{itemize}
\item an absolute coequalizer (in $\catC$) of $\catC$-arrows $f,g: A \to B$ is a $\catC$-arrow $e: B \to C$ such that for all functors $F$, $F(e)$ is a coequalizer of $F(f), F(g)$.
\item a split coequalizer (in $\catC$) of $\catC$-arrows $f,g: A \to B$ is a $\catC$-arrow $e: B \to C$ such that $e \circ f=e\circ g$ and such that there exist arrows $s: C \to B$ and $t:B\to A$ such that $e \circ s= \id_C$, $f \circ t=\id_B$ and $g\circ t=s \circ e$.
\item $U: \catD \to \catC$ strictly creates coequalizers for the $\catD$-arrows $f,g$ if for any coequalizer $e: U(B) \to C$ of $U(f), U(g)$ (in $\catC$), there are unique $D$ and $u: B\to D$ (in $\catD$) such that $U(D)=C$, $U(u)=e$ and $u$ is a coequalizer of $f,g$.
\end{itemize}
\end{prop}
For a proof of \autoref{thm_beck_absolute}, see, e.g., \cite[\S VI.7, Theorem 1]{MacLane71}.\footnote{Our \autoref{monadic:functor:def} of \emph{strict monadicity} coincides with that used in  \cite[p.~143]{MacLane71}, where however it is just called \emph{monadicity}. 
We chose to use the adjective \emph{strict} as it has become standard terminology in recent literature, where \emph{monadicity} has a different meaning (see e.g. \cite[p.~167]{Riehl}).
}

The following result is well known and its proof, which indeed relies on the characterisations of strict monadicity given by Beck's theorem (\autoref{thm_beck_absolute}), can be found in 
\cite[\S VI.8, Theorem 1]{MacLane71}.

\begin{prop}
\label{prop_monadicity_alg}
For any signature $\Sigma$ and class $\Phi$ of equations over $\Sigma$, the functor $U:\Catalg\rightarrow \Set$ is strictly monadic.
\end{prop}

\hl{We need one last result on strict monadicity due to Bourn} \cite[Proposition 5]{bourn1992low} \hl{and sometimes called ``cancellability of monadicity''} (see also \cite[Corollary 5.6]{Arkor2023}).
\begin{prop}\label{thm:cancelmonadicity}
    In the diagram of adjunctions below, if $U$ and $U \circ U'$ are strictly monadic, then so is $U'$.
    \begin{equation*}
\begin{tikzcd}
	\catC & \catD & \catE
	\arrow[""{name=0, anchor=center, inner sep=0}, "{U'}"', shift right=2, tail reversed, no head, from=1-2, to=1-1]
	\arrow[""{name=1, anchor=center, inner sep=0}, "{F'}", shift left=2, from=1-2, to=1-1]
	\arrow[""{name=2, anchor=center, inner sep=0}, "U", shift left=2, from=1-2, to=1-3]
	\arrow[""{name=3, anchor=center, inner sep=0}, "F", shift left=2, from=1-3, to=1-2]
	\arrow["\dashv"{anchor=center, rotate=90}, draw=none, from=1, to=0]
	\arrow["\dashv"{anchor=center, rotate=90}, draw=none, from=3, to=2]
\end{tikzcd}
    \end{equation*}
\end{prop}

As recalled in \autoref{ex_monad_termseq_adjunction}, the monad $\terms{\sig,\Phi}$  arises from the adjunction $F \dashv U$, where $U:\Catalg\rightarrow \Set$ and $F$ is the functor mapping sets to $U$-free objects. Hence, \autoref{prop_monadicity_alg} allows us to conclude that the category $\EM(\terms{\sig,\Phi})$ of Eilenberg-Moore algebras for $\terms{\sig,\Phi}$ is isomorphic to the category $\Catalg$ of models of $\Phi$. \hl{Moreover, thanks to \autoref{prop:monadisocatiso}, if $\terms{\sig,\Phi}$ is isomorphic to a monad $M$, then there is an isomorphism $\EM(M) \cong \EM(\terms{\sig,\Phi}) \cong \Catalg$, and we can view $M$-algebras as the models of $\th$. This leads us to define monad presentations.}

\begin{defi}[$\Set$ presentation]\label{defn:setpres}
    A \emph{presentation} of a monad $(M,\eta,\mu)$ on $\Set$ \hl{is a class} of equations $\th \subseteq \Eq$ along with a monad isomorphism $\terms{\sig,\th} \cong M$.
\end{defi}

\begin{exa}\label{exmp:monadpres}
\hl{We give two main examples of monads on $\Set$ with a presentation.}
\begin{enumerate}
    \item \hl{The finite non-empty powerset monad $\mP: \Set \to \Set$} (see, e.g., \cite[Example 5.1.3]{Jacobs2016}) \hl{is presented by the theory of semilattices (see, e.g.,} \cite[p.~13]{BSVjournal22}) \hl{comprising a binary operation $\oplus$ and the following equations stating that $\oplus$ is idempotent, commutative, and associative:}
    \begin{equation}\label{eqn:thsemilattice}
        \forall x. x\oplus x = x\text{, } \forall x,y. x\oplus y = y \oplus x \text{, and } \forall x,y,z. (x\oplus y) \oplus z= x \oplus (y\oplus z).
    \end{equation}
    \item \hl{The finitely supported distributions monad $\mD: \Set \to \Set$ (see, e.g.,} \cite[Eq.~4]{Jacobs10}\hl{) is presented by the theory of convex algebras (see, e.g.,} \cite[Theorem 4]{Jacobs10}\hl{) comprising a binary operation $+_p$ for every $p \in (0,1)$ and the following equations for every $p,q \in (0,1)$:}
    \begin{equation}\label{eqn:thconvexalg}
        \forall x. x+_p x= x \text{, }
        \forall x,y.x+_p y= y+_{1-p} x\text{, and }
        \forall x,y,z. (x+_qy)+_pz = x+_{pq}(y+_{\frac{p(1-q)}{1-pq}}z).
    \end{equation}
\end{enumerate}
\end{exa}

\subsection{Fuzzy Relations and Generalised Metric Spaces}\label{sec:gen:metric:spaces}

We define here fuzzy relation spaces, which are sets equipped with a $[0,1]$-valued function (see, e.g., \cite{Zadeh71}).

\begin{defi}\label{definition_of_fuzzy_relation}
 A \emph{fuzzy relation} on a set $A$ is a map $d: A\times A \rightarrow [0,1]$. The pair $(A,d_A)$ is called a fuzzy relation space (often, we directly call $(A,d_A)$ a fuzzy relation \hl{or $\FRel$ space} as well). A morphism between two fuzzy relation spaces $(A,d_A)$ and $(B,d_B)$ is a map $f: A \rightarrow B$ which is \emph{nonexpansive}, namely, \hl{for all $a,a' \in A$},
$$ d_B(f(a),f(a'))\leq d_A(a,a').$$ We denote by $\FRel$ the category of fuzzy relation spaces and nonexpansive maps.
\end{defi}

We denote with $U_{\FRel\rightarrow \Set}: \FRel\rightarrow \Set$ the forgetful functor defined as expected, and with $U$ when no confusion arises. We denote \hl{with $D$} %
the \emph{discrete functor} mapping a set $A\in\Set$ to the discrete fuzzy relation $(A,d^A_\bot)$ defined \hl{by letting $d^A_\bot(a,a') =1$ for all $a,a' \in A$,} 
and acting as identity on morphisms $f:A\rightarrow B \in \Set$. We indeed note that $D(f):(A,d^A_\bot)\rightarrow (B,d^B_\bot)$ is always nonexpansive, given the definition of $d^A_\bot$. 

\begin{prop}
\label{prop:frel_discrete_adjoint}
    The functor $D$ is left adjoint to $U$, that is, $D\dashv U$.
\end{prop}
This yields an isomorphism $\FRel(DA, (B,d_B)) \cong \Set(A,U(B,d_B))$, so every function between a set $A$ and $U(B,d_B)$ is nonexpansive when $A$ is endowed with the discrete fuzzy relation.

We will also be interested in full subcategories of $\FRel$ obtained by restricting to fuzzy relations that satisfy certain constraints, expressed by means of universally quantified logical implications in the language of first-order logic. \hl{These are often referred to as Horn sentences. We call them \emph{$\folang$-implications} to avoid any confusion or overloading with similar (but distinct) concepts appeared in the literature around quantitative algebras (see, e.g., discussion in Section \ref{subsec:fmscomp}).}

\begin{defi}[$\folang$-implications]\label{def_Limplications}
    Let $\folang$ be the language of first-order logic with the equality binary predicate ($\_=\_$) and with, for each $\epsilon \in [0,1]$, \hl{a  
    binary} predicate ($d(\_,\_)\leq \epsilon$). We call \emph{$\folang$-implications} all closed formulas $H$ of this language  
    that have the following shape: 
$$
H= \quad  \forall x_1,\dots, x_n. \Big( \big( \displaystyle \bigwedge_{1\leq i\leq k} G_i \big) \Rightarrow F \Big),
$$
where the subformulas $G_i$ and $F$ are atomic, i.e., $G_i$ and $F$ are either of the form $(x = x')$ or $(d(x,x')\leq \epsilon)$, for some  $\epsilon\in[0,1]$ and for $x,x'\in \{x_1,...,x_n\}$.
\end{defi}

Such formulas are interpreted on fuzzy relations $(A,d_A)$ as in standard first-order logic, with equality  ($\_=\_$) being the identity relation on $A$, and the binary predicate $d(\_,\_)\leq \epsilon$ holding true whenever $d_A$ assigns distance less than or equal to $\epsilon$.

\begin{defi}[Semantics of $\folang$-implications]
    Given a fuzzy relation $(A,d_A)$ and an $\folang$-implication $H$ of the form described in \autoref{def_Limplications}, we say that $(A,d_A)$ satisfies $H$ (notation: $(A,d_A)\fomodels H$) if for all functions $\Hmap: \{x_1,..., x_n\}\to A$, 
    \begin{center}
        if $(A,d_A)\fomodels_{\Hmap} G_i$ for all $1\leq i\leq k$, then $(A,d_A)\fomodels_{\Hmap} F$,
    \end{center}
    where $(A,d_A)\fomodels_{\Hmap} x=x'$ holds if $\Hmap(x)=\Hmap(x')$, and $(A,d_A)\fomodels_{\Hmap} d(x,x')\leq \epsilon$ holds if $d_A(\Hmap(x),\Hmap(x'))\leq \epsilon$.

    Given a (possibly infinite) set $\mathcal{H}$ of $\folang$-implications, we say that $(A,d_A)$ satisfies $\mathcal{H}$, written $(A,d_A)\fomodels \mathcal{H}$, if for all $H\in\mathcal{H}$, $(A,d_A)$ satisfies $H$.
\end{defi}

Consider, for example, the following useful $\folang$-implications $H$:
\begin{align}
\forall x.\quad  && x=x &\implies d(x,x) \leq 0 &&\label{eq:refl}\\
    \forall x,y.\quad  && d(x,y) \leq 0 &\implies x=y&&\label{eq:idofind}\\
    \forall x,y. \quad && d(x,y) \leq \epsilon &\implies d(y,x) \leq \epsilon &&\label{eq:symm}\\
    \forall x,y,z. \quad  && 
    d(x,y)\leq \epsilon  \ \wedge \ d(y,z)\leq \delta &\implies 
    d(x,z)\leq \gamma \ \ \ \textnormal{(where $\gamma = \epsilon + \delta$)} &&\label{eq:trineq}
\end{align}
and the set $\mathcal{H}_{\Met}$ consisting of all instances (for all $\epsilon,\delta \in [0,1]$) of these $\folang$-implications:
\[
\mathcal{H}_{\Met} = \{ \ref{eq:refl}, \ref{eq:idofind}, \ref{eq:symm}, \ref{eq:trineq}   \}.
\]
It is easy to see that $(A,d_A)\fomodels \mathcal{H}_{\Met}$ if and only if the fuzzy relation  $d_A$ is a metric. Indeed (1) expresses that each point is at distance zero from itself, (2) states that points at distance zero must be equal, (3) expresses symmetry ($d_A(a,b)=d_A(b,a)$) of the fuzzy relation and (4) expresses the triangular inequality property. Similarly, the subset
$
\mathcal{H}_{\mathbf{PMet}} = \{ \ref{eq:refl},  \ref{eq:symm}, \ref{eq:trineq}   \} 
$
is satisfied exactly by the fuzzy relations $(A,d_A)$ such that $d_A$ is a pseudo-metric \cite{DBLP:journals/tcs/BonsangueBR98}. In the literature, many other generalisations of metrics are defined as fuzzy relations satisfying a list of axioms expressible with $\folang$-implications. 
Important examples include: quasimetrics \cite{Wilson1931quasi}, ultrametrics \cite{DBLP:journals/tcs/BonsangueBR98}, semimetrics \cite{Wilson1931semi}, dislocated metrics \cite{Hitzler2000} also called diffuse metrics in \cite{Castro2021}, rectangular metrics \cite{Branciari2000}, and $b$-metrics \cite{Czerwik1993}.

\begin{defi}[$\GMet$ categories]\label{gmet_category_def}
Given a collection $\Hset$ of $\folang$-implications, we denote with $\HGMet$ (or just $\GMet$ if $\Hset$ is clear from the context or abstracted away) the full subcategory of $\FRel$ whose objects are fuzzy relations $(A,d_A)$ such that $(A,d_A)\fomodels \mathcal{H}$ and whose morphisms are all the nonexpansive maps between them. We call objects of $\GMet$ \emph{generalised metric spaces} \hl{or $\GMet$ spaces}.
\end{defi}

Note that, in accordance with the above definition, we have that $\FRel = \GMet_{\emptyset}$, i.e., $\FRel$ is the special case of $\Hset$ being empty.
Given its importance, we reserve the symbol $\Met$ for the category of metric spaces and nonexpansive maps, i.e., $\Met = \GMet_{\mathcal{H}_{\Met}}$.

Given any $\GMet$ category, we denote with $U_{\GMet\rightarrow \Set}: \GMet\rightarrow \Set$ the forgetful functor defined as the restriction of $U_{\FRel\rightarrow\Set}$ to $\GMet$, which we simply denote by $U$ when no confusion arises.
\begin{rem}
\label{rem_Gmet_meaning}
The terminology ``generalised metric space'' has appeared in the literature with different meanings. For instance, in \cite{DBLP:journals/tcs/BonsangueBR98}, generalised metric spaces are fuzzy relations satisfying reflexivity \eqref{eq:refl} and triangular inequality \eqref{eq:trineq}. Our definition is thereby a further generalisation, which also covers as special cases the spaces considered in \cite{DBLP:journals/tcs/BonsangueBR98}.
\end{rem}
\begin{exa}\label{exmp:monadsgmet}
    \hl{Here are several examples of monads on $\GMet$ categories.}
    \begin{enumerate}
        \item \hl{On $\FRel$, we can define a monad $\mP': \FRel \to \FRel$ inspired by the powerset monad $\mP$ from \autoref{exmp:monadpres}. Given an $\FRel$ space $(X, d)$, we define $\mP'(X,d)$ to have the carrier $\mP X$ and the fuzzy relation }
        \[d': \mP X \times \mP X \rightarrow [0,1], \quad d'(S,T) = \begin{cases}d(x,y) & \text{if }S = \{x\} \text{ and } T = \{y\}\\1 &\text{otherwise} \end{cases},\]
        \hl{and for any nonexpansive map $f: (X,d_X) \rightarrow (Y,d_Y)$, we let the underlying function of $\mP'f$ be $\mP f$, which is nonexpansive whenever $f$ is. The unit and multiplication are also defined as those of $\mP$ which makes it easy to verify their naturality and the other monad laws (they hold in $\Set$, so they must hold in $\FRel$ because $U_{\FRel \to \Set}$ is faithful).}
        \item \hl{On $\Met$, there is a similar monad sending a metric $(X,d)$ to $\mP X$ with the Hausdorff metric lifting of $d$ (see, e.g.,} \cite[Definition 18]{DBLP:conf/concur/MioV20}\hl{). There is also an analog to the distributions monad $\mD$, which sends $(X,d)$ to $\mD X$ equipped with the Kantorovich metric lifting of $d$ (see, e.g., }\cite[Definition 20]{DBLP:conf/concur/MioV20}).
        \item In \cite[\S 5.3]{DBLP:conf/lics/MioSV22}, \hl{we studied $\LK{\mD}$, another variant of $\mD$ on the category of diffuse metric spaces} \cite[Definition 4.9]{Castro2021},\hl{ namely, fuzzy relations satisfying the $\folang$-implications for symmetry \eqref{eq:symm} and triangle inequality \eqref{eq:trineq}. It sends a diffuse metric space $(X,d)$ to $\LK{\mD}(X,d)= (\mD X, \LK{d})$, which is $\mD X$ equipped with the \L ukaszyk–Karmowski (\L K for short) distance $\LK{d}$ defined on all $\Delta_1,\Delta_2 \in \mD X$ by}
        \begin{equation}\label{eqn:LKdefn}
            \LK{d}(\Delta_1,\Delta_2) = \textstyle{\sum}_{x,x' \in X} \Delta_1(x)\Delta_2(x')d(x,x').
        \end{equation}
        \hl{It is a consequence of} \cite[Theorem 5.6]{DBLP:conf/lics/MioSV22} \hl{that $\LK{\mD}$ is a monad with the action on functions, the unit, and the multiplication all defined as those of $\mD$.}
    \end{enumerate}
\end{exa}

\section{Presentation of the Framework and Results}\label{framework:presentation:section}

In this section we present our framework of Universal Quantitative Algebra. We will introduce it following the same pattern as in the background Section \ref{background_universal_algebra} on Universal Algebra. 
We begin with the central notion of this section, the concept of quantitative algebra.

\begin{defi}[Quantitative Algebra]\label{defi_quantitative_algebra}
Given a signature $\Sigma$, an \emph{$\FRel$ quantitative $\Sigma$-algebra} $\mathbb{A}$, or just a \emph{quantitative algebra} for short, is a triple $\mathbb{A}=(A,d_A, \{op^A\}_{op\in\Sigma})$ where
\begin{itemize}
    \item $(A,d_A)$ is \hl{an $\FRel$} space \hl{(\autoref{definition_of_fuzzy_relation})}, and %
    \item $(A,\{op^A\}_{op\in\Sigma})$ is a $\Sigma$-algebra \hl{(\autoref{defn:sigalg})}. %
    \end{itemize}
\end{defi}

\begin{rem}
Note that, in contrast with the definition in \cite{DBLP:conf/lics/MardarePP16} (and with much subsequent literature \cite{DBLP:conf/lics/MardarePP17,DBLP:conf/concur/MioV20,DBLP:conf/calco/BacciMPP21,DBLP:conf/lics/MardarePP21}), under our definition the distance $d_A$ is not required to satisfy the axioms of metric spaces, as it can be an arbitrary fuzzy relation, and the interpretations $op^A$ of the operations in $\Sigma$ are not required to be nonexpansive and can be arbitrary functions. See Section \ref{subsec:MPPcomp} for a more detailed comparison.
\end{rem}

\begin{defi}[Homomorphisms]
Given a signature $\Sigma$  and quantitative algebras 
\begin{center}
    $\mathbb{A}=(A,d_A, \{op^A\}_{op\in\Sigma})$ \ \ \  $\mathbb{B}=(B,d_B, \{op^B\}_{op\in\Sigma})$,
\end{center}
a \emph{homomorphism} (of quantitative algebras) is a function $f:A\rightarrow B$ such that
\begin{itemize}
\item $f:(A,d_A)\rightarrow (B,d_B)$ is \hl{nonexpansive} \hl{(Definition \ref{definition_of_fuzzy_relation}), and}%
\item $f$ is a homomorphism between $(A, \{op^A\}_{op\in\Sigma})$ and $(B, \{op^B\}_{op\in\Sigma})$ \hl{(see \autoref{defn:sigalg})}.%
\end{itemize}
\hl{An \emph{isomorphism} of quantitative algebras is, as usual, a homomorphism that has an inverse homomorphism. It can equivalently be defined as a bijective homomorphism that is an \emph{isometry} (i.e., that preserves distances: $d_B(f(a),f(a')) = d_A(a,a')$).}
\end{defi}
We denote with $\QalgFRel$, or often just $\Qalg$,  the category of $\FRel$ quantitative $\Sigma$-algebras and their homomorphisms. We denote with $U_{\Qalg\rightarrow \FRel}$ and $U_{\Qalg\rightarrow \Alg}$ the forgetful functors defined as expected.%

\begin{defi}[Equations and quantitative equations]
\label{quantitative_equation_def}
An \emph{$\FRel$ $\Sigma$-equation}, or just an \emph{equation} for short, is a judgment of the form
\[
\forall (A,d_A). s = t,
\]
where $(A,d_A)$ is \hl{an $\FRel$} space, \hl{with $A\cap \Sigma=\emptyset$}, and $s,t\in\TermsA$.
An \emph{$\FRel$ quantitative $\Sigma$-equation}, or just a \emph{quantitative equation} for short, is a judgment of the form
\[
\forall (A,d_A). s =_\epsilon t, \]
where $(A,d_A)$ is \hl{an $\FRel$} space, \hl{with $A\cap \Sigma=\emptyset$}, $s,t\in\TermsA$, and $\epsilon \in [0,1]$.

We use the letters $\phi,\psi$ to range over equations and quantitative equations, and we denote  with $\MEq(\Sigma)$, the proper class of all $\FRel$ $\Sigma$-equations and quantitative $\Sigma$-equations.

\end{defi}

\begin{defi}[Interpretations]\label{interpretation_met}
Given \hl{an $\FRel$} quantitative $\Sigma$-algebra 
$\mathbb{A}$ \hl{with underlying fuzzy relation $(A,d_A)$} 
and \hl{an $\FRel$} space $(B,d_B)$, an \emph{interpretation} of $(B,d_B)$ in $\mathbb{A}$ is a nonexpansive function
$
\imap: (B,d_B) \rightarrow (A,d_A)
$. The interpretation $\imap$ extends uniquely to a function of type $\sem{\_}_\imap^A:\Terms{B}\rightarrow A$ specified as in \autoref{interpretation_set}.
\end{defi}

In accordance with the above definition, all \hl{interpretations in a quantitative algebra} are nonexpansive. While this prevents any confusion, we will sometimes stress \hl{that fact} 
as this is often crucial in some statements and proofs.

\begin{defi}[Semantics of equations and quantitative equations]\label{semantics_quantitative_eq}
Let $\mathbb{A}=(A,d_A, \{op^A\})$ be \hl{an $\FRel$} quantitative $\Sigma$-algebra. 
Let $\phi_1$ and $\phi_2$ be the following $\FRel$ $\Sigma$-equation and quantitative $\Sigma$-equation:
\begin{center}
    $\phi_1 =\forall (B,d_B). s = t$ \  \ \ \ \ \ \ $\phi_2 =\forall (B,d_B). s =_\epsilon t$.
\end{center}
We say that  $\mathbb{A}$ \emph{satisfies} $\phi_1$, written $\mathbb{A}\models \phi_1$, if for all nonexpansive interpretations $
\imap:(B,d_B)\rightarrow (A,d_A)$ of $(B,d_B)$ in $\mathbb{A}$, $\sem{s}_\imap^A =  \sem{t}_\imap^A$ holds. Similarly, we say that  $\mathbb{A}$ satisfies $\phi_2$, written $\mathbb{A}\models \phi_2$, if for all nonexpansive interpretations $\imap:(B,d_B)\rightarrow (A,d_A)$ of $(B,d_B)$ in $\mathbb{A}$,
$d_A(\sem{s}_\imap^A, \sem{t}_\imap^A)\leq \epsilon $ holds.
\end{defi}

\begin{rem}\label{remark_inf_variables_2}
\hl{We highlight (cf.~\autoref{remark:infinitevariables1}) that the $\FRel$ space $(A,d_A)$, in $\FRel$ equations and quantitative equations, is not required to be finite and can have any cardinality.}

\hl{This flexibility is not necessary in Universal Algebra because an equation $\forall A.s=t$ is always equivalent (i.e., satisfied by the same algebras) to the equation $\forall V.s=t$, where $V$ is the set of variables occurring in $s$ and $t$, which is finite. In contrast, an $\FRel$ equation $\forall (A,d_A).s=t$ is in general not equivalent to $\forall (V,d_V).s=t$, where $V$ is the set of variables occurring in $s$ and $t$, and $d_V$ is $d_A$ restricted to $V\times V$.}

\hl{For instance, take $A = \{x_r \mid r \in [0,1]\}$ with the Euclidean metric ($d_A(x_r,x_s) = |r-s|$), and the $\FRel$ equation $\forall (A,d_A).x_0 = x_1$. Then, take the $\FRel$ equation $\forall (V,d_V).x_0 = x_1$ where $V$ contains two variables $x_0$ and $x_1$, at distance $1$ from each other and $0$ from themselves. These two $\FRel$ equations are not equivalent. To see this, take the empty signature $\Sigma = \emptyset$ and consider the quantitative $\Sigma$-algebra $\mathbb{V}=(V,d_V,\emptyset)$. We have that $\mathbb{V}$ does not satisfy $\forall (V,d_V).x_0=x_1$, since the identity map $\id_V$ is a nonexpansive interpretation of $(V,d_V)$ in $\mathbb{V}$ with $\id_V(x_0) \neq \id_V(x_1)$. In contrast, $\mathbb{V}$ satisfies $\forall (A,d_A).x_0 = x_1$ because any nonexpansive map $\tau: (A,d_A) \rightarrow (V,d_V)$ must be constant, yielding $\tau(x_0) = \tau(x_1)$. Indeed, for any two $r,s \in (0,1)$, $d_A(x_r,x_s) < 1$ means that $\tau$ must send $x_r$ and $x_s$ to the same point $x \in \mathbb{V}$ in order to be nonexpansive, and, since $d_A(x_0,x_{\sfrac{1}{2}}) =  d_A(x_{\sfrac{1}{2}},x_1) < 1$, $\tau$ must send $x_0$ and $x_1$ to $\tau(x_{\sfrac{1}{2}}) = x$ as well.} 

\hl{We do not know whether taking equations and quantitative equations with finitely many variables is enough to express everything which can be expressed with infinitely many variables. We formally state this open problem in Section \ref{conclusion_section}.}
\end{rem}

\begin{defi}[Quantitative equational theory of a class of models]
Let $\mathcal{K}\subseteq \Qalg$ be a class of $\FRel$ quantitative $\Sigma$-algebras. The \emph{quantitative equational theory} of $\mathcal{K}$ is defined as the class of $\FRel$ $\Sigma$-equations and quantitative $\Sigma$-equations satisfied by all quantitative algebras in $\mathcal{K}$, formally
\begin{center}
    $\QTheory (\mathcal{K}) = \{ \phi \in \MEq(\Sigma) \mid \forall \mathbb{A}\in \mathcal{K}.\  \mathbb{A}\models \phi \}$.
\end{center}
\end{defi}

\begin{defi}[Models and quantitative equationally defined classes]\label{eq_def_class:def}
Let $\Phi\subseteq \MEq(\Sigma)$ be a class of $\FRel$ $\Sigma$-equations and quantitative $\Sigma$-equations. The \emph{models} of $\Phi$ are the quantitative algebras that satisfy all equations and quantitative equations in $\Phi$, formally
\begin{center}
    $\QModels (\Phi) = \{ \mathbb{A}\in \Qalg \mid \forall \phi \in \Phi.\  \mathbb{A}\models \phi \}$.
\end{center}
If $\mathcal{K}\subseteq \Qalg $ is such that $\mathcal{K}=\QModels (\Phi)$ for some $\Phi\subseteq \MEq(\Sigma)$, we say that $\mathcal{K}$ is a quantitative equationally defined (by $\Phi$) class. 
\end{defi}

Note that, following the above definition, we simply have $\Qalg = \QModelsCat(\emptyset)$. With some abuse of notation, we also denote with $\QModelsCat(\Phi)$ the full subcategory of $\Qalg$ whose objects are in $\QModelsCat(\Phi)$.  In other words, $\QModelsCat(\Phi)$ is the category having as objects quantitative $\Sigma$-algebras $\mathbb{A}$ such that $\mathbb{A}\models \phi$, for all $\phi\in\Phi$, and as morphisms all their homomorphisms of quantitative algebras.

We denote with $U_{\QModelsCat(\Phi)\rightarrow \FRel}$ the forgetful functor defined as the restriction of $U_{\Qalg\rightarrow \FRel}$, \hl{or, equivalently, the composite}
$\QModelsCat(\Phi) \hookrightarrow \Qalg \xrightarrow{U} \FRel$.
As usual, this is most often just denoted by $U$ when no confusion arises.

\begin{exa}
    \hl{To give a quantitative equationally defined class of models, we can convert the equations defining semilattices from \autoref{exmp:monadpres} to $\FRel$}\hl{ equations by putting the discrete fuzzy relation $d_\bot$ (defined before \autoref{prop:frel_discrete_adjoint}) on the set of variables. We have}
    \begin{equation}\label{eqn:qthsemilattice}
    \Phi = \begin{Bmatrix}
        \forall (\{x\},d_\bot). x\oplus x = x\text{, } \forall (\{x,y\},d_\bot). x\oplus y = y \oplus x \text{, and }\\
        \forall (\{x,y,z\},d_\bot). (x\oplus y) \oplus z= x \oplus (y\oplus z)
    \end{Bmatrix}.
    \end{equation}
    \hl{Since any function from a discrete space into a quantitative algebra $\mathbb{A}$ is nonexpansive, $\mathbb{A}$ satisfies each of these equations if and only if the underlying (non-quantitative) algebra $U_{\Qalg \to \Alg}\mathbb{A}$ satisfies those of \eqref{eqn:thsemilattice}. Therefore, $\QModelsCat(\Phi)$ is the category of semilattices equipped with a fuzzy relation, with morphisms being nonexpansive semilattice homomorphisms.  Note that the equations in $\Phi$ do not impose any relationship between the semilattice structure and the fuzzy relation. As an example of such a relationship, following \cite{DBLP:conf/lics/MardarePP16}, we consider the requirement that the join operation $\oplus$ is nonexpansive. This can be expressed by the following set of quantitative equations, one for each pair $\epsilon,\epsilon'\in[0,1]$, where $X = \{x,y,x',y'\}$, $d(x,y) = \epsilon$, $d(x',y') = \epsilon'$, and all other distances are $1$:}
    \begin{equation}\label{eqn:oplusnexp}
        \forall (X,d).x\oplus x' =_{\max\{\epsilon,\epsilon'\}} y\oplus y'.
    \end{equation}
\end{exa}
\begin{exa}
    \hl{For the signature of convex algebras $\Sigma = \{+_p \mid p \in (0,1)\}$ in \autoref{exmp:monadpres}, the nonexpansiveness requirement is imposed by quantitative equations like those of \eqref{eqn:oplusnexp}, replacing $\oplus$ with $+_p$ for each $p \in (0,1)$ and $\epsilon,\epsilon' \in [0,1]$:}
    \begin{equation}\label{eqn:pluspnexp}
        \forall (X,d).x+_p x' =_{\max\{\epsilon,\epsilon'\}} y+_p y'.
    \end{equation}
    \hl{The construction of the monad $\LK{\mD}$ in \autoref{exmp:monadsgmet}.(3) relies on} \cite[Theorem 5.6]{DBLP:conf/lics/MioSV22} \hl{which characterises the quantitative algebra on $\LK{\mD}(X,d)$ where $+_p$ is interpreted as the convex combination $(\Delta_1,\Delta_2) \mapsto p\Delta_1+\bar{p}\Delta_2$, where $\bar{p} = (1-p)$. This operation is not nonexpansive in the sense of \cite{DBLP:conf/lics/MardarePP16}, so this quantitative algebra does not satisfy the quantitative equations in \eqref{eqn:pluspnexp}. For instance, taking $\Delta_1$ and $\Delta_2$ to be Dirac distributions at $x$ and $y$ respectively, and letting $d(x,x) = d(y,y) = 0$ and $d(x,y) = d(y,x) = 1$, we have}
    \[\LK{d}(p\Delta_1+\bar{p}\Delta_2, p\Delta_1+\bar{p}\Delta_2) \stackrel{\eqref{eqn:LKdefn}}{=} 2p(1-p) \not\leq 0 \stackrel{\eqref{eqn:LKdefn}}{=} \max\{\LK{d}(\Delta_1,\Delta_1),\LK{d}(\Delta_2,\Delta_2)\}.\]

    \hl{This illustrative example shows that sometimes useful operations, acting on spaces equipped with distances, fail to be nonexpansive. Our framework can handle such situations, unlike the original framework of \cite{DBLP:conf/lics/MardarePP16} where nonexpansiveness is required.}
\end{exa}

\begin{defi}[Model theoretic entailment relation]
Let $\Phi\subseteq \MEq(\Sigma)$ be a class of $\FRel$ $\Sigma$-equations and quantitative $\Sigma$-equations. We define a binary (consequence) relation\footnote{As in \autoref{defn:entailment}, both $\mathcal{P}(\MEq(\sig))$ and $\implyfrelfrel$ are conglomerates in the sense of \cite[2.3]{AdamekACC}.}
${ \implyfrelfrel}
 \subseteq \mathcal{P}(\MEq(\Sigma))\times \MEq(\Sigma)$ (or just $\implyfrel$ for short) as follows:
$$
\Phi \implyfrelfrel \phi \ \ \ \ \ \Longleftrightarrow   \ \ \ \ \  \phi \in \QTheory(\QModels(\Phi)).
$$
Thus, the meaning of $\Phi \implyfrelfrel \phi$ is that any $\FRel$ quantitative $\Sigma$-algebra that satisfies $\Phi$ (i.e., all the $\FRel$ equations and quantitative equations in $\Phi$) necessarily also satisfies $\phi$.
\end{defi}

We summarize here the introduced notions with their analogs from Universal Algebra:
\begin{center}
\begin{tabular}{@{}m{0.35\textwidth}m{0.52\textwidth}@{}}
\toprule
Universal Algebra & Universal Quantitative Algebra \\ \toprule
$\Sigma$-algebra $(A,\{op^A\}_{op\in \Sigma})$  &
Quantitative $\Sigma$-algebra $(A, d_A, \{op^A\}_{op\in \Sigma})$\\\midrule[0.1pt]
Homomorphism of $\Sigma$-algebras  &
(Nonexpansive) homomorphism of\newline quantitative $\Sigma$-algebras \\\midrule[0.1pt]
Category $\Alg$ of $\Sigma$-algebras  &
Category $\Qalg$ of quantitative $\Sigma$-algebras \\\midrule[0.1pt]
$\Sigma$-equation $\forall A. s=t$  &
$\Sigma$-equation $\forall (A,d_A).s=t$ and \newline quantitative $\Sigma$-equation $\forall (A,d_A). s=_\epsilon t$  \\\midrule[0.1pt]
Interpretation of $\Sigma$-equations &
(Nonexpansive) interpretation of $\Sigma$-equations and of quantitative $\Sigma$-equations  \\\midrule[0.1pt]
Category $\Models(\Phi)$ of \newline models of $\Phi\subseteq\Eq$ &
Category $\QModels(\Phi)$ of \newline models of $\Phi\subseteq\MEq(\Sigma)$ \\\midrule[0.1pt]
Equational theory $\Theory (\mathcal{K})$& Quantitative equational theory $\QTheory (\mathcal{K})$ \\\midrule[0.1pt]
Equationally defined class of\newline  $\sig$-algebras $\mathcal{K} = \Models(\Phi)$ & Quantitative equationally defined class of \newline quantitative $\sig$-algebras $\mathcal{K} = \QModels(\Phi)$\\\midrule[0.1pt]
Entailment relation $\imply$ & Entailment relation $\implyfrelfrel$\\ \bottomrule
\end{tabular}
\end{center}

\subsection{Summary of Contributions}
\label{results_announcement_section}

We now give an overview of the main results that we will prove in the following sections.

\begin{enumerate}[(I)]
\item \label{main_result_number_1}
The entailment relation $\implyfrel_{\FRel}$ can be axiomatised by means of a deductive system analogous to the deductive system of Birkhoff's equational logic. More formally, there is an inductively  defined relation ${\derivefrel_{\FRel}}\subseteq \mathcal{P}(\MEq(\Sigma))\times  \MEq(\Sigma)$, specified as the smallest relation containing a given set of pairs and closed under a given set of deductive rules (see Section \ref{section:result:1} for details), which is sound and complete with respect to $\implyfrel_{\FRel}$, i.e., for all $\Phi\subseteq\MEq(\Sigma)$ and  $\phi\in\MEq(\Sigma)$,
\[
  \Phi \derivefrel_{\FRel} \phi \Longleftrightarrow\Phi \implyfrel_{\FRel} \phi .
\]
The soundness of the deductive system (i.e., the implication
$\Phi \derivefrel_{\FRel} \phi \Rightarrow\Phi \implyfrel_{\FRel} \phi$) is proved in Section \ref{section:result:1}.
The completeness (i.e., the implication
$\Phi \implyfrel_{\FRel} \phi \Rightarrow\Phi \derivefrel_{\FRel} \phi $) is proved in Section \ref{subsub4:proofs}, as a consequence of our second result below \ref{main_result_number_2}.

We recall, from the introduction, that our deductive system $\derivefrel_{\FRel}$  has significant differences with the one presented \hl{in \cite{DBLP:conf/lics/MardarePP16}. We give more details in Section \ref{subsec:MPPcomp}.}

\item  \label{main_result_number_2}
For every signature $\Sigma$ and collection $\Phi\subseteq\MEq(\Sigma)$, 
the category $\QModelsCat({\Phi})$ has $U$-free objects. The $U$-free object $F(A,d_A)$ generated by $(A,d_A)\in\FRel$ can be identified (up to isomorphism of quantitative algebras) as follows:
\[
F(A,d_A)= (\TermsA/_{\equiv}, \Delta^{F(A,d_A)}, \{op^{F(A,d_A)}\}_{op\in\Sigma} ), \text{where }
\]
\begin{enumerate}
\item the equivalence relation ${\equiv} \subseteq  \TermsA\times \TermsA$ is defined as
$$
s \equiv t \Longleftrightarrow \Phi \derivefrel_{\FRel} \forall (A,d_A). s=t,
$$
\item the fuzzy relation $\Delta^{F(A,d_A)}: (\TermsA/_{\equiv})^2 \rightarrow [0,1]$ is defined as
$$
\Delta^{F(A,d_A)}( [s]_\equiv, [t]_{\equiv})\leq \epsilon \Longleftrightarrow \Phi\derivefrel_{\FRel} \forall (A,d_A). s=_\epsilon t, \text{ and}
$$

\item the interpretation $op^{F(A,d_A)}:(\TermsA/_{\equiv})^n\rightarrow (\TermsA/_{\equiv})$ of any $n$-ary operation $op\in\Sigma$, is defined as

$$
op^{F(A,d_A)}([s_1]_\equiv, \dots, [s_n]_{\equiv}) = [ op(s_1,\dots, s_n)]_\equiv \ .
$$
\end{enumerate}
It can be shown that the definitions of $\Delta^{F(A,d_A)}$ and $op^{F(A,d_A)}$ are well specified regardless of the choice of representatives $s,t$ for the classes $[s]_\equiv, [t]_{\equiv}$, and that indeed the quantitative algebra $F(A,d_A)$ belongs to $\QModels(\Phi)$.

These results are formally stated and proved in Section \ref{section:result:2}, and they give us an analog to the result mentioned in \autoref{ex_freealgebra} for Universal Algebra.

\item  \label{main_result_number_3}
As a corollary of the two results above and of \autoref{prop_freeadjunction}, there is a functor $F:\FRel \to \QModelsCat({\Phi})$ which associates to each $\FRel$ space $(A,d_A)$ the corresponding free object $F(A,d_A)$. The functor $F$ is a left adjoint of $U_{\QModelsCat(\Phi)\to \FRel}$:
\begin{equation*}
\begin{tikzcd}
	\FRel && {\QModelsCat({\Phi})}
	\arrow[""{name=0, anchor=center, inner sep=0}, "U", shift left=2, from=1-3, to=1-1]
	\arrow[""{name=1, anchor=center, inner sep=0}, "F"', shift right=2, tail reversed, no head, from=1-3, to=1-1]
	\arrow["\dashv"{anchor=center, rotate=-90}, draw=none, from=1, to=0]
\end{tikzcd}
\end{equation*}

This adjunction gives us a monad $\qterms{\Sigma,\Phi}$  on $\FRel$, which is defined similarly to the $\Set$ monad $\terms{\Sigma,\Phi}$ of quotiented terms discussed in Examples \ref{ex_monad_termseq} and \ref{ex_monad_termseq_adjunction}.
In Section \ref{section:result:3} we concretely identify this adjunction and monad, and we prove that the functor $U: \QModelsCat({\Phi})\rightarrow \FRel$ is strictly monadic, i.e., there is an isomorphism
$$
\EM(\qterms{\Sigma,\Phi}) \cong \QModelsCat (\Phi),
$$
where $\EM(\qterms{\Sigma,\Phi})$ is the category of {Eilenberg--Moore} algebras for the monad $\qterms{\Sigma,\Phi}$.

\item \label{main_result_number_4} 
We identify two relevant collections of $\FRel$ monads and of classes of equations and quantitative equations $\Phi \subseteq \MEq(\sig)$, respectively. On one side, we consider monads $M$ \hl{on} $\FRel$ that are \emph{monad liftings} of a monad $N$ \hl{on} $\Set$ having an equational presentation $\Psi\subseteq \Eq$. On the other, classes of $\FRel$ equations and quantitative equations $\Phi$ that are \emph{quantitative extensions} of $\Psi$. In Section \ref{section:lifting}, after having defined the above notions and that of \emph{quantitative equational presentation} of a monad on $\FRel$, we establish (\autoref{thm:correspondencemonliftthext}) the following correspondence:%
\begin{enumerate}[(1)]
    \item\label{item:overviewlifting1} If $M$ is a monad lifting of $N$, then \hl{a quantitative extension of $\Psi$ presents $M$.}%
    \item\label{item:overviewlifitng2} If $\Phi$ is a quantitative extension of $\Psi$, then $\Phi$ is a quantitative equational presentation of \hl{an $\FRel$} monad that lifts $N$.
\end{enumerate}

\item\label{main_result_number_5} All the results \ref{main_result_number_1}--\ref{main_result_number_4} above, stated and proved for the category $\FRel$, can be specialised and hold true for generalised metric spaces, i.e., for all full subcategories $\GMet$ of $\FRel$ defined as in Section \ref{sec:gen:metric:spaces}.  We show this in Section \ref{section:fromFRel_to_GMET}.

For example, it is possible to consider the category $\QalgMet$ of quantitative algebras whose underlying fuzzy relation space is a metric space $(A,d_A)\in\Met$. Accordingly, it is possible to define the entailment relation $\Vvdash_{\Met}$ restricted to $\QalgMet$ and have a sound and complete proof system $\vdash_\Met$. Furthermore, free quantitative algebras generated by metric spaces exist in $\QmodMet(\Phi)$ and the forgetful functor $U:\QmodMet(\Phi)\rightarrow \Met$ is strictly monadic. 
Finally, we also obtain the analogs of points \ref{item:overviewlifting1} and \ref{item:overviewlifitng2} above, relating $\Met$ monads liftings and quantitative extensions.

\end{enumerate}

\section{The Deductive System}\label{section:result:1}

In this section\hl{, we fix a signature $\Sigma$,} and we introduce a deductive system which can be used to derive judgments of the form: $\Phi\derivefrel_{\FRel} \phi$, for $\Phi\in \mathcal{P}(\MEq(\Sigma))$ and $\phi\in \MEq(\Sigma)$. Thus, formally, we define by induction a relation ${\derivefrelfrel} \subseteq \mathcal{P}(\MEq(\Sigma))\times \MEq(\Sigma) $. As standard, we often use $\derivefrelfrel$ in infix notation, i.e., we write $\Phi\derivefrelfrel \phi$ for $(\Phi,\phi)\in {\derivefrelfrel}$. 

\hl{We often write $\derivefrel$ instead of $\derivefrel_{\FRel}$, and $\Phi, \Psi \vdash\phi$ in place of  $\Phi\cup \Psi \vdash \phi$.}

\begin{defi}[Deductive System]\label{proof:system:definition}
The relation ${\derivefrelfrel}\subseteq \mathcal{P}(\MEq(\Sigma))\times \MEq(\Sigma) $ is defined as the smallest relation satisfying the following properties:
\begin{enumerate}
\item \label{INIT:rule:deductive} Closure under the INIT rule:
given any $\Phi \in \mathcal{P}(\MEq(\Sigma))$ and $\phi\in \MEq(\Sigma)$, if $\phi\in \Phi$ then $\Phi\derivefrel \phi$ holds. That is:
\begin{prooftree}
\AxiomC{$\ $}
\RightLabel{INIT  (proviso: $\phi\in\Phi$)}
\UnaryInfC{
$\Phi \derivefrel \phi$}
\end{prooftree}

\item \label{CUT:rule:deductive} Closure under the CUT rule:
given any $\Phi, \Phi' \in \mathcal{P}(\MEq(\Sigma))$ and $\psi\in \MEq(\Sigma)$, if we have that $\Phi\derivefrel \phi$ holds for all $\phi\in \Phi'$  and that $\Phi, \Phi'\derivefrel \psi$ holds, then $\Phi\derivefrel \psi$ holds. That is:
\begin{prooftree}
\AxiomC{$\{ \Phi\derivefrel \phi\}_{\phi \in \Phi'}$}
\AxiomC{$\Phi, \Phi' \derivefrel \psi$}
\RightLabel{CUT}
\BinaryInfC{
$\Phi \derivefrel \psi$}
\end{prooftree}

\item \label{WEAKENING:rule:deductive} Closure under the WEAKENING rule:
given any $\Phi, \Phi' \in \mathcal{P}(\MEq(\Sigma))$ and $\phi\in \MEq(\Sigma)$, if $\Phi\derivefrel \phi$ holds, then $\Phi, \Phi' \derivefrel \phi$ holds. That is:
\begin{prooftree}
\AxiomC{$\Phi \derivefrel \phi$}
\RightLabel{WEAKENING}
\UnaryInfC{
$\Phi , \Phi' \derivefrel \phi$}
\end{prooftree}
\item The relation $\derivefrel$ contains all the pairs $\Phi\derivefrel \phi$ listed below \eqref{REFL:axiom:deductive}--\eqref{LEFTRIGHTCONG:axiom:deductive}. These are ``axiom schemes'', meaning that pairs are obtained from \eqref{REFL:axiom:deductive}--\eqref{LEFTRIGHTCONG:axiom:deductive} by instantiating the involved fuzzy relation $(A,d_A)$, terms $s,t\in \TermsA$,  substitutions $\sigma$, \emph{etc.},  to concrete ones. 
    
\begin{enumerate}\makeatletter
\renewcommand\p@enumii{}
\makeatother

\item \label{REFL:axiom:deductive}
(REFL of $=$): $$ \emptyset \vdash \forall (A,d_A). s = s $$
\hl{expressing that equality is reflexive.}
\item \label{SYMM:axiom:deductive} (SYMM of $=$): $$  \forall (A,d_A). s = t \vdash  \forall (A,d_A). t = s $$
\hl{expressing that equality is symmetric.}
\item \label{TRANS:axiom:deductive}(TRANS of $=$): $$  \forall (A,d_A). s = t, \forall (A,d_A). t = u \vdash \forall (A,d_A). s = u $$
\hl{expressing that equality is transitive.}
\item \label{CONG:axiom:deductive}(CONG of $=$): For all $op\in\Sigma$ of arity $n$,
$$  \forall (A,d_A). s_1 = t_1, \dots, \forall (A,d_A). s_n = t_n \vdash \forall (A,d_A). op(s_1,\dots, s_n) = op(t_1,\dots, t_n)  $$
\hl{expressing that equality is a congruence relation with respect to all $op\in\Sigma$.}
\item \label{SUBST:axiom:deductive} (SUBSTITUTION for $=$ and $=_\epsilon$):\\
Given \hl{$\FRel$ spaces $(A,d_A)$ and $(B,d_B)$, and a substitution $\sigma: A\rightarrow \Terms{B}$,}
we have two similar axiom schemes: one allowing substitution on \hl{$\FRel$ equations ($=$)} and the other on \hl{$\FRel$ quantitative equations ($=_{\epsilon}$)}:%
\begin{align*}
    \Psi_\sigma,  
\forall (A,d_A). s= t &\vdash \forall (B,d_B). \sigma(s) = \sigma(t), \text{ and }\\
\Psi_\sigma, 
\forall (A,d_A). s=_\epsilon t &\vdash \forall (B,d_B). \sigma(s) =_\epsilon \sigma(t),
\end{align*}
where in both axiom schemes, the set $\Psi_\sigma$ is defined as:
\[\Psi_\sigma = \big\{ \forall (B,d_B). \sigma(a_i) =_{\epsilon_{i,j}} \sigma(a_j) \mid a_i,a_j\in A, \ \epsilon_{i,j} = d_A(a_i,a_j) \big\},\]
and where the function $\sigma: A\rightarrow \Terms{B}$ is extended to a function of type $\sigma: \Terms{A}\rightarrow \Terms{B}$ (which we denote with the same symbol) as expected by induction on terms, by letting $\sigma (op(s_1,...,s_n))= op(\sigma(s_1),...\sigma(s_n)) $.

\hl{Note that $\Psi_\sigma$ expresses that the substitution $\sigma$ is (provably) nonexpansive. The restriction to nonexpansive substitutions, implemented by this rule, constitutes a main difference with respect to the original proof system of \cite{DBLP:conf/lics/MardarePP16}, where arbitrary substitutions can be applied.}

\item \label{USEVAR:axiom:deductive}(USE VARIABLES):  For an \hl{$\FRel$ space} $(A,d_A)$ and $a,a'\in A$ and $\epsilon=d_A(a,a')$:
$$ \emptyset \vdash \forall (A,d_A). a =_{\epsilon} 
 a'.$$
 \hl{This axiom scheme allows us to derive information from the distance ($d_A$) between variables ($A$) and is justified by the fact that interpretations of variables in quantitative algebras are nonexpansive.}

\item  \label{MAX:axiom:deductive} (UP-CLOSURE): For all \hl{$\epsilon \leq \delta \in [0,1]$}:
$$   \forall (A,d_A). s =_\epsilon t \vdash  \forall (A,d_A). s =_\delta t .$$ 
\hl{This axiom scheme reflects the semantics (Definition \ref{semantics_quantitative_eq}) of the judgment $s=_\epsilon t$ as an inequality.}
\item \label{1MAX:axiom:deductive}(1-MAX):
$$  \emptyset \vdash  \forall (A,d_A). s =_1 t $$
\hl{This axiom scheme reflects that the maximal possible distance is $1$.}
\item \label{ORDER:axiom:deductive}(ORDER COMPLETENESS): For an index set $I$,
$$    \big\{ \forall (A,d_A). s =_{\epsilon_i} t\big\}_{i\in I} \vdash  \forall (A,d_A). s =_{\inf\{\epsilon_i\}_{i\in I}} t .$$

\hl{This axiom scheme reflects the fact that the distance bounds $\epsilon_i$ are values in the complete lattice $[0,1]$.}

\item \label{LEFTRIGHTCONG:axiom:deductive}(Left and Right CONGRUENCE) of $=$ with respect to $=_\epsilon$:
\begin{align*}
    \forall (A,d_A). s= t, \forall (A,d_A). t=_\epsilon u &\vdash  \forall (A,d_A). s=_\epsilon u, \text{ and}\\
    \forall (A,d_A). s= t, \forall (A,d_A). u=_\epsilon s &\vdash \forall (A,d_A). u =_\epsilon t .
\end{align*}

\hl{Finally, this axiom scheme reflects the fact that equality is a congruence, both on the left and right components, with respect to the binary $=_\epsilon$ relation.}

\end{enumerate}
\end{enumerate}

\end{defi}

\begin{rem}
\hl{The axiom schemes 
  \eqref{REFL:axiom:deductive}--\eqref{LEFTRIGHTCONG:axiom:deductive} of Item (4) can alternatively be presented as inference rules. Both versions are mutually admissible, in the proof system, by applications of the CUT and INIT rules. For example, from the axiom scheme \ruleorange{(SYMM of $=$)} \eqref{SYMM:axiom:deductive}
    we can show that the corresponding inference rule }
        \begin{prooftree}
        \AxiomC{$\Phi \derivefrel \forall (A,d_A).s=t$}
        \RightLabel{Symm Rule}
        \UnaryInfC{$\Phi \derivefrel \forall (A,d_A).t=s$}
    \end{prooftree}
    \hl{is admissible as follows:}
    \begin{prooftree}
        \AxiomC{\ruleorange{(SYMM of $=$)} \eqref{SYMM:axiom:deductive}}
        \UnaryInfC{$\forall (A,d_A).s=t \derivefrel \forall (A,d_A).t=s$}
        \AxiomC{$\Phi \derivefrel \forall (A,d_A).s=t$}
        \RightLabel{CUT}
        \BinaryInfC{$\Phi \derivefrel \forall (A,d_A).t=s$}
    \end{prooftree}
    \hl{Similarly, from the inference rule, it is possible to derive the axiom scheme as follows:}%
        \begin{prooftree}
        \AxiomC{}
        \RightLabel{INIT}
        \UnaryInfC{$ \forall (A,d_A).s=t \derivefrel \forall (A,d_A).s=t$ }
        \RightLabel{Symm Rule}
        \UnaryInfC{$ \forall (A,d_A).s=t \derivefrel \forall (A,d_A).t=s$ }
        \end{prooftree}
\end{rem}

\begin{rem}
\hl{
One can consider the following variant of the \ruleorange{(ORDER COMPLETENESS)} axiom scheme (\ref{ORDER:axiom:deductive}): for each $\epsilon \in [0,1]$,}
\begin{equation}\label{eqn:archaxiom}
    \big\{ \forall (A,d_A). s =_{\delta} t  \ \mid \delta > \epsilon\big\} \vdash  \forall (A,d_A). s =_{\epsilon} t.\tag{Arch}
\end{equation}
\hl{This version of the axiom scheme more directly corresponds to the deduction rule called ``Arch'' in } \cite[Definition 2.1]{DBLP:conf/lics/MardarePP16}. \hl{The two axioms are equivalent and each can be derived from the other, in presence of the other axioms. In one direction, \ruleorange{(Arch)} is simply a special instance of \ruleorange{(ORDER COMPLETENESS)}, by taking the index set $I = \{\delta \in [0,1] \mid \delta> \epsilon\}$ with $\inf(I) = \epsilon$. In the other direction, the \ruleorange{(ORDER COMPLETENESS)}  axiom can be derived from  the \ruleorange{(Arch)} axiom as follows: first, turn the left-side  $\big\{ \forall (A,d_A). s =_{\epsilon_i} t\big\}_{i\in I}$, which has $\epsilon=\inf\{\epsilon_i\}$ as infimum, into an upward-closed set of quantitative equations by applications of the  \ruleorange{(UP-CLOSURE)} axiom scheme and the \ruleorange{(CUT)}  rule. Secondly, since the resulting set of quantitative equations is the form
$\big\{ \forall (A,d_A). s =_{\delta} t\big\}_{\delta>\epsilon}$ we can apply the \ruleorange{(Arch)} axiom scheme to derive the right-hand side $\forall (A,d_A). s =_{\epsilon}t$.}

\hl{It is worth observing that the \ruleorange{(1-MAX)} axiom scheme (\ref{1MAX:axiom:deductive}) is admissible in the proof system, by instantiating \ruleorange{(ORDER COMPLETENESS)} with an empty index set $I$, because $\inf\emptyset = 1$ in the complete lattice $[0,1]$. Similarly, \ruleorange{(1-MAX)} is derivable from \ruleorange{(Arch)} with $\epsilon = 1$, because the set  $\{ \delta > \epsilon \}$ is empty. Even if admissible, we have opted to keep the \ruleorange{(1-MAX)} axiom in the list of Definition \ref{proof:system:definition} since it clearly expresses that the distance function is total and there is always a distance assigned to terms.}%
\end{rem}

The first basic result regarding the deductive system is the soundness theorem.
\begin{thm}[Soundness]\label{soundness_theorem_statement}
The inclusion $\derivefrelfrel \ \subseteq  \ \implyfrelfrel$ holds.
\end{thm}
\begin{proof}

We write $\derivefrel$ for $\derivefrelfrel$, as already done above, and also $\implyfrel$ for  $\implyfrelfrel$. Assume $\Phi\derivefrel \phi$. We prove that $\Phi\implyfrel \phi$ holds by induction of the derivation tree used to derive $\Phi\derivefrel \phi$.

Most cases are straightforward, including the occurrences of INIT, CUT and WEAKENING rules and most axiom schemes, so we only detail some of those. The only non-obvious case is the (SUBSTITUTION) axiom scheme \eqref{SUBST:axiom:deductive}, which we will prove
in full details.

For an instance of an easy to prove axiom scheme,
consider the (USE VARIABLES) axiom scheme \eqref{USEVAR:axiom:deductive}.
We need to show that for any quantitative algebra $\mathbb{C}=(C,d_C,\{op^C\}_{op\in \Sigma})$,
\begin{center}
    $\algC \models \forall (A,d_A). a =_{\epsilon} a'$
\end{center}
for any $\FRel$ space $(A,d_A)$ and $a,a'\in A$ with $\epsilon=d_A(a,a')$.
This means that, for any nonexpansive interpretation $\imap: (A,d_A) \to (C,d_C)$, we want to prove
\begin{center}
    $d_C(\sem{a}^\algC_\imap,\sem{a'}^\algC_\imap)\leq \epsilon$,
\end{center}
which\hl{, by the definition of $\sem{-}$,} is equivalent to
\begin{center}
    $d_C(\imap(a),\imap(a'))\leq \epsilon$.
\end{center}
This holds since $\imap$ is nonexpansive, so we have proven that the axiom scheme is sound.

For another simple example, consider the (Left CONGRUENCE) axiom scheme \eqref{LEFTRIGHTCONG:axiom:deductive}.
We need to show that a quantitative algebra $\mathbb{C}=(C,d_C,\{op^C\}_{op\in \Sigma})$ satisfying all the premises, i.e. (1) $\mathbb{C} \models \forall (A,d_A). s  = t$ and (2) $\mathbb{C} \models \forall (A,d_A). t  =_\epsilon u$, must also satisfy the conclusion: 
\[ \mathbb{C} \models \forall (A,d_A). s  =_\epsilon u. \]
Take an interpretation $\imap: (A,d_A) \to (C,d_C)$. Then by the two premises we obtain (1)
$\sem s ^\algC_\imap  = \sem t ^\algC_\imap$ and (2) $d_C(\sem t ^\algC_\imap, \sem u ^\algC_\imap)\leq \epsilon
$, which immediately imply the desired conclusion:
\[
d_C(\sem s ^\algC_\imap, \sem u ^\algC_\imap)\leq \epsilon.\]

We now prove soundness of the (SUBSTITUTION) axiom scheme \eqref{SUBST:axiom:deductive}, the non-obvious case. Let $\sigma:A\rightarrow \Terms{B}$ be an arbitrary substitution. We need to show that a quantitative algebra $\mathbb{C}=(C,d_C,\{op^C\}_{op\in \Sigma})$ satisfying all the premises of \eqref{SUBST:axiom:deductive}, i.e., 
\begin{enumerate}
\item 
$
\mathbb{C} \models \forall (A,d_A). s  =_
\epsilon t
$, and
\item 
$
\mathbb{C} \models \big\{ \forall (B,d_B). \sigma(a_i) =_{\epsilon_{i,j}} \sigma(a_j) \mid a_i,a_j\in A, \ \epsilon_{i,j} = d_A(a_i,a_j) \big\}
$,
\end{enumerate}
necessarily also satisfies the equation or quantitative equation of the conclusion (we just consider the quantitative equation case as the two are similar), i.e., 
\begin{center}
$\mathbb{C} \models \forall (B,d_B). \sigma(s) =_\epsilon \sigma(t)$.
\end{center}
Towards this end, take an arbitrary nonexpansive interpretation $\imap: (B,d_B) \rightarrow (C,d_C)$. From the interpretation $\imap$ and the substitution $\sigma$, we define a new interpretation $\hat{\sigma}$ as follows:
\begin{equation}\label{eqn:defnsubinter}
    \hat{\sigma}:(A,d_A)\rightarrow (C,d_C) \ \ \ \ \ \hat{\sigma}(a) = \sem{\sigma(a)}^\algC_{\imap}.
\end{equation}
Before proceeding further, we need to show that $\hat{\sigma}$ is nonexpansive. So, take any $a,a'\in A$ and assume $d_A(a,a')= \delta$. Since (from (2)) $\mathbb{C}$ satisfies $\forall (B,d_B). \sigma(a) =_{\delta} \sigma(a')$,%
\[d_C(\hat{\sigma}(a), \hat{\sigma}(a')) \stackrel{\eqref{eqn:defnsubinter}}{=} d_C( \sem{\sigma(a)}^\algC_{\imap} , \sem{\sigma(a')}^\algC_{\imap}  ) \stackrel{(2)}{\leq} \delta,\]
which concludes the proof that $\hat{\sigma}$ is nonexpansive. 

Next, from (1), and taking as interpretation $\hat{\sigma}$, we know that $d_C\big( \sem{s}^\algC_{\hat{\sigma}} , \sem{t}^\algC_{\hat{\sigma}} \big) \leq \epsilon$. Since $\hat{\sigma}$ satisfies $\sem{\sigma(s)}^\algC_{\imap} = \sem{s}^\algC_{\hat{\sigma}}$ and $\sem{\sigma(t)}^\algC_{\imap} = \sem{t}^\algC_{\hat{\sigma}}$ (recall from \eqref{SUBST:axiom:deductive} how $\sigma$ is extended to terms),
we get
$
d_C\big( \sem{\sigma(s)}^\algC_{\imap} , \sem{\sigma(t)}^\algC_{\imap} \big) \leq \epsilon
$. %
We conclude that $\mathbb{C} \models \forall (A,d_A). \sigma(s) =_\epsilon \sigma(t)$.
\end{proof}

The above proof of soundness is rather direct and simple. Proving the opposite direction, i.e., the completeness theorem, \hl{requires more} work. Indeed, we will use the proof system $\derivefrelfrel$ to construct free objects in $\QModelsCat(\Phi)$, prove some results about such free objects and finally derive the completeness result.

Our results regarding free objects in $\QModelsCat(\Phi)$ are presented in Section \ref{section:result:2}. The completeness \autoref{completeness_of_proof_system_qa} is also established in that section, as a corollary.

\section{Free Quantitative Algebras}
\label{section:result:2}

In the following, \hl{we fix a class of} $\FRel$ (quantitative) $\Sigma$-equations $\Phi \subseteq \MEq(\Sigma)$, \hl{and we write $U$ for the forgetful functor $\QModelsCat({\Phi})\rightarrow \FRel$ previously defined.}

We are going to prove the following statement using (the soundness of) the deductive system $\derivefrelfrel$ \hl{as a main tool}. Recall that free objects are defined as in \autoref{free:object:def}.

\begin{thm}
For all $(A,d_A)\in \FRel$, the $U$-free object generated by $(A,d_A)$ exists.%
\end{thm}

The proof of this statement occupies the rest of this section.

Let us \hl{fix $(A,d_A)\in\FRel$.} We are going to explicitly construct the $U$-free object in $\QModelsCat({\Phi})$, denoted by $F(A,d_A)$, with respect to a nonexpansive map 
\begin{center}
    $\freemap: (A,d_A) \rightarrow U(F(A,d_A))$
\end{center}
defined later on in \autoref{definition_of_iota}. \hl{As we will see, the carrier of the free object will be a set of equivalence classes of terms over $A$, and $\alpha: a \mapsto [a]_{\equiv}$ the function mapping an element of $A$ to its equivalence class in the set of quotiented terms.} 
We proceed as follows:
\begin{enumerate}
\item (Subsection \ref{subsub1:proofs}) Formally define the quantitative algebra $F(A,d_A)\in \QalgFRel$.
\item (Subsection \ref{subsub2:proofs}) Prove that, indeed, $F(A,d_A)$ belongs to $\QModelsCat({\Phi})$. In other words, we show that $F(A,d_A)$ satisfies all the $\FRel$ equations and quantitative equations in $\Phi$.
\item (Subsection \ref{subsub3:proofs}) Finally, define the map 
$
\freemap: (A,d_A) \rightarrow U(F(A,d_A)) 
$
and show that $F(A,d_A)$ satisfies the universal property (from \autoref{free:object:def}) defining the (unique, up to isomorphism) free algebra generated by $(A,d_A)$.
\end{enumerate}

\subsection{Definition of \texorpdfstring{$F(A,d_A)$}{}}\label{subsub1:proofs}

We recall that, by the soundness (\autoref{soundness_theorem_statement}) of the deductive system,  whenever 
$\Phi \derivefrelfrel \phi 
$ holds, also $\Phi \implyfrelfrel \phi 
$ holds. %
We start by defining a binary relation $\equiv$ and a fuzzy relation $d$ on the set of terms $\Terms{A}$ built from $A$.

\begin{defi}\label{defn:equivrelterms}
We define $\equiv$ as follows, for all $s,t\in \Terms{A}$:
\begin{center}
$s \equiv t  \ \ \ \Leftrightarrow \ \ \  \Phi \derivefrel \forall (A,d_A). s = t $.
\end{center}
We define $d$ as follows, for all $s,t\in \Terms{A}$:
\begin{center}
$d(s,t) = \inf_{\epsilon}\big\{  \Phi \derivefrel \forall (A,d_A). s =_\epsilon t  \big\}$.
\end{center}
\end{defi}
The axiom schemes \eqref{REFL:axiom:deductive}--\eqref{CONG:axiom:deductive} ensure that $\equiv$ is an equivalence relation and a congruence with respect to the operations in $\Sigma$. Moreover, it follows from \eqref{1MAX:axiom:deductive} that $d$ is a fuzzy relation (bounded by $1$), and from \eqref{LEFTRIGHTCONG:axiom:deductive} that $\equiv$ is a congruence with respect to $d$ (i.e. $d(s,t) = d(s',t')$ whenever $s\equiv s'$ and $t \equiv t'$).%

The following technical lemma relates the proof system $(\derivefrel)$ with the definition of the fuzzy relation $d$.

\begin{lem}\label{lemma_connection_syntax_semantics_1}
\hl{For any $\epsilon$,} $\Phi\derivefrel \forall (A,d_A). s=_\epsilon t  \ \Longleftrightarrow  \ d(s,t)\leq \epsilon$.
\end{lem}
\begin{proof}
The ($\Rightarrow$) direction follows immediately from the definition of $d$ as an infimum.
\hl{For the ($\Leftarrow$) direction, assume $d(s,t)\leq \epsilon$. 
As an instance of the (ORDER COMPLETENESS) axiom scheme \eqref{ORDER:axiom:deductive} we have}
\begin{equation*}
   \{\forall (A,d_A).s=_\delta t\mid \Phi \derivefrel \forall (A,d_A).s=_\delta t \} \derivefrel \forall (A,d_A).s=_{d(s,t)} t.
\end{equation*}
\hl{By the CUT rule we then derive $\Phi \derivefrel \forall (A,d_A).s=_{d(s,t)} t$. By  our hypothesis that $d(s,t) \leq \epsilon$, the \ruleorange{(UP-CLOSURE)} axiom scheme \eqref{MAX:axiom:deductive} yields $\forall (A,d_A). s=_{d(s,t)} t \derivefrel \forall (A,d_A).s=_{\epsilon} t$.  By the CUT rule, we then conclude $\Phi \derivefrel \forall (A,d_A). s=_{\epsilon} t$ as desired.}
\qedhere

\end{proof}

Since we have established that $\equiv$ is an equivalence relation, the quotient $\Terms{A}/_{\equiv}$, consisting of $\equiv$-equivalence classes, is well-defined. Furthermore, the equivalence $\equiv$ being a congruence for the fuzzy relation $d$ implies that the following is a good definition, regardless of the choice of representatives.

\begin{defi}\label{cor:distancequotientedterms}
The fuzzy relation $\Delta: ( \TermsA/_{\equiv} \times \TermsA/_{\equiv}) \rightarrow [0,1]$ is given by
$$
\Delta([s]_{\equiv}, [t]_{\equiv}) = d(s,t).
$$
\end{defi}

 Moreover, since we have already established %
 that $\equiv$ is a congruence on $\TermsA$,  the interpretation $op^{F(A,d_A)}$ of each operation $op\in\Sigma$ specified as
$$
op^{F(A,d_A)}\big( [s_1]_{\equiv},\dots, [s_n]_{\equiv} \big) = [op(s_1,\dots, s_n)]_{\equiv}
$$
is well-defined and does not depend on a specific choice of representatives. 

We can collect the results of this subsection as follows:

\begin{cor}
The triple $(\TermsA/_{\equiv}, \Delta, \{op^{F(A,d_A)}\}_{op\in\Sigma})$ is a quantitative $\Sigma$-algebra.
\end{cor}

The quantitative algebra identified above is our definition of $F(A,d_A)$.

\begin{defi}\label{def:free_algebra}
The quantitative $\Sigma$-algebra $F(A,d_A)$ is defined as
\begin{center}
    $F(A,d_A) = (\TermsA/_{\equiv}, \Delta, \{op^{F(A,d_A)}\}_{op\in\Sigma})$.
\end{center}
\end{defi}

\subsection{Proof that \texorpdfstring{$F(A,d_A)\in \QModelsCat(\Phi)$}{} }
\label{subsub2:proofs}

We show in \autoref{lem_free_models} below that the quantitative algebra $F(A,d_A)$ constructed from $(A,d_A)$, in Subsection \ref{subsub1:proofs}, 
satisfies all $\FRel$ equations and quantitative equations in $\Phi$.
The proof exploits the following lemma.

\begin{lem}
\label{lem_free_interpret}   
Let $\imap: X \to \TermsA/_{\equiv}$ be a function. Let $c: \TermsA/_{\equiv}\rightarrow \TermsA$ be a choice function, i.e., such that $c([s]_\equiv) \in [s]_\equiv$.\footnote{\hl{Equivalently, $c$ chooses an element $s\in\TermsA$ for each $\equiv$-equivalence class $[s]_\equiv \in \TermsA/_{\equiv}$. Note that one such $c$ exists by the axiom of choice.}} Define $\sigma_\imap: X \to \TermsA$ as: $\sigma_\imap(x) = c(\tau(x))$.
    Then, for all $s\in\TermsA$, it \hl{holds that}
    \[\sem s ^{F(A,d_A)}_\imap = [\sigma_\imap(s)]_\equiv,\]
    where $\sigma_\imap(s)$ is the term obtained by applying the substitution $\sigma_\tau$ to $s$. 
\end{lem}
\begin{proof}
    The proof is by induction on $s$. For $s=x$, the result is immediate by definition of $\sigma_\imap$:%
    \[\sem x ^{F(A,d_A)}_\imap=\imap(x)=[\sigma_\imap(x)]_\equiv .\]   
    For $s=op(s_1,...s_n)$, we have
    \begin{align*}
        \sem s ^{F(A,d_A)}_\imap
        &= op^{F(A,d_A)}(\sem{s_1}^{F(A,d_A)}_\imap,...,\sem{s_n}^{F(A,d_A)}_\imap)&&\text{(by definition of $\sem{\_}$)}\\
        &= op^{F(A,d_A)} ([\sigma_\imap(s_1)]_\equiv,...,[\sigma_\imap(s_n)]_\equiv)&&\text{(by inductive hypothesis)}\\
        &=[op (\sigma_\imap(s_1),...,\sigma_\imap(s_n)]_\equiv &&\text{(by definition of $op^{F(A,d_A)}$)}\\
        &=[\sigma_\imap(op (s_1,...,s_n))]_\equiv. &&\text{(by definition of $\sigma_\imap$ on terms)}\qedhere
    \end{align*}
\end{proof}

\begin{lem}
\label{lem_free_models}
It \hl{holds that} $F(A,d_A) \in \QModels(\Phi)$.
\end{lem}
\begin{proof}
    We need to show that if $\phi\in \Phi$ then $F(A,d_A) \models \phi$. Suppose that $\phi$ is an $\FRel$ quantitative equation of the form (the case of $\phi$ being an equation is similar)
    \[\phi = \ \  \forall (X,d_X). s=_\epsilon t \]
    for some $\FRel$ space $(X,d_X)$, terms $s,t\in\Terms{X}$ and $\epsilon \in [0,1]$. %
    Therefore, we need to show that for every nonexpansive interpretation $\imap: (X,d_X) \rightarrow (\TermsA/_{\equiv}, \Delta)$\hl{,}%
    \[ \Delta\Big( \sem s ^{F(A,d_A)}_\imap ,  \sem t ^{F(A,d_A)}_\imap\Big)\leq \epsilon, \]
    or equivalently, by applying \autoref{lem_free_interpret} where we take $\sigma_\imap$ defined as in the lemma, \hl{that}%
    \[ \Delta\Big([\sigma_\imap(s)]_{\equiv}, [\sigma_\imap(t)]_{\equiv} \Big)\leq \epsilon,\]
    which in turn, \hl{by the definition of $\Delta$ (Definition \ref{cor:distancequotientedterms}) and \autoref{lemma_connection_syntax_semantics_1}} 
    \hl{is equivalent to showing}
    \[\Phi\vdash \forall (A,d_A). \sigma_\imap(s) =_\epsilon \sigma_\imap(t).\]

    The assumption that $\imap$ is nonexpansive means that for all $x_i, x_j \in X$,
    \[ d_X(x_i,x_j)\leq \epsilon_{ij} \ \ \ \Longrightarrow \ \ \ \Delta( \sem {x_i} ^{F(A,d_A)}_\imap, \sem {x_j} ^{F(A,d_A)}_\imap) \leq \epsilon_{ij} 
    \]
    which by \autoref{lem_free_interpret} is equivalent to
    \[
    d_X(x_i,x_j)\leq \epsilon_{ij} \ \ \ \Longrightarrow \ \ \ \Delta( [\sigma_\imap(x_i)]_\equiv, [\sigma_\imap(x_j)]_\equiv) \leq \epsilon_{ij} 
    \]
    \hl{Since $\Delta( [\sigma_\imap(x_i)]_\equiv, [\sigma_\imap(x_j)]_\equiv) = d(\sigma_\imap(x_i),\sigma_\imap(x_j))$, it follows from} \autoref{lemma_connection_syntax_semantics_1} that
    \begin{equation}
    \label{eq_nonexp_free_models}
    d_X(x_i,x_j)\leq \epsilon_{ij} \ \ \ \Longrightarrow \ \ \ \Phi\vdash  \forall (A,d_A). \sigma_\imap(x_i) =_{\epsilon_{ij}} \sigma_\imap(x_j) .
    \end{equation}
    Now, we know that:
    \begin{enumerate}[(I)]
    \item\label{item:substitution1} Since $\forall (X,d_X). s =_\epsilon t $ belongs to $\Phi$, by the \ruleorange{INIT} rule we have
    $
    \Phi \vdash \forall (X,d_X). s =_\epsilon t
    $.
    \item\label{item:substitution2} By \eqref{eq_nonexp_free_models} we have all the following judgments:
    \[
    \big\{ \Phi\vdash  \forall (A,d_A). \sigma_\imap(x_i) =_{\epsilon_{ij}} \sigma_\imap(x_j) \mid x_i,x_j\in X , \ \  \epsilon_{ij}=d_X(x_i,x_j)  \big\}.
    \]
    \end{enumerate}
    Note that, using $\sigma_\imap$ as substitution, \hl{the conclusions of \ref{item:substitution1} and \ref{item:substitution2} are the premises of the \ruleorange{(SUBSTITUTION)} axiom scheme \eqref{SUBST:axiom:deductive} for quantitative equations,} namely, we have
    \[\Psi \vdash \forall (A,d_A). \sigma_\imap(s) =_\epsilon \sigma_\imap(t), \]
    where $\Psi=\big\{ \forall (A,d_A). \sigma_\imap(x_i) =_{\epsilon_{ij}} \sigma_\imap(x_j) \mid x_i,x_j\in X, \ \epsilon_{ij} = d_X(x_i,x_j) \big\}\cup  
    \{\forall (X,d_X). s=_\epsilon t\}$.
    Then, by \ref{item:substitution1} and \ref{item:substitution2}, $\Phi$ entails all of $\Psi$, so we can use CUT to get the desired
    \[
    \Phi\vdash \forall (A,d_A). \sigma_\imap(s) =_\epsilon \sigma_\imap(t).\qedhere\]
\end{proof}

\subsection{Proof of freeness of \texorpdfstring{$F(A,d_A)$}{}}
\label{subsub3:proofs}

We now show that the quantitative algebra $F(A,d_A)$
\hl{described in \autoref{def:free_algebra} is indeed the free object (see Definition \ref{free:object:def}) in $\QModelsCat(\Phi)$ generated by $(A,d_A)$ with respect to the universal map $\alpha: (A, d_A) \to UF(A, d_A)$ given by the assignment $a \mapsto [a]_{\equiv}$, which is nonexpansive by the following:}

\begin{lem}\label{definition_of_iota}
The map $\freemap: (A,d_A) \rightarrow (\TermsA/_{\equiv}, \Delta)$ is nonexpansive. 
\end{lem}
\begin{proof}
\hl{Let $d_A(a_1,a_2)=\epsilon$. We want to prove that $\Delta(\alpha(a_1),\alpha(a_2)) \leq \epsilon$.  
 By the (USE VARIABLES) axiom scheme \eqref{USEVAR:axiom:deductive}, $d_A(a_1,a_2)=\epsilon$ implies $\emptyset\derivefrel\forall (A,d_A). a_1 =_\epsilon a_2$. By WEAKENING we derive $\Phi\derivefrel\forall (A,d_A). a_1 =_\epsilon a_2$. 
 Therefore $\alpha$ is nonexpansive because, by definition,}
 \[\Delta(\alpha(a_1),\alpha(a_2))= \Delta([a_1]_\equiv,[a_2]_{\equiv})= d(a_1,a_2) =
 \inf_{\epsilon}\big\{  \Phi \derivefrel \forall (A,d_A). a_1 =_\epsilon a_2  \big\}.\qedhere\]
\end{proof}

\begin{rem}
We note that the map $\freemap$ is generally not an isometry (i.e., distance preserving) nor an injection, although it is in many interesting cases. For example, consider the case when the generating fuzzy relation space $(A,d_A)$ is defined as follows:
\[
A=\{ a_1,a_2\} \ \ \ \ d_A(a_1,a_2) = d_A(a_2,a_1)= \tfrac{1}{2}\ \ \ \ d_A(a_1,a_1)= d_A(a_2,a_2)=0,
\]
i.e., it is a metric space consisting of two points at distance $\frac{1}{2}$, the signature $\sig$ is empty, and the set $\Phi=\{ \forall (A,d_A). a_1 = a_2  \}$ consists of just one $\FRel$ equation. In this case it is easy to check that $\TermsA/_{\equiv}$ has a single element (i.e., all terms in $\TermsA$ are $\equiv$-equivalent), and thus the map $\freemap$ is neither an injection nor an isometry.
\end{rem}

\begin{thm}\label{free_algebra_theorem}
Let $U: \Catqalg \to \FRel$.
The quantitative algebra $F(A,d_A)$ is the $U$-free object generated by $(A,d_A)$ relative to the map $\freemap:(A,d_A) \rightarrow (\TermsA/_{\equiv}, \Delta)$.
\end{thm}

\begin{proof}

We need to show that for every quantitative algebra $\mathbb{B} = (B, d_B, op^{\mathbb{B}}) \in \QModelsCat(\Phi)$ and nonexpansive map $f: (A,d_A) \to (B,d_B)$\hl{,} there is a unique homomorphism of quantitative algebras 
$\hat{f}:F(A,d_A) \to \mathbb{B}$ which extends $f$, i.e., which satisfies $f=U(\hat f) \circ \freemap$. In the \hl{rest} of the proof we will generally omit the explicit use of the forgetful functors on morphisms.%

\hl{We define $\hat f$ for all terms $s\in \Terms{A}$ by $\hat f([s]_{\equiv})=\sem{s}^\mathbb{B}_f$.} To see this is well-defined, observe that if $s \equiv t$ then, by definition of the relation ($\equiv$), it holds that $\Phi\vdash \forall (A,d_A). s = t$. By the soundness (\autoref{soundness_theorem_statement}) of the deductive system, we have that
$$
\forall (A,d_A). s = t  \ \in \ \QTheory(\QModels(\Phi)).
$$
Since $\mathbb{B}\in \QModels(\Phi)$ by hypothesis, this means that 
$$
\mathbb{B}\models \forall (A,d_A). s = t,
$$
and in particular, \hl{by viewing $f:(A,d_A)\rightarrow (B,d_B)$ as a (nonexpansive) interpretation}, it holds that $\sem{s}^\mathbb{B}_f = \sem{t}^\mathbb{B}_f$. Hence $\hat{f}$ is well-defined as a function.

It remains to show that $\hat{f}$ is a homomorphism and nonexpansive. The first follows from the interpretation $op^{F(A,d_A)}$ of the operations in $F(A,d_A)$ as follows:
 \begin{align*}
        \hat f(op^{F(A,d_A)} ([s_1]_{\equiv },..., [s_n]_{\equiv }))
        &= \hat f ([op (s_1,..., s_n)]_{\equiv })\\
        &= \sem {op (s_1,..., s_n)}_f^{\mathbb B}\\
        &= op^{\mathbb B} (\sem{s_1}_f^{\mathbb B},..., \sem{s_n}_f^{\mathbb B})\\
        & = op^{\mathbb B} (\hat f([s_1]_{\equiv }),..., \hat f ([s_n]_{\equiv })).
     \end{align*}
Regarding the second point (nonexpansiveness), take two arbitrary $[s]_\equiv, [t]_\equiv\in \TermsA/_{\equiv}$ and let $\Delta([s]_\equiv, [t]_\equiv) = \epsilon$ be their distance in $F(A,d_A)$. We need to show that 
$$
d_B\big(\hat{f}([s]_\equiv), \hat{f}([t]_\equiv)\big)\leq \epsilon.
$$
As established in \autoref{lemma_connection_syntax_semantics_1}, the hypothesis $\Delta([s]_\equiv, [t]_\equiv) = \epsilon$ implies
$
\Phi \vdash \forall (A,d_A). s=_\epsilon t
$.
From the soundness of the deductive system (\autoref{soundness_theorem_statement}), we therefore know that
$
\forall (A,d_A). s=_\epsilon t  \ \ \in\ \  \QTheory(\QModels(\Phi))
$,
and since $\mathbb{B}\in \QModels(\Phi)$ by hypothesis, we find
$$
\mathbb{B}\models \forall (A,d_A). s=_\epsilon t.
$$
Taking as nonexpansive interpretation $f:(A,d_A)\rightarrow (B,d_B)$, we therefore obtain that
$d_B(\sem{s}^\algB_f, \sem{t}^\algB_f) \leq \epsilon$. By the definition of $\hat{f}$ we have $\hat{f}([s]_\equiv)= \sem{s}^\algB_f$ and $\hat{f}([t]_\equiv)= \sem{t}^\algB_f$, so we conclude as desired that
\[ d_B(\hat{f}([s]_\equiv), \hat{f}([t]_\equiv)) \leq \epsilon.\]

It follows by unrolling definitions that $\hat{f}$ extends $f$:
\[\hat{f}( \freemap(a)) = \hat{f}([a]_\equiv) = \sem{a}^{\mathbb{B}}_f = f(a),\]
and a simple induction shows any homomorphism that extends $f$ must be equal to $\hat{f}$.\qedhere

\end{proof}

\subsection{Completeness of the Deductive System}
\label{subsub4:proofs}

By exploiting the existence of free objects, we can now establish the completeness of the deductive system $\derivefrel_{\FRel}$, \autoref{completeness_of_proof_system_qa} below.

The proof relies on the following property of $F(A,d_A)$, which, as we have seen, is the $U$-free object (here $U$ is $U:
\Catqalg \to \FRel$) generated by the $\FRel$ space $(A,d_A)$ relative to the nonexpansive map $\freemap:(A,d_A) \to F(A,d_A)$.

\begin{lem}
\label{lem_completeness}
For all $s,t\in \TermsA$,
\begin{enumerate}[(i)]
    \item if $\sem s ^{F(A,d_A)}_\freemap = \sem t ^{F(A,d_A)}_\freemap$ then $\Phi  \derivefrel_{\FRel} \forall (A, d_A). s=t$.
    \item if $\Delta(\sem s ^{F(A,d_A)}_\freemap,\sem t ^{F(A,d_A)}_\freemap)\leq \epsilon$, then $ \Phi \derivefrel_{\FRel} \forall (A, d_A). s=_\epsilon t$.
\end{enumerate}
\end{lem}
\begin{proof}
For \hl{(i)}, suppose that $\sem s ^{F(A,d_A)}_\freemap = \sem t ^{F(A,d_A)}_\freemap$.

     Note that by \autoref{lem_free_interpret}, where we instantiate $\imap$ with $\freemap$, we have
     $\sem s ^{F(A,d_A)}_\freemap=[\sigma_{\freemap}(s)]_\equiv$, where $\sigma_\freemap:A\to \TermsA$ is a choice function for $\freemap$ as required in the lemma.
     
     \hl{Since $\sigma_\freemap$ substitutes occurrences of elements of $A$ with terms in the $\equiv$-equivalence class, we will prove that $\Phi \derivefrel \forall (A, d_A). \sigma_{\freemap}(s)=s$ by induction on the structure of $s$, and it then follows, by the definition of $\equiv$} (\autoref{defn:equivrelterms}), that
     $$\sem s ^{F(A,d_A)}_\freemap=[\sigma_{\freemap}(s)]_\equiv=[s]_{\equiv}.$$
     
     \hl{If $s$ is an element of $A$, we already know that $\sigma_{\freemap}(a)$ will be in its equivalence class, so $\Phi \derivefrel \forall (A, d_A). \sigma_{\freemap}(s)=s$ by the definition of $\equiv$. Suppose now that $s = op(s_1,\dots, s_n)$ and that for each $i$, $\Phi \derivefrel \forall (A, d_A). \sigma_{\freemap}(s_i)=s_i$. Since}
     \[\sigma_{\freemap}(s) = \sigma_{\freemap}(op(s_1,\dots, s_n)) = op(\sigma_{\freemap}(s_1),\dots, \sigma_{\freemap}(s_n)),\]
     \hl{we can apply the (CONG of $=$) axiom scheme \eqref{CONG:axiom:deductive} and the CUT rule to derive that
     $\Phi \derivefrel \forall (A, d_A). \sigma_{\freemap}(s)=s$.
     
     Analogously, we derive that 
     $\sem t ^{F(A,d_A)}_\freemap= [t]_\equiv$.
     Hence,   $[s]_\equiv= [t]_\equiv$, which by definition of $\equiv$  means $ \Phi \vdash \forall (A, d_A). s=t$.
    }

     For \hl{(ii)}, we analogously have that
     $\Delta(\sem s ^{F(A,d_A)}_\freemap,\sem t ^{F(A,d_A)}_\freemap)\leq \epsilon$ implies (by \autoref{lem_free_interpret}, as above) $\Delta([s]_\equiv,[t]_\equiv)\leq \epsilon$. Then, by the definition of $\Delta$ as $d$ (\autoref{cor:distancequotientedterms} and \autoref{defn:equivrelterms}) and by \autoref{lemma_connection_syntax_semantics_1}, we conclude $\Phi \vdash \forall (A, d_A). s=_\epsilon t$.
\end{proof}

We can now prove the completeness theorem.

\begin{thm}[Completeness of the deductive system]\label{completeness_of_proof_system_qa}
Fix a signature $\Sigma$ and a class  $\Phi\subseteq \MEq(\Sigma)$ of equations and quantitative equations. For all  $\phi\in\MEq(\Sigma)$,
\begin{center}
if $\Phi\implyfrel_{\FRel}  \phi$ then 
$\Phi\derivefrel_{\FRel}  \phi$.
\end{center}
\end{thm}

\begin{proof}
    Let us consider first the case of $\phi$ being an equation of the form $\forall (A,d_A). s = t$, for some $\FRel$ space $(A,d_A)$ and terms $s,t\in\TermsA$.

By definition of the entailment relation ($\implyfrel_{\FRel}$), the hypothesis $\Phi\implyfrel_{\FRel}  \phi$ implies that for all $\algB \in \QModels(\Phi)$ and for all nonexpansive interpretations $\imap: (A,d_A) \to (B,d_B)$, it holds that $\sem s^\algB_\imap=\sem t ^\algB_\imap$. 
Hence, since $F(A,d_A)\in \QModels(\Phi)$ \hl{(\autoref{lem_free_models})} and $\freemap$ is nonexpansive \hl{(\autoref{definition_of_iota})}, we have $\sem s ^{F(A,d_A)}_\freemap = \sem t ^{F(A,d_A)}_\freemap$. Then by \autoref{lem_completeness} we conclude that $\Phi \derivefrelfrel \forall (A, d_A). s= t$.

Analogously, for quantitative equations, if $\phi$ is of the form $\forall (A, d_A). s=_\epsilon t$ then we derive from $\Phi\implyfrelfrel  \phi$ that
$\Delta(\sem s ^{F(A,d_A)}_\freemap,\sem t ^{F(A,d_A)}_\freemap)\leq \epsilon$. By \autoref{lem_completeness} we conclude 
$\Phi \derivefrel_{\FRel} \forall (A, d_A). s=_\epsilon t$.
\end{proof}

\begin{cor}[Soundness and Completeness]\label{soundnesscompleteness}
    Fix a signature $\Sigma$ and a class $\Phi\subseteq \MEq(\Sigma)$. For all equations and quantitative equations $\phi\in\MEq(\Sigma)$,
    \begin{center}$\Phi \implyfrelfrel\phi \quad \Longleftrightarrow \quad  \Phi \derivefrelfrel \phi$.\end{center}
\end{cor}

\section{The Free-Forgetful Adjunction and Strict Monadicity}
\label{section:result:3}

By relying on the construction of free objects shown in Section \ref{section:result:2}, we identify in this section the free-forgetful adjunction arising from it, together with the associated monad. We then proceed to prove the strict monadicity of the adjunction.

Recall from \autoref{prop_freeadjunction} %
that, given a functor $U: \catD \to \catC$  such that $\catD$ has 
$U$-free objects, there is a functor $F: \catC \to \catD$ which assigns to each element of $\catC$ its corresponding $U$-free object,
and which gives an adjunction $F \dashv U$ and a monad with functor $U\circ F$.

We have seen in Section \ref{section:result:2}
that the forgetful functor $U:\QModelsCat({\Phi})\rightarrow \FRel$ has $U$-free objects, identified (up to isomorphism) as the quantitative algebras of quotiented terms. Hence, using the recipe from \autoref{prop_freeadjunction}, we obtain the adjunction $F \dashv U$, where $F$ is the functor mapping each $\FRel$ space $(A,d_A)$ to the free quantitative algebra of quotiented terms $F(A,d_A)= (\TermsA/_{\equiv_A}, \FreedistA, \{op^{F(A,d_A)}\}_{op\in \Sigma})$.\footnote{\hl{We sometimes use superscripts to specify which $\FRel$ space generates the free algebra.}} %
For a nonexpansive function $f: (A,d_A) \to (B,d_B)$, the functor $F$ gives the nonexpansive homomorphism of quantitative algebras $F(f): F(A,d_A)\to F(B,d_B) $ which is the unique homomorphic extension of $f$. It can be defined inductively as follows:
\[F(f)[a]_{\equiv_A} = [f(a)]_{\equiv_B} \qquad F(f)[op(t_1,\dots,t_n)]_{\equiv_A} = op^{F(B,d_B)}(F(f)[t_1]_{\equiv_A}, \dots, F(f)[t_n]_{\equiv_A}).\]

We denote the resulting $\FRel$ monad on $UF$ (by \autoref{prop_adjunction_monad}) by $\qterms{\sig,\Phi}$, and we can describe it concretely as follows.
\begin{itemize}
    \item The functor $\qterms{\sig,\Phi}=U\circ F$ maps an object $(A,d_A)$ to \[\qterms{\sig,\Phi}(A,d_A)=(\TermsA/_{\equiv_{A}}, \FreedistA),\] and a morphism $f:(A,d_A) \to (B,d_B)$ to \[\qterms{\sig,\Phi}(f): (\TermsA/_{\equiv_{A}}, \FreedistA) \to (\TermsB/_{\equiv_{B}}, \FreedistB),\] where $\qterms{\sig,\Phi}(f)([t]_{\equiv_A})= U F (f)([t]_{\equiv_A})$ is the nonexpansive homomorphism which can be specified by induction on terms $t$ as above.
    
    \item The unit $\eta$ is given by the unit of the adjunction, i.e., for every $(A,d_A)\in\FRel$ we have that $\eta_{(A,d_A)}$ is the function $\freemap_{(A,d_A)}$ \hl{proven nonexpansive in \autoref{definition_of_iota}}. %
    Concretely, this is the function assigning to $a\in A$ the equivalence class $[a]_{\equiv_A}$.

    \item The multiplication of $\qterms{\sig,\Phi}$ is given by $\mu_{(A,d_A)} = U(\varepsilon_{F(A,d_A)})$, where $\varepsilon$ is the counit of the adjunction. Concretely, this is the function substituting each occurrence of an equivalence class of terms with a representative of the class, thus ``flattening'' the term as follows:
    \[\mu_{(A,d_A)}\Big( \big[ t\big([t_1]_{\equiv_A},\dots, [t_n]_{\equiv_A}  \big) \big]_{\equiv_{F(A,d_A)}}\Big) = [t(t_1,\dots, t_n)]_{\equiv_A} .\]

\end{itemize}

We now proceed to prove that the forgetful functor $U: \QModelsCat({\Phi})\rightarrow \FRel$ is strictly monadic (see \autoref{monadic:functor:def}), i.e., that there is an isomorphism of categories
\begin{center}
$\EM(\qterms{\sig,\Phi}) \cong \QModelsCat (\Phi)$,
\end{center}
where $\EM(\qterms{\sig,\Phi})$ is the \emph{Eilenberg--Moore} category of 
 $\qterms{\sig,\Phi}$ (see \autoref{def:algebra-of-a-monad}).

We will prove in \autoref{thm_monadicity_QA} that $U: \Catqalg \to \FRel$ satisfies condition (3) of Beck's theorem (\autoref{thm_beck_absolute}), from which we conclude that it is strictly monadic.
In order to do so, we first consider the special case when $\Phi=\emptyset$. In this case, we recall that $\QModelsCat(\emptyset)=\Qalg$ consists of the category of \textit{all} quantitative $\Sigma$-algebras.

\begin{thm}%
\label{thm_monadicity_Uempty}
The forgetful functor $U_\emptyset: \Catqalgempty \to \FRel$ is strictly monadic.
\end{thm}
\begin{proof}
    Instantiating the results of the previous sections to the case $\Phi = \emptyset$, we get that $U_{\emptyset}$ has a left adjoint ($U_{\emptyset}$-free objects exist by \autoref{free_algebra_theorem}). Therefore, by Beck's theorem (condition (2) in \autoref{thm_beck_absolute}), it is enough to show that $U_{\emptyset}$ strictly creates coequalizers for all $\Catqalgempty$-arrows $f,g$ such that $U_\emptyset(f), U_\emptyset(g)$ has an absolute coequalizer (in $\FRel$). We follow the structure of the proof of \cite[\S VI.8, Theorem 1]{MacLane71}, i.e., of the analogous result for $\Set$. To ease notation, in this proof we write $U$ instead of $U_\emptyset$.

    Let $f,g:\algA \to \algB$ be $\Catqalgempty$-arrows such that $U(f), U(g): (A,d_A) \to (B,d_B)$ have an absolute coequalizer $e: (B,d_B) \to (C,d_C)$. We show that there exists a quantitative algebra $\algC$ in $\Catqalgempty$ and a homomorphism $u:\algB \to \algC$ such that \ref{item:monadicity_carrier} $U(\algC)=(C,d_C)$, \ref{item:monadicity_morphism} $e=U(u)$, \ref{item:monadicity_unique} $\algC$ and $u$ are unique satisfying the previous two items, and \ref{item:monadicity_coeq} $u$ is a coequalizer of $f, g$.
    \begin{equation*}
\begin{tikzcd}[column sep=1.3em, row sep=1.3em]
	\Catqalgempty & {\mathbb{A}} && {\mathbb{B}} && {\mathbb{C}} \\
	\FRel & {(A,d_A)} && {(B,d_B)} && {(C,d_C)}
	\arrow["u", from=1-4, to=1-6]
	\arrow["g"', curve={height=6pt}, from=1-2, to=1-4]
	\arrow["f", curve={height=-6pt}, from=1-2, to=1-4]
	\arrow["e = U(u)", from=2-4, to=2-6]
	\arrow["{U(f)}", curve={height=-6pt}, from=2-2, to=2-4]
	\arrow["{U(g)}"', curve={height=6pt}, from=2-2, to=2-4]
	\arrow["U"', from=1-1, to=2-1]
\end{tikzcd}
    \end{equation*}
    \begin{enumerate} [i)]
        \item\label{item:monadicity_carrier} The carrier of $\algC$ is $(C,d_C)$, we just need to define the interpretations $\{ op^\mathbb{C}\}_{op\in\Sigma}$ of the operations in $\Sigma$. Fix an $n$-ary $op \in \Sigma$, \hl{and let $L^n : \FRel \to \FRel$ be the discrete $n$-ary product functor on $\FRel$ defined as}
        \begin{equation}\label{eq:discreteproductlifting}
        L^n = \FRel \xrightarrow{U_{\FRel\rightarrow\Set}} \Set \xrightarrow{(\_)^n} \Set \xrightarrow{D} \FRel,
        \end{equation}
        \hl{where $(\_)^n$ is the $n$-ary product functor in $\Set$ and $D$ is the discrete functor defined above Proposition \ref{prop:frel_discrete_adjoint}.}
         Concretely, $L^n$ sends $(A,d_A)$ to $(A^n,L^n(d_A))$, where $L^n(d_A) = d^{A^n}_\bot$ is the discrete fuzzy relation that assigns distance $1$ to all pairs of elements of $A^n$, and $f: (A,d_A) \to (B,d_B)$ to $f^n: (A^n, L^n(d_A)) \to (B^n, L^n(d_B))$.
        
        By \autoref{prop:frel_discrete_adjoint}, any function out of a discrete space is nonexpansive since $\FRel(DX,(Y,d_Y))\simeq \Set(X,Y)$. Therefore, there are nonexpansive functions
        \[ \hat{op}^\mathbb{A}: (A^n,L^n(d_A)) \rightarrow (A,d_A) \quad \text{ and } \quad  \hat{op}^\mathbb{B}: (B^n,L^n(d_B)) \rightarrow (B,d_B) \]
        whose underlying functions are $op^{\algA}$ and $\op^{\algB}$ respectively. We then have %
        \begin{align*}
            e \circ \hat{op}^\algB \circ L^nU(f) &= e \circ U(f) \circ \hat{op}^\algA &&\text{(by $f$ a homomorphism: $op^{\algB} \circ f^n = f \circ op^{\algA}$)}\\
            &= e \circ U(g) \circ \hat{op}^\algA &&\text{(by $e$ a coequalizer)}\\
            &=  e \circ \hat{op}^\algB \circ L^nU(g). &&\text{(by $g$ a homomorphism: $op^{\algB} \circ g^n = g \circ op^{\algA}$)}
        \end{align*}

        Note that, \hl{since the arrow $e$ is an absolute coequalizer of the parallel pair $U(f),U(g)$ in $\FRel$,}  $L^n(e): (B^n,L^n(d_B)) \to (C^n, L^n(d_C))$ is a coequalizer of $L^nU(f),L^nU(g)$ in $\FRel$. Hence we have the following diagram in $\FRel$:
        \begin{equation}\label{diag:defopfromh}
            \begin{tikzcd}[sep=small]
                {(A^n, L^n(d_A))} & {} & {(B^n, L^n(d_B))} && {(C^n, L^n(d_C))} \\
                \\
                && {(B,d_B)} && {(C,d_C)}
                \arrow["{L^nU(g)}"', curve={height=6pt}, from=1-1, to=1-3]
                \arrow["{L^nU(f)}", curve={height=-6pt}, from=1-1, to=1-3]
                \arrow["{L^n(e)}", from=1-3, to=1-5]
                \arrow["{\hat{op}^{\algB}}"', from=1-3, to=3-3]
                \arrow["e"', from=3-3, to=3-5]
            \end{tikzcd}
        \end{equation}
        The universal property of the coequalizer means there exists a unique morphism $h: (C^n, L^n(d_C)) \to (C, d_C)$ that completes the square in \eqref{diag:defopfromh}, namely, such that
        \begin{equation}
        \label{eqn:defopc}
        h \circ L^n(e) = e \circ \hat{op}^\algB.
        \end{equation}
        Let the interpretation of $op$ in $\algC$ be the function underlying $h$, i.e. $op^{\algC} = U_{\FRel \to \Set}(h)$.

        \item\label{item:monadicity_morphism} Our construction of $\algC$ (carrier and operations) ensures that $e$ is a homomorphism $\algB \to \algC$. Indeed, $e$ is a nonexpansive function $(B,d_B) \to (C,d_C)$, and we can show it satisfies the homomorphism property by applying $U_{\FRel \to \Set}$ to \eqref{eqn:defopc} to obtain $op^{\algC} \circ (U_{\FRel \to \Set}e)^n = U_{\FRel \to \Set}e \circ op^{\algB}$.
        
        \item\label{item:monadicity_unique} Let $\algC'$ be an algebra and $u': \algB \to \algC'$ be a homomorphism satisfying the previous two items. We show that $\algC = \algC'$ and $u = u'$. By \ref{item:monadicity_carrier}, the carrier of $\algC'$ is $(C,d_C)$, so $op^{\algC'}$ has type $C^n \to C$ for each $n$-ary $op \in \Sigma$. We can construct (as we did $\hat{op}^{\algA}$ and $\hat{op}^{\algB}$) the nonexpansive function $\hat{op}^{\algC'}: (C^n,L^n(d_C)) \to (C,d_C)$ that satisfies $U_{\FRel \to \Set}\hat{op}^{\algC'} = op^{\algC'}$.
        
        Now, since $u'$ is a homomorphism and $U(u') = e$, we have, for all $op \in \Sigma$,
        \[op^{\algC'} \circ (U_{\FRel \to \Set}e)^n = U_{\FRel \to \Set}e \circ op^{\algB}.\]
        \hl{By applying the definitions and by functoriality, we can rewrite the left hand side:}
        \begin{align*}
        op^{\algC'} \circ (U_{\FRel \to \Set}e)^n 
        &= op^{\algC'} \circ U_{\FRel \to \Set}D((U_{\FRel \to \Set}e)^n)\\
        &= op^{\algC'} \circ U_{\FRel \to \Set}(L^n(e))\\
        &= U_{\FRel \to \Set}(\hat{op}^{\algC'}) \circ  U_{\FRel \to \Set}(L^n(e))\\
        &= U_{\FRel \to \Set}(\hat{op}^{\algC'} \circ L^n(e)).
        \end{align*}
        \hl{Analogously, we can rewrite the right hand side:}
        $$
        U_{\FRel \to \Set}e \circ op^{\algB}
        = 
        U_{\FRel \to \Set}e \circ U_{\FRel \to \Set}(\hat{op}^{\algB})
        =
        U_{\FRel \to \Set}(e \circ \hat{op}^{\algB}),
        $$
        and since $U_{\FRel \to \Set}$ is faithful, we conclude that $\hat{op}^{\algC'} \circ L^n(e) = e \circ \hat{op}^{\algB}$. Namely, $\hat{op}^{\algC'}$ also completes the square in \eqref{diag:defopfromh}. By uniqueness, we get $\hat{op}^{\algC'} = h$, hence $op^{\algC'} = op^{\algC}$ and $\algC = \algC'$. The equality $u=u'$ follows from faithfulness of $U$ as $U(u) = e = U(u')$.
        \item\label{item:monadicity_coeq} It remains to prove that $u$ is a coequalizer of $f,g$ in $\Catqalgempty$, i.e., that for any quantitative algebra $\algD$ in $\Catqalgempty$ and for any homomorphism $k: \algB \to \algD$, if $k \circ f=k\circ g$ then $k$ factorises uniquely through $u$ (i.e. $\exists ! l. k = l \circ u$).
        
        Given such a $k: \algB \to \algD$, since $e = U(u)$ is the coequalizer of $U(f)$ and $U(g)$, $U(k)$ factorises uniquely through $e$, say via $U(k) = m \circ e$ for $m: (C,d_C) \to (D,d_D)$. It remains to prove that $m$ is a homomorphism from $\algC$ to $\algD$ (uniqueness as a homomorphism follows because $U$ is faithful). This amounts to showing the following identity in $\FRel$:%
        \begin{equation*}
        m \circ \hat{op}^\algC= \hat{op}^\algD \circ L^n(m), 
        \end{equation*}
        where $\hat{op}^{\algC}: (C^n,L^n(d_C)) \to (C,d_C)$ and $\hat{op}^{\algD}: (D^n, L^n(d_D)) \to (D,d_D)$ are constructed as above. That follows from the derivation below, after noting that, \hl{since $e$ is an absolute coequalizer,} $L^n(e)$ is a coequalizer for $L^nU(f),L^nU(g)$, and thus it is an epimorphism.
        \begin{align*}
        m \circ \hat{op}^\algC \circ L^n(e) &= m \circ e \circ \hat{op}^\algB &&\text{(by $e$ a homomorphism, see point (ii))}\\
        &= U(k) \circ \hat{op}^\algB &&\text{(by construction of $m$)}\\
        &= \hat{op}^\algD \circ L^n(U(k)) &&\text{(by $k$ a homomorphism)}\\
        &= \hat{op}^\algD \circ L^n(m) \circ L^n(e) &&\text{(by $L^n$ a functor and $U(k)=m\circ e$)}\qedhere
        \end{align*}
    \end{enumerate}
\end{proof}

We are now going to use the above result, which deals with the special case $\Phi=\emptyset$, to prove \autoref{thm_monadicity_QA} in its full generality, showing that the functor 
$U:\Catqalg \to \FRel$ is strictly monadic for arbitrary $\Phi$. The proof is a generalisation of the analogous result in \cite[Theorem 2.17]{DBLP:conf/lics/Adamek22}, which proves strict monadicity in the  framework of \cite{DBLP:conf/lics/MardarePP16}.

In particular, we prove that $U$ satisfies condition (3) of Beck's theorem, i.e., in contrast with \autoref{thm_monadicity_Uempty}, we use split coequalizers instead of absolute coequalizers. We do so since split coequalizers guarantee the existence of a right inverse of the coequalizer, which allows us to apply the following fact (\autoref{lem_hom_inverse}): $\Catqalg$ is closed under the images of homomorphisms that have a right inverse in $\FRel$.

\begin{lem}
\label{lem_hom_inverse}
Let $\algA=(A,d_A,\{op^\mathbb{A}\}_{op\in\Sigma}) \in \Catqalg$, for some class of $\FRel$ equations and quantitative equations $\Phi\subseteq\MEq(\Sigma)$, and let $\algB=(B,d_B,\{op^\mathbb{B}\}_{op\in\Sigma}) \in \Catqalgempty$.
If there is a homomorphism of quantitative algebras $f:\algA \to \algB$ such that $U(f)$ has a nonexpansive right inverse $g: (B,d_B) \to (A,d_A)$ then $\algB$ is in $\Catqalg$.
\end{lem}

\begin{proof}

We show that, under the hypothesis of the statement, for every $\phi\in\Phi$, $\mathbb{B}\models \phi$ holds. We first consider the case of $\phi$ being an $\FRel$ equation
$$
\phi = \ \ \ \forall (X,d_X). s=t 
$$
for some $\FRel$ space $(X,d_X)$ and terms $s,t\in \Terms{X}$.
By definition (of $\mathbb{B}\models \phi$), we need to show that $\sem s_\imap^\algB= \sem t_\imap^\algB$, for all nonexpansive interpretations $\imap: (X,d_X)\rightarrow (B,d_B)$.

    Let $\imap: (X,d_X) \to (B,d_B)$ be such an interpretation. Then $g\circ \imap: (X,d_X) \to (A,d_A)$ is an interpretation in $\algA$ (which is nonexpansive as the composition of nonexpansive functions is nonexpansive).
    Moreover, one can show by induction that for any term $r\in \Terms{X}$ (and so, in particular, $s$ and $t$),%
    \begin{equation}\label{assignment_inverse_lemma}
    U(f)(\sem r_{g \circ \imap}^\algA) = \sem r_\imap^\algB.
    \end{equation}

        Then, since, by hypothesis, $\algA$ is a model of  $\Phi$, we know that $\mathbb{A}\models \phi$ and therefore
    \begin{equation}
        \label{equationAgi}
        \sem s_{g \circ \imap}^\algA= \sem t_{g \circ \imap}^\algA.
    \end{equation}
    We conclude that $\algB$ satisfies the equation $\phi$ under the interpretation $\imap$:
    \[ \sem s_{\imap}^\algB \stackrel{\eqref{assignment_inverse_lemma}}{=} U(f)(\sem s_{g \circ \imap}^\algA) \stackrel{\eqref{equationAgi}}{=} U(f)(\sem t_{g \circ \imap}^\algA)  \stackrel{\eqref{assignment_inverse_lemma}}{=} \sem t_{\imap}^\algB.\]

We now consider the case of $\FRel$ quantitative equations $\phi\in\Phi$ of the form 
$$\phi = \ \ \  \forall (X,d_X). s=_\epsilon t .$$
Since, by hypothesis, $\algA \models \phi$, we have 
$
d_A( \sem{s}_{g \circ \imap}^{\mathbb{A}} , \sem{t}_{g \circ \imap}^{\mathbb{A}} )\leq \epsilon
$, from which we derive
\begin{align*}
    d_B(\sem s_{\imap}^\algB, \sem t_{\imap}^\algB)
    &=d_B(U(f)(\sem s_{g\circ\imap}^\algB), U(f)(\sem t_{g \circ \imap}^\algB)) &&\text{(by \eqref{assignment_inverse_lemma})}\\
    &\leq d_A(\sem s_{g\circ \imap}^\algA, \sem t_{g\circ \imap}^\algA) &&\text{(by $U(f)$ nonexpansive)}\\
    &\leq \epsilon. &&\text{(by $\algA$ a model of $\Phi$)}\qedhere
\end{align*}
\end{proof}

\begin{thm}%
\label{thm_monadicity_QA}
The forgetful functor $U: \Catqalg \to \FRel$ is strictly monadic.
\end{thm}

\begin{proof}
\hl{As discussed at the beginning of Section \ref{section:result:3} (see also result (III) in Section \ref{results_announcement_section})}, we know that there is an adjunction $F \dashv U$.
    We now prove that the forgetful functor $U: \Catqalg \to \FRel$ strictly creates coequalizers \hl{of $U$-split pairs, i.e.,} of all $\Catqalg$-arrows $f,g$ such that $U(f), U(g)$ has a split coequalizer (in $\FRel$). From this, it immediately follows by Beck's theorem (\autoref{thm_beck_absolute}) that $U$ is strictly monadic. 

Let $f,g:\algA \to \algB$ be $\Catqalg$-arrows such that $U(f), U(g): (A,d_A) \to (B,d_B)$ have a split coequalizer $e: (B,d_B) \to (C,d_C)$. We show that there exists a unique algebra $\algC$ in $\Catqalg$ such that $U(\algC)=(C,d_C)$, such that $e=U(u)$ for $u:\algB \to \algC$ an arrow in $\Catqalg$, and such that $u$ is a coequalizer of $f, g$ in $\Catqalg$.

Recall that $U_\emptyset$ is the forgetful functor $U_\emptyset:\Catqalgempty \to \FRel$ from \autoref{thm_monadicity_Uempty}. Since $U_\emptyset$ is strictly monadic (by \autoref{thm_monadicity_Uempty}), it satisfies condition (3) of \autoref{thm_beck_absolute}.
Since $\Catqalgempty$-arrows between objects in $\Catqalg$ coincide with $\Catqalg$-arrows, i.e., they are both defined as nonexpansive homomorphisms of quantitative $\Sigma$-algebras, condition (3) of \autoref{thm_beck_absolute} implies that there is a unique algebra $\algC$ in $\Catqalgempty$ such that $U_\emptyset(\algC)=C$, such that $e=U_\emptyset(u)$ for $u:\algB \to \algC$ an arrow in $\Catqalgempty$, and such that $u$ is a coequalizer of $f,g$ in $\Catqalgempty$.

Now, $e = U_\emptyset(u)$ is a split coequalizer, so it has a right inverse $r: (C,d_C) \to (B,d_B)$. \hl{Therefore the quantitative algebra homomorphism $u$ satisfies the requirements of \autoref{lem_hom_inverse} and thus we know that $\algC$ satisfies all the equations and quantitative equations satisfied by $\algB$}. In particular, $\algC$ is a model of $\Phi$ because $\algB$ is a model of $\Phi$, i.e., $\algC$ and $u$ are in $\Catqalg$.

We conclude by noting that the uniqueness and universal property (being a coequalizer of $f,g$) of $u$ that were true in $\Catqalgempty$ are also true in $\Catqalg$ because the latter is a full subcategory of the former.
\end{proof}

\section{Lifting Presentations from \texorpdfstring{$\Set$}{Set} to \texorpdfstring{$\FRel$}{FRel}}
\label{section:lifting}

We recall, from  \autoref{defn:setpres}, that a $\Set$ monad $M$ has an \emph{equational presentation} if there exists a class of equations $\Phi\subseteq \Eq$ over some signature $\Sigma$ such that $\tmonad{\Phi} \cong M$. A well known result, dating back to the seminal works of Lawvere \cite{Lawvere1963} connecting the theory of monads with Universal Algebra, states that a $\Set$ monad $M$ has an equational presentation if and only if it is \emph{finitary} (see, e.g., \cite[Chapter 3]{AR1994}). 
This result provides a useful correspondence between a logical notion (definability by equations) and a categorical one (finitary monad).

In the context of our theory of quantitative algebras, it is natural to give the following definition of $\FRel$ monads having a \emph{quantitative equational presentation.}

\begin{defi}[Quantitative Equational Presentation]\label{defn:frelpresentation}
\hl{An $\FRel$} monad $M$ has a quantitative equational presentation if there is a class of equations and quantitative equations $\th\subseteq \MEq(\sig)$ over some signature $\Sigma$ such that $\qterms{\sig,\th} \cong M$.
\end{defi}
\begin{lem}\label{lem:presisogmet}
    \hl{An $\FRel$ monad $M$ is presented by $\th\subseteq \MEq(\sig)$ if and only if there is an isomorphism $\QModelsCat(\Phi) \cong \EM(M)$ that commutes with the forgetful functors to $\FRel$.}
\end{lem}
\begin{proof}
    \hl{By \autoref{prop:monadisocatiso}, a presentation is equivalent to an isomorphism $\EM(M) \cong \EM(\qterms{\sig,\qth})$ commuting with the forgetful functors. Since we can compose that with the isomorphism $\EM(\qterms{\sig,\qth}) \cong \QModelsCat(\Phi)$ resulting from \autoref{thm_monadicity_QA}, which also commutes with the forgetful functors, we get the desired equivalence.}
\end{proof}
\begin{exa}\label{exmp:preseasypset}
    \hl{The $\FRel$ monad $\mP'$ of \autoref{exmp:monadsgmet} is presented by the quantitative equational theory of semilattices $\Phi$ in \eqref{eqn:qthsemilattice}. Indeed, we already noted that $\QModelsCat(\Phi)$ is the category of semilattices equipped with a fuzzy relation, and one can show that $\EM(\mP')$ is the category of $\mP$-algebras equipped with a fuzzy relation. Then, since $\mP$ is presented by the theory of semilattices (\autoref{exmp:monadpres}.(1)), i.e., there is an isomorphism between semilattices and $\mP$-algebras, it follows that $\QModelsCat(\Phi) \cong \EM(\mP')$. Therefore, by \autoref{lem:presisogmet} we conclude that $\mP'$ is presented by $\Phi$.}
\end{exa}
    
The problem of characterising which $\FRel$ monads have a quantitative equational presentation in terms of categorical properties seems to be hard. For instance, Ad{\'{a}}mek provides in \cite[Example 4.1]{DBLP:conf/lics/Adamek22} an example of a class $\th\subseteq \MEq(\sig)$ such that $\qtmonad{\Phi}$ is not finitary. This example is formulated in the context of the framework of Mardare, Panangaden and Plotkin \cite{DBLP:conf/lics/MardarePP16}, but can be reformulated in our setting (see Section \ref{subsec:MPPcomp}). %

In this section we establish (\autoref{thm:correspondencemonliftthext}) a correspondence between $\FRel$ monads that are \emph{liftings of} $\Set$ monads (\autoref{defn:monadlifting}) having an equational presentation $\Psi\subseteq \Eq$ (i.e., finitary $\Set$ monads) and \emph{quantitative equational presentations} $\Phi\subseteq \MEq(\Sigma)$ that are \emph{extensions} of $\Psi\subseteq \Eq$ (\autoref{defn:liftingth}).

Before proceeding with the formal definitions, since we have to deal with both $\Set$ and $\FRel$ monads, and with both classes of equations in $\mathcal{P}(\Eq)$ and classes of $\FRel$ equations and quantitative equations in $\mathcal{P}(\MEq(\sig))$, in the rest of this section we adopt the following notational convention:
\begin{enumerate}
\item We reserve the letters $M$ for $\Set$ 
monads and $\Phi$ for classes of equations in $\Eq$.
\item We use the letters $\lift{M}$ and $\qth$, with the ``hat'' notation, for $\FRel$ monads $\lift{M}$ and classes of $\FRel$ equations and quantitative equations $\qth\subseteq \MEq(\Sigma)$, respectively.
\end{enumerate}

\noindent
We first give the standard (see, e.g., \cite[p.~121]{Beck69}) definition of lifting of a monad.

\begin{defi}[Lifting]\label{defn:monadlifting} \hl{An $\FRel$} monad $(\lift{M},\lift{\eta},\lift{\mu})$ is a \emph{lifting} of a $\Set$ monad $(M,\eta,\mu)$ if
\begin{equation*}
\begin{tikzcd}[row sep=1.1em, column sep=1.3em]
	\FRel & \FRel \\
	\Set & \Set
	\arrow["{\lift{M}}", from=1-1, to=1-2]
	\arrow["U"', from=1-1, to=2-1]
	\arrow["U", from=1-2, to=2-2]
	\arrow["M"', from=2-1, to=2-2]
\end{tikzcd}, \quad U\lift{\eta} = \eta U, \text{ and } U\lift{\mu} = \mu U.
\end{equation*}
\end{defi}
More explicitly,
\begin{enumerate}[(i)]
    \item\label{prop:monliftobjects} the action of $\lift{M}$ on objects is an assignment $(A,d_A) \mapsto (MA,\lift{d}_A)$ that lifts every fuzzy relation $d_A$ on $A$ to a fuzzy relation $\lift{d}_A$ on $MA$,
    \item\label{prop:monliftfunctor} the actions of $M$ and $\lift{M}$ on morphisms coincide set-theoretically,
    \item\label{prop:monliftunit} the units $\eta$ and $\lift{\eta}$ coincide set-theoretically. This means that for any $(A,d_A)\in\FRel$, the function $\eta_{A}:A\rightarrow MA$ is a nonexpansive map $ \lift{\eta}_{(A,d_A)}:(A,d_A) \to (MA,\lift{d}_A)$.
    
    \item\label{prop:monliftmult} the multiplications $\mu$ and $\lift{\mu}$ coincide set-theoretically, i.e., for any $(A,d_A)\in\FRel$, the function $\mu_{A}:MMA\rightarrow MA$ is a nonexpansive map $\lift{\mu}_{(A,d_A)}:(MMA,\skew{1.2}\doublewidehat{d}_A) \to (MA,\lift{d}_A)$.
\end{enumerate}

\begin{exa}\label{exmp:monliftmP}
    \hl{It is clear from the definition that the $\FRel$ monad $\mP'$ (\autoref{exmp:monadsgmet}) and the $\Set$ powerset monad $\mP$ satisfy the four items above, and hence $\mP'$ is a lifting of $\mP$.}
\end{exa}

We now formally define when a class $\qth\subseteq \MEq(\sig)$ is an extension of a class $\Phi \subseteq \Eq$.

\begin{defi}[Quantitative Extension]\label{defn:liftingth}
Let $\Sigma$ be a signature. A class  $\lift{\Phi}\subseteq \MEq(\Sigma)$ is a \emph{quantitative extension} of a class $\Phi\subseteq \Eq$ if
    \begin{equation}\label{eqn:defnliftingth}
        \begin{matrix}
            \text{for all $(A,d_A) \in \FRel$ and $s,t \in \TermsA$,}\\
            \th \imply \forall A.s=t  \Longleftrightarrow \qth \implyfrelfrel \forall (A,d_A).s=t.
        \end{matrix}
    \end{equation}
\end{defi}

This guarantees that the equations entailed by $\lift{\Phi}$ ``coincide'' with those of $\Phi$, in the sense that $\forall A.s=t$ follows from $\Phi$ if and only if $\forall (A,d_A).s=t$ follows from $\lift{\Phi}$, for all possible fuzzy relations $d_A$ on $A$.

We are now ready to state our main result of this section.
\begin{thm}\label{thm:correspondencemonliftthext}
Let $(M,\eta,\mu)$ be a monad on $\Set$ presented by $\th\subseteq \Eq$.
\begin{enumerate}[(1)]
    \item\label{thm:exttolifting} For any quantitative extension $\qth$ of $\th$, there is a monad lifting $\lift{M}$ of $M$ presented by $\qth$.
    \item\label{thm:liftingtoext} For any monad lifting $\lift{M}$ of $M$, there is a quantitative extension $\qth$ of $\th$  presenting $\lift{M}$. 
\end{enumerate}
\end{thm}

The goal of the rest of this section is to prove the above theorem. We first give a proof sketch illustrating the main ideas. All technical details are delayed to Subsection \ref{sec:liftingdetails}.

For \ref{thm:exttolifting}, we are a given a class $\qth\subseteq\MEq(\Sigma)$ extending $\Phi\subseteq \Eq$, and a $\Set$ monad $M$ presented by $\Phi$ (with a given monad isomorphism $\rho:\tmonad{\th} \cong M$). Our goal is to exhibit \hl{an $\FRel$} monad $\lift{M}$ that lifts $M$ and is presented by $\qth$.

As a first step, we establish that, from the assumption that $\qth$ extends $\Phi$, it follows that $\qterms{\sig,\qth}$ is a monad lifting of $\terms{\sig,\th}$ (\autoref{lem:extensionislifting}). Hence, diagrammatically, the assumptions can be depicted as below (left) and our goal is to complete the diagram as in the (right):
\[\begin{tikzcd}[column sep=1.3em, row sep=1.2em]
	& {\qterms{\sig,\qth}} && {\lift{M}} & {\qterms{\sig,\qth}} \\
	M & {\terms{\sig,\th}} && M & {\terms{\sig,\th}}
	\arrow["{\stackrel{\rho}{\cong}}"{description}, shift left=1, draw=none, from=2-4, to=2-5]
	\arrow["{\stackrel{\lift{\rho}}{\cong}}"{description}, shift left=1.5, draw=none, from=1-4, to=1-5]
	\arrow["U"', from=1-4, to=2-4]
	\arrow["U", from=1-5, to=2-5]
	\arrow["U", from=1-2, to=2-2]
	\arrow["{\stackrel{\rho}{\cong}}"{description}, draw=none, shift left=1, from=2-1, to=2-2]
\end{tikzcd}\]
We thus need to define \hl{an $\FRel$} monad $\lift{M}$ lifting $M$. We remark that, from \autoref{defn:monadlifting} of monad lifting, the unit, the multiplication and the action on morphisms on any such $\lift{M}$ are fully determined by $M$. We therefore just need to specify the action of $\lift{M}$ on objects $(A,d_A)\in\FRel$, respecting the constraint of \autoref{defn:monadlifting}:
$$(A,d_A) \mapsto (MA, \lift{d}_A).$$
To define the fuzzy relation $\lift{d}_A: (MA)^2 \rightarrow [0,1]$, we use the monad isomorphism $\rho:  \terms{\sig,\th} \cong M$ to get a bijection
$$
\rho_A^{-1}: MA \rightarrow \TermsA/_{\equiv^A_{\Phi}}
$$
between $MA$ and the set $\TermsA/_{\equiv^A_{\Phi}}$ underlying $\terms{\sig,\th}$ and, as we have already established, also $\qterms{\sig,\qth}$. We can now define $ \lift{d}_A$ as follows:
\[\forall m,m' \in MA.\ \  \lift{d}_A(m,m') = \FreedistA(\rho_A^{-1}(m),\rho_A^{-1}(m')),\]
where $\FreedistA$ is the distance on quotiented terms obtained in \autoref{cor:distancequotientedterms}.

This completes the definition of the $\FRel$ monad $\lift{M}$. The verification that all these definitions are valid in $\FRel$ (i.e., that the unit, the multiplication and the action of morphisms yield nonexpansive maps) is straightforward. The fact that $\lift{M}$ is a lifting of $M$ follows directly from its construction.

Finally, we define the components of the monad isomorphism $\lift{\rho}$
$$\lift{\rho}_{(A,d_A)}: \qterms{\sig,\qth}(A,d_A) \rightarrow \lift{M}(A,d_A)
$$
to coincide with $\rho_A:\terms{\sig,\th}A \to MA$, for every $(A,d_A)\in\FRel$. Checking that $\lift{\rho}_{(A,d_A)}$ is indeed a map in $\FRel$ (i.e., it is nonexpansive) and that $\lift{\rho}$ satisfies the constraints of a monad isomorphism is also straightforward.

For \ref{thm:liftingtoext}, we are given \hl{an $\FRel$} monad $\lift{M}$ which is a lifting of a $\Set$ monad $M$ presented by some class of equations $\Phi\subseteq \Eq$ (with a given monad isomorphism $\rho:\tmonad{\Phi} \cong M$). Our goal is to find a class of $\FRel$ equations and quantitative equations $\qth\subseteq\MEq(\Sigma)$ such that:~(i) $\qth$ extends $\Phi$ and (ii) there is a monad isomorphism $\qtmonad{\qth} \cong \lift{M}$. 

We define $\qth$ to be the union of a class of $\FRel$ equations $\qth_{\mathrm{EQ}}$ and a class of $\FRel$ quantitative equations $\qth_{\mathrm{QEQ}}$, given as follows.

The class $\qth_{\mathrm{EQ}}$ consists of all equations $\forall X.s=t$ entailed by $\Phi$, transformed to $\FRel$ equations $\forall (X,d).s=t$, for all possible fuzzy relations $d$ on $X$:
\begin{equation}\label{eqn:liftingintotheq}
\qth_{\mathrm{EQ}} =  \left\{ \forall(X,d).s=t \mid \th \imply \forall X.s=t  \text{ and }(X,d) \in \FRel\right\}.
\end{equation}
The class $\qth_{\mathrm{QEQ}}$ contains quantitative equations of the form $\forall (X,d).s=_\epsilon t$, for all possible $\FRel$ spaces $(X,d)$ and $s,t\in\Terms{X}$. The $\epsilon\in[0,1]$, expressing the distance between $s$ and $t$, is obtained by:
\begin{enumerate}[(a)]
\item using the monad isomorphism $\rho:\terms{\sig,\th}\cong M$ to get a bijection
$$
\rho_X: \Terms{X}/_{\equiv_{\Phi}} \to MX,
$$
where the equivalence $\equiv_{\Phi}$ is defined as: $s \equiv_\Phi t \Leftrightarrow \Phi\imply \forall X. s=  t$ (see \autoref{ex_monad_termseq}),
\item using the distance provided by the given $\FRel$ monad $\lift{M}$, $\lift{M}(X,d) = \big( MX, \lift{d} \big)$, to obtain the required value for $\epsilon$:
$$
\epsilon =  \lift{d}(\rho_X\big([s]_{\equiv_\Phi}\big), \rho_X\big([t]_{\equiv_\Phi}\big)).
$$
\end{enumerate}
Thus, formally,
\begin{equation}\label{eqn:liftingintothqeq}
\qth_{\mathrm{QEQ}} = \left\{ \forall(X,d). s=_\epsilon t\ \middle|\  \begin{matrix}(X,d)\in \FRel, s,t \in \Terms{X},\\\text{and } \epsilon = \lift{d}\left( \rho_X([s]_{\equiv_\Phi}),\rho_X([t]_{\equiv_\Phi}) \right)\end{matrix} \right\}.
\end{equation}
The rest of the proof consists in verifying that the defined $\qth = \qth_{\mathrm{EQ}} \cup \qth_{\mathrm{QEQ}}$ satisfies the desired properties: (i) $\qth$ extends $\th$ (\autoref{lem:qthliftmextends}) and (ii)  $\lift{\rho}$, defined set-theoretically as $\rho$, is a monad isomorphism $\lift{\rho}: \qtmonad{\qth} \cong \lift{M}$ (\autoref{lem:liftrhoiso}).

\begin{rem}
    \hl{In \autoref{exmp:preseasypset}, we provided a presentation of $\mP'$. \autoref{thm:correspondencemonliftthext} automatically tells us that there is a quantitative equational presentation for this monad that extends the theory of semilattices, simply based on the fact that $\mP'$ is a lifting of $\mP$ (\autoref{exmp:monliftmP}).}
    
    \hl{We stress that \autoref{thm:correspondencemonliftthext} does not imply that all $\FRel$ monads with a presentation are liftings of finitary $\Set$ monads. See Remark \ref{final_remark_section_8} for a simple example based on the theory of metric spaces $\Met$.}

\end{rem}

This proof sketch of \autoref{thm:correspondencemonliftthext} is expanded in full details in the following subsection.

\subsection{Detailed Proof of \autoref{thm:correspondencemonliftthext} Following the Proof Sketch}\label{sec:liftingdetails}
In this section, we consider a fixed a monad $(M,\eta,\mu)$ on $\Set$, a class of equations $\th \subseteq \Eq$ and a monad isomorphism $\rho: \terms{\sig,\th} \cong M$ witnessing the fact that $\th$ presents $M$.
We start by showing statement \ref{thm:exttolifting} of \autoref{thm:correspondencemonliftthext}.

Given a class of $\FRel$ equations and quantitative equations $\qth \subseteq \MEq(\sig)$ extending $\th$, we construct a monad $(\lift{M},\lift{\eta},\lift{\mu})$ on $\FRel$ that lifts $M$ and is presented by $\qth$. As explained in the sketch, the lifting properties \ref{prop:monliftobjects}--\ref{prop:monliftmult} from \autoref{defn:monadlifting} enforce the action of $\lift{M}$ on morphisms, its unit, its multiplication and the fact that its action on objects must be an assignment $(A,d_A) \mapsto (MA, \lift{d}_A)$. Hence, we only need to define the fuzzy relation $\lift{d}_A$ on $MA$ and prove that the other (determined) parts of the monad are valid, i.e. it sends nonexpansive maps to nonexpansive maps and the units and multiplications are nonexpansive.

We first prove that, from the assumption that $\qth$ is a quantitative extension of $\th$, we can derive that $\qterms{\sig,\qth}$ is a monad lifting of $\terms{\sig,\th}$.
\begin{lem}\label{lem:extensionislifting}
    If $\qth \subseteq \MEq(\sig)$ be a quantitative extension of $\th \subseteq \Eq$, then $\qterms{\sig,\qth}$ is a monad lifting of $\terms{\sig,\th}$.
\end{lem}
\begin{proof}
    Recall (from \autoref{ex_freealgebra}) that, for any  $A\in\Set$, $\terms{\sig,\th}A$ is the set of terms $\TermsA$ quotiented by the relation $\equiv_{\th}$ defined by
    \[s \equiv_{\th} t \Leftrightarrow \th \imply \forall A.s=t , \]
    and (from Subsection \ref{subsub1:proofs}) that, for any $(A,d_A)\in\FRel$, the underlying set of $\qterms{\sig,\qth}(A,d_A)$ is the set of terms $\TermsA$ quotiented by the relation $\equiv_{\qth}$ defined by
    \[ s \equiv_{\qth} t \Leftrightarrow \qth \derivefrelfrel \forall (A,d_A).s=t \stackrel{*}\Leftrightarrow \qth \implyfrelfrel \forall (A,d_A).s=t,\]
    where the ($ \stackrel{*}\Leftrightarrow$) implication follows from the soundness and completeness of $\derivefrelfrel$ (\autoref{soundnesscompleteness}). 
    It thus follows form the assumption that $\qth$ is a quantitative extension of $\th$ (see \autoref{eqn:defnliftingth}) that $\equiv_{\th}$ coincides with $\equiv_{\qth}$, for any $(A,d_A)$, and this implies that $U\qterms{\sig,\qth}(A,d_A) = \terms{\sig,\th}U(A,d_A)$. Moreover, the action on morphisms, the units, and the multiplications of both monads are defined set-theoretically in the same way. This implies that all properties \ref{prop:monliftobjects}--\ref{prop:monliftmult} from \autoref{defn:monadlifting} of monad lifting are satisfied. We conclude that $\qterms{\sig,\qth}$ is a monad lifting of $\terms{\sig,\th}$.
\end{proof}

In what follows, we just write $[t]$ in place of $[t]_{\equiv_{\th}}$ (or equivalently $[t]_{\equiv_{\qth}}$, as shown above) for the equivalence class of a term $t \in \TermsA$ in $\terms{\sig,\th}A$ defined as in \autoref{ex_monad_termseq}.

As explained in the proof sketch, this allows us to define a distance $\lift{d}_A(m,m')$ between any two elements $m,m' \in MA$ by viewing them as elements of $\terms{\sig,\th}A$ (i.e., equivalence classes $[t]$ of terms) via the monad isomorphism $\rho_A^{-1}$, and using the distance on those elements given by $\FreedistA$, the distance on $\qterms{\sig,\qth}(A,d_A)$ as specified in \autoref{cor:distancequotientedterms}. We prove that this indeed defines a monad lifting of $M$.
\begin{lem}
    Let $\lift{M}$ be defined on objects by
    \[\lift{M}(A,d_A) = (MA, \lift{d}_A) \text{ where } \lift{d}_A(m,m') = \FreedistA(\rho_A^{-1}(m),\rho_A^{-1}(m')).\]
    This assignment becomes \hl{an $\FRel$} monad $\lift{M}$ when the action on morphisms, unit, and multiplication are specified as in $M$ (see properties \ref{prop:monliftfunctor}--\ref{prop:monliftmult} of monad liftings in \autoref{defn:monadlifting}). Furthermore, it is presented by $\qth$.
\end{lem}
\begin{proof}

    It follows from the definition of $\lift{d}_A$, that for any fuzzy relation $(A,d_A)$, the function $\lift{\rho}_{(A,d_A)}$ defined set-theoretically like $\rho_A$ is an isomorphism $\lift{\rho}_{(A,d_A)}: \qterms{\sig,\qth}(A,d_A) \cong \lift{M}(A,d_A)$. Indeed, we know that $\rho_A$ is bijective, and for any $[s],[t] \in \qterms{\sig,\qth}(A,d_A)$,
    \[\lift{d}_A(\rho_A[s],\rho_A[t]) = \FreedistA(\rho_A^{-1}(\rho_A[s]), \rho_A^{-1}(\rho_A[t])) = \FreedistA([s],[t]),\]
    so $\lift{\rho}_{(A,d_A)}$ is a bijective isometry, i.e. an isomorphism in $\FRel$.
    
    Moreover, by \autoref{lem:extensionislifting}, we know that $\qterms{\sig,\qth}$ is a monad lifting of $\terms{\sig,\th}$, hence by \ref{prop:monliftfunctor}--\ref{prop:monliftmult} in \autoref{defn:monadlifting}, we know that for any $(A,d_A)$ and nonexpansive map $\hat{f}:(A,d_A) \to (B,d_B)$,
    \begin{enumerate}[(a)]
        \item\label{item:qtermsliftingfunctor} the function $\qterms{\sig,\qth}\hat{f}$ is set-theoretically defined like $\terms{\sig,\th}U\hat{f}$ and it is nonexpansive from $\qterms{\sig,\qth}(A,d_A)$ to $\qterms{\sig,\qth}(B,d_B)$,
        \item\label{item:qtermsliftingunit} the component $\eta^{\qth}_{(A,d_A)}$ of the unit of $\qterms{\sig,\qth}$ is defined set-theoretically like the component $\eta^\th_A$ of the unit of $\terms{\sig,\th}$, and $\eta^{\qth}_{(A,d_A)}:(A,d_A)\to\qterms{\sig,\qth}(A,d_A)$ is nonexpansive, and 
        \item\label{item:qtermsliftingmult} the component $\mu^{\qth}_{(A,d_A)}$ of the multiplication of $\qterms{\sig,\qth}$ is  defined set-theoretically like the component $\mu^\th_A$ of the multiplication of $\terms{\sig,\th}$, and $\mu^{\qth}_{(A,d_A)}: \qterms{\sig,\qth}\qterms{\sig,\qth}(A,d_A)\to\qterms{\sig,\qth}(A,d_A)$ is nonexpansive.
    \end{enumerate}
    Since $\rho$ is a monad isomorphism from $\terms{\sig,\th}$ to $M$ (\autoref{defn:monadmorph}), we have the following equations for any $\hat{f}:(A,d_A) \to (B,d_B)$, %
    \begin{align*}
        MU\hat{f} &= \rho_B\circ \terms{\sig,\th}U\hat{f} \circ \rho_A^{-1} &&\text{(by naturality of $\rho$)}\\
        \eta_A &= \rho_A \circ \eta^\th_A &&\text{(by (1) in \autoref{defn:monadmorph})} \\
        \mu_A &= \rho_A \circ \mu^\th_A \circ \rho_{\terms{\sig,\th}A}^{-1} \circ M\rho_A^{-1}. &&\text{(by (2) in \autoref{defn:monadmorph})}
    \end{align*}
    Combining these equations with the information on nonexpansiveness we have derived in \ref{item:qtermsliftingfunctor}-\ref{item:qtermsliftingunit}-\ref{item:qtermsliftingmult} above, we find that for any $(A,d_A)$ and nonexpansive function $\hat{f}:(A,d_A) \to (B,d_B)$,
    \begin{itemize}
        \item $\lift{M}\hat{f}$ defined set-theoretically like $MU\hat{f}$ is nonexpansive from $\lift{M}(A,d_A)$ to $\lift{M}(A,d_B)$,
        \item $\lift{\eta}_{(A,d_A)}$ defined set-theoretically like $\eta_A$ is nonexpansive from $(A,d_A)$ to $\lift{M}(A,d_A)$, and
        \item $\lift{\mu}_{(A,d_A)}$ defined set-theoretically like $\mu_A$ is nonexpansive from $\lift{M}\lift{M}(A,d_A)$ to $\lift{M}(A,d_A)$.
    \end{itemize}
    We conclude that $(\lift{M},\lift{\eta},\lift{\mu})$ is a monad lifting of $(M,\eta,\mu)$, and it is presented by $\qth$ via the monad isomorphism $\lift{\rho}: \qterms{\sig,\qth} \Rightarrow \lift{M}$. All the conditions that need to be checked (e.g. naturality, preservation of composition, monad laws, etc.) hold because they hold in $\Set$ after applying $U: \FRel \to \Set$ and $U$ is faithful.
\end{proof}

Next, we give the details for the proof sketch we gave for statement \ref{thm:liftingtoext} of \autoref{thm:correspondencemonliftthext}. Consider a monad lifting $(\lift{M},\lift{\eta},\lift{\mu})$ of $(M,\eta,\mu)$, and the class $\qth \subseteq \MEq(\sig)$ comprising the $\FRel$ equations in $\qth_{\mathrm{EQ}}$ from \eqref{eqn:liftingintotheq} and the $\FRel$ quantitative equations in $\qth_{\mathrm{QEQ}}$ from \eqref{eqn:liftingintothqeq}. We need to show that $\qth$ is a quantitative extension of $\th$, and that it presents $\lift{M}$ via an isomorphism $\lift{\rho}$ defined set-theoretically like $\rho$.

Given a set $A$, the free model of $\th$ generated by $A$ can be seen, thanks to \autoref{thm_monadicity_QA} and the fact that monad isomorphisms correspond to isomorphisms between their Eilenberg--Moore categories (by \autoref{prop:monadisocatiso}), as the image of the free $M$-algebra $(MA,\mu_A)$ under the composite isomorphism $\EM(M) \cong \EM(\terms{\sig,\th}) \cong \ModelsCat(\th)$. Denote the resulting $\sig$-algebra by $(MA,\{\op^{MA}\}_{\op \in \sig})$. Given a fuzzy relation $(A,d_A)$, the lifting $\lift{M}$ yields a fuzzy relation $\lift{d}_A$ on $MA$, so we obtain a quantitative $\sig$-algebra $\mathbb{M}_{(A,d_A)} = (MA,\lift{d}_A,\{\op^{MA}\}_{\op \in \sig})$. We can show that $\mathbb{M}_{(A,d_A)}$ is a model of $\qth$.

\begin{lem}\label{lem:monliftfreealg}
    The quantitative $\sig$-algebra $\mathbb{M}_{(A,d_A)} = (MA,\lift{d}_A,\{\op^{MA}\}_{\op \in \sig})$ constructed above belongs to $\mathbf{QMod}_{\sig}(\qth)$.
\end{lem}
\begin{proof}
    \hl{We first want to identify the extended interpretation $\sem{-}_{\imap}^{MA}$ in $\mathbb{M}_{(A,d_A)}$. Starting with the $M$-algebra $(MA,\mu_A)$, we apply the isomorphism $\EM(M)\cong \EM(\terms{\sig,\th})$ obtained from  $\rho$ by \autoref{prop:monadisocatiso}, which is given by post-composition by $\rho$ (see} \cite[Theorem 6.3]{TTT}) \hl{and yields the $\terms{\sig,\th}$-algebra $(MA, \mu_A \circ \rho_{MA})$. Next, the isomorphism $\EM(\terms{\sig,\th}) \cong \ModelsCat(\th)$ mentioned after \autoref{thm:cancelmonadicity} is applied, and gives the following operations on $MA$:}%
    \[op^{MA}(x_1,\dots, x_n) = \mu_A(\rho_{MA}([op(x_1,\dots, x_n)]_{\equiv})).\]
    \hl{Finally, after a simple induction using the fact that $\mu_A \circ \rho_{MA}$ is a $\terms{\sig,\th}$-algebra, we find that for any interpretation $\imap:(X,d_X) \rightarrow (MA,\lift{d}_A)$, the extended interpretation $\sem{-}_{\imap}^{MA}$ is}
    \[\Terms{X} \xrightarrow{[-]_{\equiv}} \terms{\sig,\th}X \xrightarrow{\terms{\sig,\th}\imap} \terms{\sig,\th}MA \xrightarrow{\rho_{MA}} MMA \xrightarrow{\mu_A} MA. \]
    \hl{For later use, we apply the naturality of $\rho$ to rewrite the composite as}%
    \begin{equation}\label{eqn:interpretationmu}
        \sem{-}_{\imap}^{MA} = \Terms{X} \xrightarrow{[-]_{\equiv_{\th}}} \terms{\sig,\th}X \xrightarrow{\rho_{X}} MX \xrightarrow{M\imap}   MMA \xrightarrow{\mu_A} MA.
    \end{equation}
    Now, we show $\mathbb{M}_{(A,d_A)}$ satisfies the $\FRel$ equations in \eqref{eqn:liftingintotheq}. If $\th \imply \forall X.s=t$, then the $\sig$-algebra underlying $\mathbb{M}$ satisfies $\forall X.s=t$ because it is a model of $\th$, hence for any nonexpansive interpretation $\imap: (X,d) \rightarrow (MA,\lift{d}_A)$,
    we have $\sem{s}_{\imap}^{MA} = \sem{t}_{\imap}^{MA}$. We conclude $\mathbb{M}_{(A,d_A)} \models \forall (X,d).s=t$ for all those equations in $\qth_{\mathrm{EQ}}$.

    Next, we show that $\mathbb{M}_{(A,d_A)}$ satisfies the $\FRel$ quantitative equations in \eqref{eqn:liftingintothqeq}.
    Let $\forall (X,d). s=_\epsilon t\in\qth_{\mathrm{QEQ}}$ with $\epsilon = \lift{d}(\rho_X[s]_{\equiv_\th},\rho_X[t]_{\equiv_\th})$, and let $\imap: (X,d) \rightarrow (MA,\lift{d}_A)$ be nonexpansive. We have the following derivation:
    \begin{align*}
        \lift{d}_A\left( \sem{s}_{\imap}^{MA},\sem{t}_{\imap}^{MA} \right) &= \lift{d}_A(\mu_A(M\imap(\rho_X([s]_{\equiv_\th}))),\mu_A(M\imap(\rho_X([t]_{\equiv_\th})))) &&\text{(using \eqref{eqn:interpretationmu})}\\
        &\leq \skew{1.2}\doublewidehat{d}_A(M\imap(\rho_X([s]_{\equiv_\th})),M\imap(\rho_X([t]_{\equiv_\th}))) &&\text{(by \autoref{defn:monadlifting}\ref{prop:monliftmult})}\\
        &\leq \lift{d}(\rho_X([s]_{\equiv_\th}),\rho_X([t]_{\equiv_\th})) &&\text{(by \autoref{defn:monadlifting}\ref{prop:monliftfunctor})}\\
        &= \epsilon.
    \end{align*}
    We conclude that $\mathbb{M}_{(A,d_A)} \models \forall (X,d). s=_{\epsilon} t$. Since $\forall (X,d). s=_{\epsilon} t$ has been chosen arbitrarily in $\qth_{\mathrm{QEQ}}$, we get that $\mathbb{M}_{(A,d_A)} \models \qth_{\mathrm{QEQ}}$. Hence we conclude that $\mathbb{M}_{(A,d_A)} \in \QModelsCat(\qth)$.
\end{proof}
By exploiting Lemma \ref{lem:monliftfreealg} can now show that $\qth$ is a quantitative extension of $\th$.
\begin{lem}\label{lem:qthliftmextends}
    The class $\qth = \qth_{\mathrm{EQ}} \cup \qth_{\mathrm{QEQ}}$ is a quantitative extension of $\th$.
\end{lem}
\begin{proof}
    Fix $(A,d_A) \in \FRel$ and $s,t \in \TermsA$. We only need to show that $\qth \implyfrelfrel \forall (A,d_A). s=t$ implies $\th \imply \forall A.s=t$. The converse implication holds because $\qth \supseteq \qth_{\mathrm{EQ}}$.
    
    By definition of $\implyfrelfrel$, and since $\mathbb{M}_{(A,d_A)}$ is a model of $\qth$, we know that $\mathbb{M}_{(A,d_A)} \models \forall (A,d_A). s=t$. Then, taking the assignment $\lift{\eta}_{(A,d_A)}:(A,d_A) \to (MA,\lift{d}_A)$ which is nonexpansive since it is the unit of the monad $\lift{M}$, and recalling that $\lift{\eta}_{(A,d_A)}$ is defined as the unit $\eta_A$ of $M$, we find using \eqref{eqn:interpretationmu} and the monad law $\mu_A \circ M\eta_A = \id_{MA}$ that
    \[\rho_A([s]_{\equiv_\th}) = \sem{s}^{MA}_{\eta_A} = \sem{t}^{MA}_{\eta_A} = \rho_A([t]_{\equiv_\th}).\]
    Since $\rho_A$ is a bijection, we find $[s]_{\equiv_\th} = [t]_{\equiv_\th}$. This means, by definition of the equivalence relation (see \autoref{ex_monad_termseq}), that  $\th \imply \forall A.s=t$.
\end{proof}
By \autoref{lem:extensionislifting}, this means that $\qterms{\sig,\qth}$ is a monad lifting of $\terms{\sig,\th}$. In particular, the equivalence class of any term $t\in \Terms{A}$ in $\terms{\sig,\th}A$ and $\qterms{\sig,\qth}(A,d_A)$ coincide, and we can just write $[t]$ for both 
$[t]_{\equiv_\th}$ and $[t]_{\equiv_{\qth}}$. Moreover, it means that for any fuzzy relation $(A,d_A)$, the function $\rho_A$ goes from the carrier of $\qterms{\sig,\qth}(A,d_A)$ to the carrier of $\lift{M}(A,d_A)$. We show this assembles into a monad isomorphism.
\begin{lem}\label{lem:liftrhoiso}
    For any $(A,d_A) \in \FRel$, $\lift{\rho}_{(A,d_A)}: \qterms{\sig,\qth}(A,d_A) \to \lift{M}(A,d_A)$, defined set-theoretically like $\rho_A$, is an isomorphism in $\FRel$. Hence, there is a monad isomorphism $\lift{\rho}: \qterms{\sig,\qth} \cong \lift{M}$ defined by $U\lift{\rho}_{(A,d_A)} = \rho_A$.
\end{lem}
\begin{proof}
    To prove the first part, we need to show that $\lift{\rho}_{(A,d_A)}$ is a bijective isometry. We know it is bijective because $\rho_A$ is a $\Set$ isomorphism between the carriers of $\qterms{\sig,\qth}(A,d_A)$ and $\lift{M}(A,d_A)$. To prove it is an isometry, we need to show that for any $s,t \in \TermsA$,
    \[\FreedistA([s],[t]) = \lift{d}(\rho_A[s],\rho_A[t]),\]
    where $\FreedistA$ is the distance between classes of terms in $\qterms{\sig,\qth}$ obtained in \autoref{cor:distancequotientedterms}. By \autoref{lemma_connection_syntax_semantics_1}, \autoref{soundness_theorem_statement}, and \autoref{completeness_of_proof_system_qa}, it is equivalent to show %
    \[\qth \implyfrelfrel \forall (A,d_A).s=_\epsilon t \Longleftrightarrow  \lift{d}_A(\rho_A[s],\rho_A[t])\leq \epsilon.\]
    
    ($\Rightarrow$) \autoref{lem:monliftfreealg} says that $\mathbb{M}_{(A,d_A)}$ satisfies $\forall (A,d_A). s=_\epsilon t$ because it is a model of $\qth$. As for \autoref{lem:qthliftmextends}, we take the assignment $\lift{\eta}_{(A,d_A)}: (A,d_A) \rightarrow \lift{M}(A,d_A)$ which is nonexpansive by \ref{prop:monliftunit}, and we obtain
    \[\lift{d}_A\left(\rho_A[s],\rho_A[t]\right) = \lift{d}_A\left(\sem{s}_{\eta_A}^{MA},\sem{t}_{\eta_A}^{MA}\right) \leq \epsilon.\]

    ($\Leftarrow$) By definition of $\qth$, writing $\epsilon_0 = \lift{d}_A(\rho_A[s],\rho_A[t])$, we know $\forall (A,d_A).s=_{\epsilon_0} t \in \qth$, which means $\qth \implyfrelfrel \forall (A,d_A).s=_{\epsilon_0} t$. Now, if $\epsilon_0 \leq \epsilon$, then by the (UP-CLOSURE) axiom scheme \eqref{MAX:axiom:deductive}, $\qth \implyfrelfrel \forall (A,d_A).s=_\epsilon t$.

    We conclude that $\lift{\rho}_{(A,d_A)}$ is a bijective isometry, thus an isomorphism in $\FRel$. In order to prove that $\lift{\rho}$ is a monad isomorphism, we need to show the equations required by monad morphisms hold. We note that applying the forgetful functor $U: \FRel \to \Set$ to those equations yields equations that hold because $\rho$ is a monad morphism between $\terms{\sig,\th} $ and $ M$. This is because applying $U$ to $\lift{\rho}$ yields $\rho$, and $\qterms{\sig,\qth}$ and $\lift{M}$ are monad liftings of $\terms{\sig,\th}$ and $M$, respectively. Since $U$ is faithful, the original equations in $\FRel$ must also hold, and we get that $\lift{\rho}$ is a monad isomorphism $\qterms{\sig,\qth} \cong \lift{M}$.
\end{proof}

\section{From Fuzzy Relations to Generalised Metric Spaces}\label{section:fromFRel_to_GMET}
Most of the literature on quantitative algebras \hl{follows \cite{DBLP:conf/lics/MardarePP16}} (see, e.g., \cite{DBLP:conf/lics/BacciMPP18,DBLP:conf/calco/BacciMPP21,DBLP:conf/lics/MioSV21,DBLP:conf/lics/Adamek22})\hl{ and considers} quantitative algebras whose carriers are metric spaces. Up to this point, our results have been stated for quantitative algebras (in the sense of \autoref{defi_quantitative_algebra}) whose 
carriers are arbitrary fuzzy relations.

In this section, we show that all the results proved so far also hold when, instead of $\FRel$, we take as base category an arbitrary category $\GMet$ of generalised metric spaces (see Section \ref{sec:gen:metric:spaces}), such as the category $\Met$ of metric spaces.

In what follows, we fix a category of generalised metric spaces $\GMet$ defined by a set $\Hset$ of $\folang$-implications (see \autoref{def_Limplications}) and a signature $\Sigma$. We denote by $\QalgGMet$ the full subcategory of $\QalgFRel$ comprising only quantitative algebras whose underlying fuzzy relations satisfy the $\folang$-implications defining $\GMet$
\begin{equation}\label{defn:gmetalg}
    \QalgGMet = \{(A,d_A, \{op^A\}_{op \in \sig}) \mid (A,d_A)\fomodels \Hset \}\subseteq \QalgFRel.
\end{equation}
Given a class of $\FRel$ equations and quantitative equations $\Phi \subseteq \MEq(\sig)$, we \hl{define $\QmodGMet(\Phi)$ as} the full subcategory of $\QmodFRel(\Phi)$ comprising only quantitative algebras that belong to $\QalgGMet$:
\begin{equation}\label{defn:gmetmod}
    \QmodGMet(\Phi) = \QalgGMet \cap \QmodFRel(\Phi) .
\end{equation}
Note that since we are taking full subcategories, homomorphisms of $\GMet$ quantitative $\sig$-algebras are still nonexpansive homomorphisms of the underlying $\sig$-algebras.

We first show  that $\QalgGMet$ is a quantitative equationally definable class of quantitative $\sig$-algebras in the sense of \autoref{eq_def_class:def}. In other words, we show (\autoref{corollary:translation:2}) that there is a class $\HPhi\subseteq \MEq(\Sigma)$ of $\FRel$ equations and quantitative equations such that
$$
\QalgGMet = \QmodFRel(\HPhi).
$$
We will prove this fact by giving an explicit procedure to translate any $\folang$-implication $H\in\Hset$ to \hl{an $\FRel$} equation or quantitative equation $\Hphi\in \MEq(\Sigma)$ such that, for any quantitative algebra $(A,d_A, \{ op^A\}_{op\in\Sigma})\in \QalgFRel$, 
$$
(A,d_A) \fomodels H \ \ \ \Longleftrightarrow \ \ \ \mathbb{A}\models \Hphi.
$$
In fact, the terms in $\Hphi$ will not be built using any of the operations $op\in\Sigma$, so this translation is independent of the signature $\sig$.

\begin{defi}[Translation]\label{translation:def}
Let
\[H = \forall x_1,\dots, x_n.\Big( \ \big( G_1\wedge \dots \wedge G_m \wedge G'_1\dots \wedge G'_k \big) \ \Rightarrow F \Big)\] 
be an $\folang$-implication, where
\begin{itemize}
\item We denote with $X=\{x_1,\dots, x_n\}$ the set of variables occurring in $H$. Note that this set cannot be empty as the atomic formulas in $H$ (not empty because $F$ is one of them) are predicates ($x=y$ or $d(x,y)\leq \epsilon$) which must use variables.
\item All atomic formulas $G_i$, $1\leq i\leq m$ (possibly an empty set when $m=0$), are of the form
$$
x=y
$$
for some $x,y\in X$.
\item All atomic formulas $G_j'$, $1\leq j\leq k$ (possibly an empty set when $k=0$), are of the form
$$d(x,y)\leq \epsilon$$
for some $x,y\in X$ and $\epsilon \in [0,1]$.
\end{itemize}
We are going to define an $\FRel$ (quantitative) equation $\Hphi \in \MEq(\Sigma)$ constructed from $H$. 
We first use the premises ($G_1,\dots ,G_m, G'_1\dots, G'_k$) of $H$ to construct a fuzzy relation space $(X_H,d_H)$. Let $\sim\ \subseteq X\times X$ be the smallest equivalence relation on $X$ containing %
$$
\big\{ (x,x') \mid \text{there is a formula $G_i$ in $H$ of the form: $x=x'$}
\big\}.
$$
Hence, $\sim$ consists of exactly all pairs $(x,y)$ of variables in $X$ such that $x=y$ is logically implied by the conjunction of all formulas $G_i$.
Let us denote with $X_H$ the quotient $X/{\sim}$, i.e., the set of all $\sim$-equivalence classes. Finally, let $d_H: X_H\times X_H\rightarrow [0,1]$ be the following fuzzy relation on $X_H$:
\[ d_H([x]_{\sim}, [x']_{\sim}) = \min \left\{ \epsilon \in [0,1]\ \middle| \begin{matrix} 
\text{ there is a formula $G_j'$ in $H$ of the form $d(y,y')\leq \epsilon$}\\
\text{ with $y\in [x]_\sim$ and $y'\in [x']_\sim$}
\end{matrix}\right\},\]
with the convention $\min(\emptyset) = 1$. We have thereby defined the fuzzy relation space $(X_H, d_H)$.

Now we use the conclusion $F$ of $H$ to construct $\Hphi$ which can be either \hl{an $\FRel$} equation or \hl{an $\FRel$} quantitative equation depending on $F$:
\begin{itemize}
\item If $F$ is of the form $x=y$, for some $x,y\in X$, then
$$
\textnormal{$\Hphi$ is defined as } \ \ \forall (X_H,d_H). [x]_{\sim} = [y]_{\sim}.
$$
\item If $F$ is of the form $d(x,y)\leq \epsilon$, for some $x,y\in X$ and $\epsilon \in [0,1]$, then
$$
\textnormal{$\Hphi$ is defined as } \ \ \forall (X_H,d_H). [x]_{\sim} =_\epsilon [y]_{\sim}.
$$
\end{itemize}
We note that the two terms ($[x]_{\sim}$ and $[y]_{\sim}$) appearing in $\Hphi$ belong to $\Terms{X_H}$ for any signature because they both belong to $X_H$. %
Hence the translation is well-defined for all $\Sigma$.
\end{defi}

Before proving the main result regarding this translation (\autoref{translation:lemma1}) we provide some illustrative examples.%

\begin{exa}
Consider the $\folang$-implication $H$ (a logically equivalent variant of \eqref{eq:refl})
\begin{equation}\label{eqn:variantrefl}
    \forall x_1,x_2.\  \big(  x_1=x_2 \ \Rightarrow\  d(x_1,x_2)\leq 0 \big).
\end{equation}
We are in the case where $X=\{x_1,x_2\}$, $n=2$ (two variables), $m=1$ (one atomic equation among the premises), and $k=0$ (no atomic formula of the form $d(x,y)\leq \epsilon$ among the premises). Since the only premise ($G_1$) of the formula is $x_1=x_2$, we have that $x_1\sim x_2$, and thus $X_H$ consists of only one element $X_H = \{\ [x_1]_\sim\}$. By the definition of $d_H$ we have that $d_H([x_1]_\sim, [x_1]_\sim) = 1$. Finally, since the conclusion is of the form $d(x_1,x_2)\leq 0$, we have that
$$
\textnormal{$\Hphi$ is defined as } \ \ \forall \big(\{ [x_1]_\sim \}, d_H \big).\ \  [x_1]_\sim =_0 [x_1]_\sim.
$$
Now, we note that a nonexpansive interpretation $\imap: (X_H,d_H) \to (A,d_A)$ is simply a choice of an element $a = \imap([x_1]_{\sim}) \in A$, and $\Hphi$ holds under $\tau$ if and only if $d_A(a,a) = 0$. Therefore, an algebra satisfies $\Hphi$ if and only if all its elements have self-distance $0$. This is indeed also the meaning of the $\folang$-implication \eqref{eqn:variantrefl} and of the logically equivalent variant 
\eqref{eq:refl}.
\end{exa}

\begin{exa}
Consider the $\folang$-implication $H$ (cf.~\eqref{eq:symm} in Section \ref{sec:gen:metric:spaces})
\begin{equation}\label{eq:symm:variant}
    \forall x_1,x_2.\  \big(  d(x_1,x_2)\leq \epsilon \ \Rightarrow\  d(x_2,x_1)\leq \epsilon \big).
\end{equation}
We are in the case where $X=\{x_1,x_2\}$, $n=2$, $m=0$, and $k=1$. The equivalence $\sim$ is the identity relation on $X$, hence $X_H = X$, and the fuzzy relation $d_H$ is given by
$$
d_H(x_1,x_1) = 1 \ \ \ d_H(x_1,x_2) = \epsilon \ \ \ d_H(x_2,x_1) = 1 \ \ \ d_H(x_2,x_2) = 1.  
$$
Finally, since the conclusion is of the form $d(x_2,x_1)\leq \epsilon$, we have that
$$
\textnormal{$\Hphi$ is defined as } \ \ \forall \big(\{ x_1,x_2 \}, d_H \big).\ \  x_2 =_\epsilon x_1.
$$
One can check that a quantitative $\sig$-algebra satisfies $\Hphi$ if and only if the underlying fuzzy relation is symmetric, which is exactly what satisfaction of \eqref{eq:symm:variant} means.
\end{exa}

\begin{lem}\label{translation:lemma1}
If $\algA=(A,d_A,\{op^A\}_{op\in\Sigma}) \in \QalgFRel$%
, $H$ is an $\folang$-implication, and $\Hphi$ is the corresponding $\FRel$ (quantitative) equation constructed in \autoref{translation:def}, then
\begin{center}
$(A,d_A)\fomodels H$ \ \ \ \ $\Longleftrightarrow$ \ \ \ \ $\mathbb{A}\models \Hphi$.
\end{center}
\end{lem}
\begin{proof}
Let $H$ be of the form described in \autoref{translation:def}
\[
\forall x_1,\dots, x_n.\Big( \ \big( G_1\wedge \dots \wedge G_m \wedge G'_1\dots \wedge G'_k \big) \ \Rightarrow F \Big),
\]
and let $X=\{x_1,\dots,x_n\}$. We consider in parallel the two cases when the conclusion $F$ is
\begin{center}
$x =y$ \ \ \ \ \ \ \  or \ \ \ \ \ \ \ $d(x,y)\leq \epsilon$
\end{center}
for some $x,y\in X$. Let $\Hphi$ be defined as in \autoref{translation:def} and be of the form
$$
\forall (X_H, d_H).\  [x]_\sim =  [y]_\sim  \ \ \ \ \ \ \  \text{or}  \ \ \ \ \ \ \  \forall (X_H, d_H).\  [x]_\sim =_\epsilon  [y]_\sim.
$$

Assuming $(A,d_A)\fomodels H$ and given an interpretation of the fuzzy relation $(X_H,d_H)$ in $\mathbb{A}$, i.e., a nonexpansive map $\tau:(X_H,d_H)\rightarrow (A,d_A)$, we need to show
\begin{equation}\label{eqn:goalsatisfaction}
    \sem{[x]_\sim}^{\algA}_\tau=\sem{[y]_\sim}^{\algA}_\tau \ \ \ \ \ \ \  \text{or} \ \ \ \ \ \ \ d_A\big( \sem{[x]_\sim}^{\algA}_\tau, \sem{[y]_\sim}^{\algA}_\tau \big) \leq \epsilon
\end{equation}
We define an interpretation $\foi_\tau: X\rightarrow A$ of the variables $X$ as $\foi_\tau(x) = \tau([x]_{\sim})$. We can show $\foi_\tau$ satisfies the premises of $H$, and, by $(A,d_A) \fomodels H$, it satisfies the conclusion too, namely,
\begin{center}
    $\foi_\tau(x) = \foi_\tau(y)$   \ \ \ \ \ \ \  or  \ \ \ \ \ \ \   $d_A(\foi_\tau(x) , \foi_\tau(y))\leq \epsilon$.
\end{center}
By definition of $\foi_\tau$, this is equivalent to
\begin{center}
    $\tau([x]_\sim) = \tau( [y]_\sim)$ \ \ \ \ \ \ \  or  \ \ \ \ \ \ \ $d_A\big(\tau([x]_\sim) , \tau( [y]_\sim) \big) \leq \epsilon$,
\end{center}
and by definition of $\sem{-}^{\algA}_\tau$ this is equivalent to \eqref{eqn:goalsatisfaction}. Hence, we conclude that $\mathbb{A}\models \Hphi$.

Assuming $\mathbb{A}\models \Hphi$ and given an interpretation $\foi: X \to A$ that satisfies all premises $G_i$ and $G'_j$ of $H$, we need to show that $\foi$ satisfies the conclusion $F$, i.e.:
\begin{equation}\label{subgoal2:translation}
\foi(x) = \foi(y)   \ \ \ \ \ \ \   \text{or}  \ \ \ \ \ \ \   d_A(\foi(x),\foi(y))\leq \epsilon.
\end{equation}
Let $\tau_\foi: (X_H,d_H)\rightarrow (A,d_A)$ be the interpretation of $(X_H,d_H)$ in $\mathbb{A}$ defined by $\tau_\foi( [x]_\sim) = \foi(x)$. After checking that it is well-defined and nonexpansive, we can apply our hypothesis ($\mathbb{A}\models \Hphi$) and obtain that the following holds:
\begin{center}
$\tau_\foi([x]_\sim) = \tau_\foi([y]_\sim)$
 \ \ \ \ \ \ \   or  \ \ \ \ \ \ \   
 $d_A\big(\tau_\foi([x]_\sim) , \tau_\foi([y]_\sim)\big)\leq \epsilon$.
\end{center}
By definition of $\tau_\foi$, this is equivalent to \eqref{subgoal2:translation}, hence we established that $(A,d_A)\fomodels H$.
\end{proof}

We can obtain a few useful corollaries from \autoref{translation:lemma1}.
The first extends the result of  \autoref{translation:lemma1} from one $\folang$-implication $H$ to a set $\Hset$ of $\folang$-implications.
In what follows, we define the set $\HPhi$ as the set of $\FRel$ (quantitative) equations
$$\HPhi=\{\Hphi \mid H\in \Hset \text{ and } \Hphi \text{ is a (quantitative) equation translating } H\} \subseteq \MEq(\Sigma),$$
where the translation is the one specified in \autoref{translation:def}.

\begin{cor}\label{corollary:translation:1}
    If $\mathbb{A}=(A,d_A,\{op^A\}_{op\in\Sigma})\in \QalgFRel$ %
    and $\Hset$ is a set of $\folang$-implications, then
\begin{center}
    $(A,d_A)\fomodels \Hset$ \ \ \ \ $\Longleftrightarrow$ \ \ \ \ $\mathbb{A}\in \QModels^\FRel(\HPhi)$.
\end{center}
\end{cor}

Hence, the class of quantitative algebras $\QalgGMet$, which contains exactly those algebras satisfying the $\folang$-implications in $\Hset$, is quantitative equationally definable.

\begin{cor}\label{corollary:translation:2}
    For any signature $\Sigma$ and any $\GMet$ category defined by a set $\Hset$ of $\folang$-implications,
    \[\QmodFRel(\HPhi) = \QalgGMet.\]

\end{cor}

\begin{rem}\label{rem:gmetisqmod}
\hl{When $\sig$ is empty, the category $\QalgGMet$ is simply the category of fuzzy relations that satisfy $\Hset$, i.e., it is $\GMet$. Thus, we have shown that $\GMet$ is a quantitative equationally definable class of fuzzy relations, namely, $\GMet = \QModelsCatNoOp^\FRel(\Phi_{\Hset})$.}
\end{rem}

The next corollary is a further generalisation of the previous one, showing that for any class of $\FRel$ equations and quantitative equations $\Phi\subseteq \MEq(\Sigma)$, $\QmodGMet(\Phi)$ is a quantitative equationally definable class of quantitative $\sig$-algebras. Namely, we show the full subcategories $\QmodFRel(\HPhi\cup \Phi)$ and $\QmodGMet(\Phi)$ of $\QalgFRel$ coincide.
\begin{cor}\label{corollario_importante}
For any signature $\Sigma$, for any $\GMet$ category defined by a set $\Hset$ of $\folang$-implications, and for any class $\Phi\subseteq\MEq(\Sigma)$ of $\FRel$ equations and quantitative equations,
$$
\QmodFRel(\HPhi\cup \Phi) = \QmodGMet(\Phi).
$$
\end{cor}
\begin{proof}
For any quantitative algebra $\algA\in \QalgFRel$, we have
\begin{align*}
    &\algA\in \QmodFRel( \HPhi\cup \Phi) \\
    &\Leftrightarrow \algA\in \QmodFRel( \Phi) \text{ and } \algA\in \QmodFRel(\Phi_\Hset)\\
    &\Leftrightarrow \algA\in \QmodFRel( \Phi) \text{ and } \algA \in \QalgGMet &&\text{(by \autoref{corollary:translation:2})}\\
    &\Leftrightarrow \algA\in \QmodGMet(\Phi). &&\text{(by \eqref{defn:gmetmod})}
\end{align*}
Hence, $\QmodFRel(\HPhi\cup\Phi)$ and $\QmodGMet(\Phi)$ have the same objects, and, since they are full subcategories of $\QalgFRel$, they also have the same morphisms.
\end{proof}

We can now show that all the results proved for $\FRel$ in Sections 
\ref{section:result:1},\ref{section:result:2},\ref{section:result:3},\ref{section:lifting} also hold when specialised for a category $\GMet$ defined by a set $\Hset$ of $\folang$-implications. 

Starting from the relation $\vdash_\FRel$ for $\FRel$ defined in \autoref{proof:system:definition}, we define a relation $\derivegmet$ for $\GMet$ as follows:
$$
\Phi \derivegmet \phi \Longleftrightarrow \Phi_\Hset, \Phi\vdash_\FRel \phi.
$$

\autoref{thm_gmetsoundcomplete} shows that the relation $\derivegmet$ is sound and complete for the relation $\implygmet$, which is the restriction of $\implyfrel_\FRel$ to $\GMet$ defined as follows:
\[\Phi \implygmet \phi \Longleftrightarrow \forall \algA\in \QmodGMet(\Phi).\ \algA\models \phi.\]

\begin{thm}[Soundness and Completeness for $\GMet$]
\label{thm_gmetsoundcomplete}
    $\Phi \derivegmet \phi \Longleftrightarrow \Phi \implygmet \phi$.
\end{thm}
\begin{proof}
We have
\begin{align*}
    \Phi \derivegmet \phi &\Leftrightarrow \HPhi , \Phi \derivefrelfrel \phi &&\text{(definition of $\derivegmet$)}\\
    &\Leftrightarrow \HPhi , \Phi \implyfrelfrel \phi &&\text{(by \autoref{soundnesscompleteness})}\\
    &\Leftrightarrow \forall \algA\in \QmodFRel(\HPhi \cup\Phi).\ \algA\models \phi &&\text{(definition of $\implyfrelfrel$)}\\
    &\Leftrightarrow \forall \algA\in \QmodGMet(\Phi).\ \algA\models \phi  &&\text{(by \autoref{corollario_importante})}\\
    &\Leftrightarrow \Phi \implygmet \phi. &&\text{(definition of $\implygmet$)}\qedhere
\end{align*}
\end{proof}

Now fix a class of equations and quantitative equations $\Phi\subseteq\MEq(\Sigma)$. We have
\begin{equation}
\label{diag_Fgmet}
\begin{tikzcd}[row sep=1.0em, column sep=0.5em]
	& {\QmodGMet(\Phi)} \\
	\\
	\GMet && \FRel
	\arrow["E"', hook, from=3-1, to=3-3]
	\arrow["{U_{\GMet}}", tail reversed, no head, from=3-1, to=1-2]
	\arrow[""{name=0, anchor=center, inner sep=0}, "F", shift left=2, from=3-3, to=1-2]
	\arrow[""{name=1, anchor=center, inner sep=0}, "U", shift left=2, from=1-2, to=3-3]
	\arrow["\dashv"{anchor=center, rotate=60}, draw=none, from=0, to=1]
\end{tikzcd}
\end{equation}
where
\begin{itemize}
    \item $\funE$ is the (full and faithful) functor embedding $\GMet$ into $\FRel$, \hl{equivalently thanks to \autoref{rem:gmetisqmod}, it is the forgetful functor $\QModelsCatNoOp(\Phi_{\Hset}) \to \FRel$};
    \item $U_\GMet$ is the forgetful functor of type $U_\GMet: \QmodGMet(\Phi) \rightarrow \GMet$;
    \item $U$ is the forgetful functor of type $U: \QmodFRel(\HPhi \cup\Phi) \rightarrow \FRel$, which indeed also has type $U: \QmodGMet(\Phi) \rightarrow \FRel$ by \autoref{corollario_importante}, \hl{and such that $U=E \circ U_\GMet$};
    \item $F$ is the left adjoint of the forgetful functor $U$, as given in Section \ref{section:result:3}.
\end{itemize}
Since the image of $U$ is contained in the full subcategory $\GMet$, we can restrict the adjunction to $\GMet$ by defining the functor $F_\GMet: = F \circ E:  \GMet \to \QmodGMet(\Phi)$. %

\begin{thm}
\label{lem_adjoint_gmet}
    The functor $F_\GMet:\GMet \to \QmodGMet(\Phi)$ is a left adjoint of $U_\GMet$.
\end{thm}
\begin{proof}
\hl{Let $\eta: \id_{\FRel} \Rightarrow U \circ F$ be the unit of the adjunction $F\dashv U$, and define} 
\begin{center}
    for all $\GMet$ spaces $(A,d_A)$,\ \  $\eta'_{(A,d_A)}=\eta_{E(A,d_A)}: E(A,d_A) \to UFE(A,d_A)$.
\end{center}
\hl{By \eqref{diag_Fgmet} we have  $UFE(A,d_A) = EU_\GMet FE(A,d_A)= EU_\GMet F_\GMet(A,d_A)$, and since $E(A,d_A)$ is the $\GMet$ space $(A,d_A)$ seen as an $\FRel$ space, we can see $\eta'_{(A,d_A)}$ as a morphism} 
\[\eta'_{(A,d_A)}: (A,d_A)\to U_\GMet F_\GMet(A,d_A).\]
\hl{Since $\eta$ is a natural transformation and $E$ acts like identity on morphisms, we also obtain a natural transformation}
\[\eta': \id_{\GMet}\Rightarrow U_\GMet F_\GMet.\]

\hl{Now take a $\GMet$ space $(A,d_A)$, a quantitative algebra}
\[ \algB=(B,d_B,\{op^\algB\}_{op\in \Sigma})\in \QmodGMet(\Phi),\]
\hl{and a nonexpansive map $f: (A,d_A)\to (B,d_B)$. Since $F \dashv U$ is an adjunction with unit $\eta$, by seeing $f$ as the $\FRel$ morphism $E(f)$ we obtain that there is a unique quantitative algebra homomorphism $g: F(E(A,d_A))\to \algB$ such that $E(f)= U (g)\circ \eta_{E(A,d_A)}$.

By definition of $E$ and $\eta'_{(A,d_A)}$, this implies that there is a unique quantitative algebra homomorphism $g: F_\GMet(A,d_A)\to \algB$ such that $f= U_\GMet (g)\circ \eta'_{(A,d_A)}$.
Hence, $F_\GMet\dashv U_\GMet$.}
\end{proof}
Note that, by definition, the functor $F_\GMet$ acts as the functor $F$ on $\GMet$ spaces, and thus the free $U_\GMet$-object generated by a generalised metric space $(A,d_A)$ is the quantitative algebra of quotiented terms built as in Section \ref{section:result:2}. The monad on $\GMet$ obtained from the composite $U_{\GMet} \circ F_{\GMet}$ will be denoted $\qtermsgmet{\sig,\Phi}$. \hl{Using \autoref{thm:cancelmonadicity}, and strict monadicity of $U$ and $E$, we obtain strict monadicity of $U_{\GMet}$.} %
\begin{thm}\label{thm:monadicitygmet}
 The functor $U_{\GMet}: \QmodGMet(\Phi)\to \GMet$  is strictly monadic. 
\end{thm} 
\begin{proof}
    \hl{We observed in \autoref{rem:gmetisqmod} and \autoref{corollario_importante} that both $\GMet$ and $\QmodGMet(\Phi)$ are categories of $\FRel$ quantitative algebras, hence their forgetful functors $E$ and $U$ are strictly monadic by \autoref{thm_monadicity_QA}. Moreover, since the composite of the right adjoints in \eqref{diag:cancelmonadic} is $U = E \circ U_{\GMet}$, we can conclude by \autoref{thm:cancelmonadicity} that $U_{\GMet}$ is strictly monadic.}
    \begin{equation}\label{diag:cancelmonadic}
\begin{tikzcd}
	{\QmodGMet(\Phi)} & \GMet & \FRel
	\arrow[""{name=0, anchor=center, inner sep=0}, "{{U_{\GMet}}}"', shift right=2, tail reversed, no head, from=1-2, to=1-1]
	\arrow[""{name=1, anchor=center, inner sep=0}, "{F_{\GMet}}", shift left=2, from=1-2, to=1-1]
	\arrow[""{name=2, anchor=center, inner sep=0}, "E", shift left=2, hook, from=1-2, to=1-3]
	\arrow[""{name=3, anchor=center, inner sep=0}, shift left=2, from=1-3, to=1-2]
	\arrow["\dashv"{anchor=center, rotate=90}, draw=none, from=1, to=0]
	\arrow["\dashv"{anchor=center, rotate=90}, draw=none, from=3, to=2]
\end{tikzcd}\qedhere
    \end{equation}    
\end{proof}

Finally, we adapt the results of Section \ref{section:lifting} to $\GMet$. The three central notions of quantitative equational presentations (\autoref{defn:frelpresentation}), monad liftings (\autoref{defn:monadlifting}) and quantitative extensions (\autoref{defn:liftingth}) just need to be modified in a straightforward way by replacing all instances of $\FRel$ to $\GMet$.\footnote{Note that the notion of equation and quantitative equation remains as in \autoref{quantitative_equation_def}, i.e., $(A,d_A)$ in  $\forall (A,d_A).s=t$ and $\forall (A,d_A).s=_\epsilon t$ is an arbitrary $\FRel$ space.}

First, a \emph{quantitative equational presentation} of a monad $M$ on $\GMet$ is a class of $\FRel$ equations and quantitative equations $\qth \subseteq \MEq(\sig)$ along with a monad isomorphism $\qtermsgmet{\sig,\qth} \cong M$, where $\qtermsgmet{\sig,\qth}$ is the monad obtained from going around the triangle in \eqref{diag_Fgmet}. Second, a $\GMet$ monad $(\lift{M},\lift{\eta},\lift{\mu})$ is a $\GMet$ \emph{lifting} of a $\Set$ monad $(M,\eta,\mu)$ if
\begin{center}
   $U\lift{M} = MU$, $U\lift{\eta} = \eta U$, and  $U\lift{\mu} = \mu U$,
\end{center}
where $U$ is now the forgetful functor $U: \GMet \to \Set$. The explicit description of what it means to be a monad lifting (\hl{after} \autoref{defn:monadlifting}) is still valid after replacing fuzzy relations with generalised metric spaces.
\begin{exa}
    \hl{All the monads in \autoref{exmp:monadsgmet} are liftings of the powerset monad or the distributions monad. Furthermore, the Hausdorff, Kantorovich, and \L K liftings were axiomatised in} \cite[Theorem 9.3]{DBLP:conf/lics/MardarePP16}, \cite[Theorem 10.5]{DBLP:conf/lics/MardarePP16}, and \cite[Theorem 5.6]{DBLP:conf/lics/MioSV22} \hl{respectively, which yields a quantitative equational presentation for these monads (we will detail in Section \ref{comparison_section} how the results from these different frameworks can be translated in our framework). For instance, for the \L K lifting of the distributions monad $\LK{\mD}$, we can adapt the proof in} \cite[\S 5.3]{DBLP:conf/lics/MioSV22} \hl{to show that it is presented by the set of quantitative equations $\qth$ containing both the equations of convex algebras in \eqref{eqn:thconvexalg} converted to $\FRel$ equations as in \autoref{exmp:preseasypset}, and the following $\FRel$ quantitative equations, one for each $p \in (0,1)$ and fuzzy relation $d:X^2 \to [0,1]$:}
    \begin{gather*}
        \forall (\{x,y,x',y'\},d). x+_p y =_{\epsilon} x'+_p y',\\
        \text{with } \epsilon = p^2d(x,x')+p(1-p)d(x,y')+(1-p)pd(y,x')+(1-p)^2d(y,y').
    \end{gather*}
\end{exa}

Third, a class of $\FRel$ equations and quantitative equations $\qth \subseteq \MEq(\sig)$ is a \emph{$\GMet$ quantitative extension} of $\th \subseteq \Eq$ if:
\begin{equation}\label{eqn:defnliftingthgmet}
    \begin{matrix}
        \text{for all $(A,d_A) \in \GMet$ and $s,t \in \TermsA$,}\\
        \th \imply \forall A.s=t  \Longleftrightarrow \qth \implygmet \forall (A,d_A).s=t.
    \end{matrix}
\end{equation}
\hl{Once these definitions are in place, it is immediate to adapt the proofs of the previous section and obtain the following theorem:}
\begin{thm}\label{thm:correspondencemonliftthextgmet}
    Let $(M,\eta,\mu)$ be a monad on $\Set$ presented by $\th\subseteq \Eq$.
\begin{enumerate}[(1)]
    \item\label{thm:exttoliftinggmet} For any $\GMet$ quantitative extension $\qth$ of $\th$, there is a $\GMet$ monad lifting $\lift{M}$ of $M$ presented by $\qth$.
    \item\label{thm:liftingtoextgmet} For any $\GMet$ monad lifting $\lift{M}$ of $M$, there is a $\GMet$ quantitative extension $\qth$ of $\th$ that presents $\lift{M}$. 
\end{enumerate}
\end{thm}

\begin{rem}\label{final_remark_section_8}
While the proof of \autoref{thm:correspondencemonliftthextgmet} above is essentially identical to that of \autoref{thm:correspondencemonliftthext}, the results that the two theorems state may present some subtle differences. For instance, there are classes of equations and quantitative equations $\qth\subseteq \MEq(\Sigma)$ such that $\qtermsgmet{\sig,\qth}$ is a $\GMet$ monad lifting (of some monad $M$ on $\Set$) but $\qterms{\sig,\qth}$ is not. For a concrete example, let $\sig =\emptyset$ and $\qth$ be the class $\Phi_{\Hset_{\Met}}$ resulting from the translation (as in \autoref{translation:def}) of the set of $\folang$-implications $\Hset_{\Met}$ defining the category $\Met$ (see \autoref{gmet_category_def}). It is readily seen that the monad $\qtermsgmet{\sig,\qth}$ on $\Met$ is a lifting of the identity monad on $\Set$. However, the monad $\qterms{\sig,\qth}$ on $\FRel$ is not a lifting of any monad on $\Set$. It sends $(A,d_A)$ to $(A,d_A)$ when $d_A$ is a metric, but when e.g. $d_A(a,b) = 0$ for $a \neq b \in A$, the carrier set of $\qterms{\sig,\qth}(A,d_A)$ will be a quotient of $A$ where $a$ and $b$ are identified. This means $\qterms{\emptyset,\qth}$ cannot lift a monad on $\Set$ because it sends two fuzzy relations with identical carrier set to fuzzy relations with different carriers.

\end{rem}

\section{Comparison with Relevant Literature}%
\label{comparison_section}
In Section \ref{framework:presentation:section} we have formally introduced our theory of quantitative algebras and in Sections \ref{section:result:1}, \ref{section:result:2}, \ref{section:result:3}, \ref{section:lifting} and \ref{section:fromFRel_to_GMET} we have stated and proved the main results.

\hl{In this section we compare our work with other frameworks of quantitative algebras in the literature. We first consider, in Section \ref{subsec:MPPcomp}, the original paper on quantitative algebras \cite{DBLP:conf/lics/MardarePP16}. We then consider, in Section \ref{subsec:fmscomp}, the generalised framework  proposed in \cite{DBLP:conf/calco/FordMS21} dealing with arbitrary relational structures. Finally, in Section \ref{subsec:msvcomp}, we compare the present work with an earlier conference paper \cite{DBLP:conf/lics/MioSV22} by the authors.}

\subsection{Comparison with \cite{DBLP:conf/lics/MardarePP16} by Mardare et al.}\label{subsec:MPPcomp}
\hl{The original work by \cite{DBLP:conf/lics/MardarePP16} presented a theory of quantitative algebras different from ours in three key ways: the carriers are required to be metric spaces, interpretations of operations are nonexpansive (with respect to the product metric) and the logical judgments are not just quantitative equations ($s=_\epsilon t$) but implications between quantitative equations.} In what follows, we refer to our theory as ``MSV theory'' and to that of \cite{DBLP:conf/lics/MardarePP16} as ``MPP theory''.

\hl{The MPP theory} deals with algebras over metric spaces formally specified as follows:\footnote{A technical difference between \autoref{MPP-algebra} and \cite[Definition 3.1]{DBLP:conf/lics/MardarePP16} is that the metrics of the latter are actually extended metrics, i.e. $d_A(a,b)$ ranges in $[0,\infty]$ instead of $[0,1]$. %
}
\begin{defi}[MPP Quantitative Algebra]\label{MPP-algebra} \cite[Definition 3.1]{DBLP:conf/lics/MardarePP16}
Given a signature $\Sigma$, \hl{an MPP} quantitative $\Sigma$-algebra is a triple $(A,d_A, \{op^A\}_{op\in\Sigma})$ such that $(A,d_A)\in\Met$ is a metric space and such that all interpretations of operation symbols in the signature
$
op^A:  A^n \rightarrow A$ (where $ar(op)=n$) are nonexpansive functions 
$$op^A:  (A^n,d^n_A) \rightarrow (A,d_A),$$
where $d^n_A$ is the product metric defined as: $d_A^n(( a_1,\dots, a_n), ( a'_1,\dots, a'_n)) = \displaystyle\max_{i=1\dots n} \{d_A(a_i, a'_i) \}$. 
\end{defi}
As a result, it only makes sense to compare the MPP theory to the MSV theory restricted to the category $\Met$ of metric spaces. This restriction is done by first seeing $\Met$ as the category of fuzzy relations satisfying the $\folang$-implications in $\mathcal{H}_{\Met}$ as explained in Section \ref{sec:gen:metric:spaces}, and then instantiating the results of Section \ref{section:fromFRel_to_GMET} with $\GMet = \Met$.

Note that \hl{an MPP} quantitative $\Sigma$-algebra is \hl{an MSV} quantititative $\Sigma$-algebra $\algA\in \QalgMet$ such that all the interpretations $op^A$ are nonexpansive in the sense of \autoref{MPP-algebra}. It is straightforward to see that, for a given $op\in\Sigma$ of arity $n$, the interpretation $op^A$ of a quantitative algebra $\algA\in \QalgMet$  is nonexpansive if and only if $\algA$ satisfies all the following $\FRel$ quantitative equations $\phi^{op}_{d}$, one for each \hl{fuzzy relation $d$ on $X=\{x_1,\dots, x_n, x'_1, \dots, x'_n\}$:}
\begin{equation}\label{eqn:opnexpasqeq}
\phi^{op}_{d} = \,\forall(X,d). op(x_1,\dots, x_n) =_{\epsilon} op(x'_1,\dots, x'_n)  \ \ \ \ \ \ \ \ \ \ \ \textnormal{where }\epsilon = \displaystyle\max_{i=1\dots n} \{d(x_i, x'_i) \}.
\end{equation}
In other words, $op^A$ is nonexpansive in the sense of \autoref{MPP-algebra} if and only if $\algA$ belongs to $\QmodMet(\Phi^{op}_{NE})$, where $\Phi^{op}_{NE}$ is the class of quantitative equations containing $\phi^{op}_{d}$ for all $d: X^2 \to [0,1]$.

Therefore the class of MPP quantitative $\Sigma$-algebras is a quantitative equationally definable class of MSV quantitative algebras in $\QalgMet$, and its theory is generated by the quantitative equations in $\Phi^{op}_{NE}$ for all $op \in \Sigma$.

Furthermore, it is an immediate consequence of \autoref{corollario_importante} that the class of MPP quantitative algebras can be quantitative equationally defined in $\QalgFRel$ as the class $\QmodFRel(\Phi_{NE}\cup \Phi_{\Hset_\Met})$, \hl{where $\Phi_{NE}$ is the union of all $\Phi_{NE}^{op}$.}
\begin{rem}\label{remark_Lipschitz}
Note that the MSV theory allows for other interesting properties of $op^A$ to be defined by quantitative equations.
For example, by using
$$
\epsilon = \alpha\cdot \displaystyle\max_{i=1\dots n} \{d(x_i, x_i) \} \  \ \ \ \  \ \ \textnormal{for some } \alpha>0
$$
in the definition of $\phi^{op}_d$ one expresses the  property of being Lipschitz with constant $\alpha$ (nonexpansiveness being the case $\alpha=1$). \hl{We further discuss this in Section \ref{subsec:msvcomp}.} %
\end{rem}

We now proceed to compare the logical expressiveness of the MPP and MSV frameworks.

First, we observe that in the MSV theory restricted to $\Met$, equations of the form $\forall (X,d). s=t$ and quantitative equations $\forall (X,d). s=_0 t$ are semantically equivalent, and in fact mutually derivable in the deductive system $\vdash_{\Met}$. This just reflects the fact that metric spaces satisfy the property $x=y\Leftrightarrow d(x,y)=0$ (cf.~the set $\mathcal{H}_\Met$ of $\folang$-implications defining the category $\Met$ in Section \ref{sec:gen:metric:spaces}). Hence in the MSV theory for $\Met$, as far as expressiveness is concerned, we can just restrict our attention to quantitative equations.

Secondly, \hl{a} logical judgment in the MPP framework is a form of implication (possibly with infinitely many premises), called \emph{quantitative inference}, of the form\footnote{The notation used in \cite[Definition 2.1]{DBLP:conf/lics/MardarePP16} is $\{ s_i =_{\epsilon_i} t_i\}_{i\in I} \vdash s=_\epsilon t$, but it clashes with our use of the turnstile $\vdash$, so we write $\Rightarrow$ instead. \hl{They also require $I$ to be finite, we will not.}}
$$
\{ s_i =_{\epsilon_i} t_i\}_{i\in I} \Rightarrow s=_\epsilon t,
$$
where $s_i,t_i,s,t\in \Terms{X}$, for some set $X$, and $\epsilon_i,\epsilon\in [0,1]$. %
\hl{An MPP} quantitative algebra $\algA=(A,d_A,\{op^A\}_{op\in\Sigma})$ (or, alternatively, \hl{an MSV} quantitative algebra in $\QmodMet(\Phi_{NE})$, as noted above) satisfies such a judgment $J$, written $\algA\models_{\textnormal{MPP}} J$, if for all functions $j:X\rightarrow A$,
\begin{center}
if, for all $i\in I$,  $d_A(\sem{s_i}^\algA_j, \sem{t_i}^\algA_j)\leq\epsilon_i$ holds, then $d_A(\sem{s}^\algA_j, \sem{t}^\algA_j)\leq\epsilon$.
\end{center}
A judgment $J$ is called a \emph{basic quantitative inference} if all the terms $s_i,t_i$ appearing on the left-side of the implication are variables in $X$, i.e., $J$ is of the form
$$
\{ x_i =_{\epsilon_i} x'_i\}_{i\in I} \Rightarrow s=_\epsilon t .
$$

One can verify that for every basic quantitative inference $J$ of the above shape and for any MSV quantitative algebra $\algA=(A,d_A,\{op^A\}_{op\in\Sigma})$ in $\QmodGMet(\Phi_{NE})$,
\[
\algA\models_{\textnormal{MPP}} J \ \ \ \ \Longleftrightarrow  \ \ \ \ \ 
 \algA\models \phi_J \ \ \ \ \ \ \textnormal{with $\phi_J$ defined as } \forall (X,d_X). s=_\epsilon t,
\]
\hl{where, in the definition of $\phi_J$, $X$ is the set of variables appearing in the premises of $J$, and $d_X:X^2\rightarrow [0,1]$ is defined as follows:}
$$
d_X(x,x') =
\inf\{ \epsilon_{i} \mid (x =_{\epsilon_i} x')\  \text{is among the premises of $J$}\},
$$
where the infimum of the empty set is $1$.
Note that, in\hl{ }$\phi_J$, the fuzzy relation $(X,d_X)$ is constructed (in a similar fashion to the translation of \autoref{translation:def}) to ensure that nonexpansive interpretations $\imap: (X,d_X)\to (A,d_A)$ correspond to set-theoretic interpretations $j: X \to A$ satisfying the premises of $J$ (i.e. $\forall i \in I. d_X(j(x_i),j(x_i')) \leq \epsilon_i$).

In the opposite direction, for any $\FRel$ quantitative equation $\phi = \forall (X,d_X).s=_\epsilon t$,
\begin{center}
$\algA\models_{\textnormal{MPP}} J_\phi$ \ \ \ \ $\Longleftrightarrow$   \ \ \ \ \ 
 $\algA\models \phi$,
\end{center}
where $J_\phi =
\{ x =_{d_X(x,x')} x' \}_{x,x'\in X} \Rightarrow s=_\epsilon t$ \hl{is a basic quantitative inference}. %

We can therefore conclude that the expressive power of MPP basic inferences $J$ and MSV quantitative equations $\phi$ is the same. This means that our MSV theory, restricted to $\Met$, coincides with the MPP theory where only \emph{basic} quantitative inferences are used. We note that this is a mild restriction, as most interesting results and application instances of the MPP framework only use basic quantitative inferences \cite{DBLP:conf/lics/MardarePP16, DBLP:conf/lics/MardarePP17, DBLP:conf/lics/BacciMPP18, DBLP:conf/calco/BacciMPP21, DBLP:conf/concur/MioV20, DBLP:conf/lics/MioSV21, DBLP:conf/lics/MioSV22}.

As a further point of comparison, we now discuss the proof systems. 
Note that our MSV proof system $\vdash_\Met$, which we have proved to be sound and complete, is not obtained by simply ``restricting'' the MPP proof system of \cite{DBLP:conf/lics/MardarePP16} to basic quantitative inferences (via the translation $J\mapsto \phi_J$). Indeed the MPP proof system is not ``closed under basic quantitative inferences''. The reason is the presence in the MPP proof system of the following \emph{substitution rule} (see \cite[Definition 2.1]{DBLP:conf/lics/MardarePP16}):
\begin{prooftree}
\AxiomC{$ \big\{\  s_i =_{\epsilon_i} t_i\ \big\}_{i\in I} \Rightarrow s =_\epsilon t$}
\RightLabel{Substitution by $\sigma$}
\UnaryInfC{$ \big\{ \ \sigma(s_i) =_{\epsilon_i} \sigma(t_i) \ \big\}_{i\in I} \Rightarrow \sigma(s) =_\epsilon \sigma(t)$}
\end{prooftree}
where all terms have variables ranging over a set $X$ and $\sigma: X\to \Terms{X}$ is a substitution, which is homomorphically extended to a function of type $\sigma: \Terms{X}\to \Terms{X}$.
Note that even in the case where the premise of the substitution rule is a basic quantitative inference (i.e., all the terms $s_i$ and $t_i$ are variables), the conclusion of the rule is generally not a basic quantitative inference, because the substitution is also applied to the premises. 
This highlights the novelty in the design of our MSV proof system $\vdash_\Met$ (the new substitution rule), which in turn also proves a novel result applicable to the MPP theory: a sound and complete proof system for basic quantitative inference\hl{s} exists (via the translation $\phi\mapsto J_\phi$). 

To conclude, we now compare the expressiveness of the MPP theory and of the MSV theory in terms of which $\Met$ monads can be presented, respectively, by a class of basic quantitative inferences and by a class of $\FRel$ (quantitative) equations. By exploiting the correspondence between monad liftings and quantitative extensions proved in Section \ref{section:lifting}, instantiated to the category $\Met$ via \autoref{thm:correspondencemonliftthextgmet}, we show the following result:

\begin{quote}
    \emph{There exist monads on $\Met$ which can be presented by a class of quantitative equations in the MSV theory, but which cannot be presented by a class of basic quantitative inferences in the MPP theory.}
\end{quote}

To see this, consider the \hl{finite, non-empty} powerset monad $(\mP,\eta, \mu)$ on $\Set$, which is presented by the equations $\th$ of semilattices (\hl{\autoref{exmp:monadpres}.(1)}). Define the monad $(\lmP,\lift{\eta},\lift{\mu})$ on $\Met$ where the functor $\lmP:\Met\to \Met$ is such that 
\[\lmP(X,d) = (\mP X, \lift{d}) \quad \text{with}\quad  \lift{d}(S,S') = \begin{cases} 0 & S = S'\\d(x,y) & S= \{x\} \text{ and } S' = \{y\}\\1 & \text{otherwise}\end{cases},\]
and where the unit $\lift{\eta}$ and multiplication $\lift{\mu}$ coincide, as $\Set$ functions, with the unit $\eta$ and multiplication $\mu$ of the monad $\mP$.
The $\Met$ monad $\lmP$ is a monad lifting of the $\Set$ monad $\mP$, and this implies by \autoref{thm:correspondencemonliftthextgmet} that there is a $\Met$ quantitative extension $\qth$ of the equations of semilattices $\th$ which is a presentation of $\lmP$ in the MSV theory.

In contrast, there is no class of basic quantitative inferences presenting the monad $\lmP$ in the MPP theory. This is a consequence of the fact that all monads which can be presented by a class of basic quantitative inferences in the MPP theory  are enriched (\hl{see} \cite[\href{https://arxiv.org/abs/2210.01565}{Full version}, after Corollary 4.19]{Adamek2023}), and that the monad $\lmP$ is not enriched.
Following \cite[Example 7.(1)]{Adamek2023}, \hl{$\lmP$ being enriched is equivalent to satisfying, for all nonexpansive maps $f,g:(X,d_X) \to (Y,d_Y)$,}
\[\sup_{x \in X} d_Y(f(x),g(x)) \geq \sup_{S \in \mP X} \lift{d_Y}(f(S),g(S)).\]
\hl{To see that this does not hold, let $f$ be the identity function on $[0,\frac{1}{2}]$ with the Euclidean distance, and $g$ be the squaring function (both are nonexpansive). Then the left hand side is at most $\frac{1}{2}$ ($d_Y$ is bounded by $\frac{1}{2}$), and the right hand side is $1$ as witnessed by $S = \{0,\frac{1}{2}\}$: $f(S) = S$ and $g(S) = \{0,\frac{1}{4}\}$, so $\lift{d_Y}(f(S),g(S)) = 1$.}

\subsection{Comparison with \cite{DBLP:conf/calco/FordMS21} by Ford et al.}\label{subsec:fmscomp}
The class of metric spaces $(X,d_X:X^2\rightarrow[0,1])$ can be defined, in the standard  language of first order logic, as  the relational structures over the  signature $\{ R_\epsilon \mid \epsilon \in [0,1] \}$, containing a binary relation $R_\epsilon$ for each $\epsilon\in[0,1]$, that satisfy a \hl{set of Horn sentences} analogous to our notion of $\folang$-implications (see Definition \ref{def_Limplications}, where we denoted $R_\epsilon$ with $d(\_, \_)\leq \epsilon$). The \hl{Horn sentences} enforce the correspondence
\begin{equation}\label{eqn:correspstrspa}
d_X(x,y)\leq \epsilon \ \ \  \Leftrightarrow \ \ \ (x,y)\in R_\epsilon.
\end{equation}
Under this correspondence, a nonexpansive function $f:(X,d_X)\rightarrow (Y,d_Y)$ acting on a metric space corresponds to a  relation preserving function acting on a relational structure:
$$ f: (X, \{R^X_\epsilon\}_{\epsilon\in[0,1]}) \rightarrow (Y, \{R^Y_\epsilon\}_{\epsilon\in[0,1]}) \ \ \ \ \  \textnormal{if $(x,x')\in R^X_\epsilon$ then $(f(x),f(x'))\in R^Y_\epsilon$.}
$$

In \cite{DBLP:conf/calco/FordMS21}, based on this observation, a framework of quantitative algebras \hl{over} arbitrary relational signatures (rather than using the specific signature  $\{ R_\epsilon \mid \epsilon \in [0,1] \}$) and an arbitrary \hl{Horn theory (rather than the specific set of Horn sentences} needed to obtain \eqref{eqn:correspstrspa}) has been proposed. \hl{A $\Sigma$-algebra} in \cite{DBLP:conf/calco/FordMS21} is a structure
\[
(A, \{R^\mathbb{A}_i\}_{i\in I}, \{ op^{\mathbb{A}}\}_{op\in \Sigma}),
\]
where $(A, \{R^\mathbb{A}_i\}_{i\in I})$ is a relational structure \hl{modelling} a certain Horn theory (e.g., a metric space), and equipped with relation preserving (e.g., nonexpansive) interpretations of all operation symbols $op\in \Sigma$.\footnote{\hl{The arities in \cite{DBLP:conf/calco/FordMS21} are more general, meaning algebras can have infinitary and partial operations.}}

On the one hand, it is possible to define $\FRel$ as a class of relational structures over the signature  $\{ R_\epsilon \mid \epsilon \in [0,1] \}$ axiomatised by the following (infinitary) \hl{Horn sentences} \hl{(cf.~the axiom schemes UP-CLOSURE, 1-MAX, and ORDER COMPLETENESS)}.
\begin{align*}
    \forall x,y. &\ (x,y)\in R_\epsilon \Rightarrow  (x,y)\in R_\delta &&\text{for all $\epsilon < \delta$}\\
    \forall x,y. &\ (x,y)\in R_1 \\
    \forall x,y. &\ \Big( \displaystyle\bigwedge_{\epsilon\in S}\{ (x,y)\in  R_{\epsilon} \}_{\epsilon \in S}\Big) \Rightarrow (x,y)\in R_{\inf(S)} &&\text{for all $S \subseteq [0,1]$}
\end{align*}
\hl{Hence,} our framework is strictly less expressive than that of \cite{DBLP:conf/calco/FordMS21}. On the other hand, since the framework \cite{DBLP:conf/calco/FordMS21} restricts \hl{interpretations of operations} %
to be relation preserving (cf.~nonexpansive), our framework is more general in this regard.

At the logical level, there are strong similarities between \cite{DBLP:conf/calco/FordMS21} and this work. While we consider quantitative equations of the form
$
\forall (A,d_A). s=_\epsilon t,
$
in \cite{DBLP:conf/calco/FordMS21} the authors use judgments of the form
\begin{center}
$M\vdash R_i(s,t)$ \ \ \ \ \ with $M$ a model of the Horn theory, 
\end{center}
with the same intended meaning of expressing that the relation $R_i$ holds for all relation preserving  interpretations of the elements in $M$ (cf.~nonexpansive interpretations of the elements in $A$).  The substitution rule we adopt in our proof system (see Definition \ref{proof:system:definition}) finds an analogous rule \texttt{Ax} in \cite[p.~14]{DBLP:conf/calco/FordMS21}: they both require that the substitution to be applied is relation preserving (cf.~nonexpansive).

\hl{One element of novelty in our logical framework, besides the fact that we drop the requirement that operations are relation preserving (cf.~nonexpansive), is that we have described in Section \ref{section:fromFRel_to_GMET} how to translate Horn sentences defining $\GMet$ spaces (which we have called $\folang$-implications, see \autoref{def_Limplications}) directly to quantitative equations.}

Another novelty of our work with respect to \cite{DBLP:conf/calco/FordMS21} is  given by our results of Section \ref{section:lifting} on lifting monad presentations from $\Set$ to $\FRel$ (and, in Section \ref{section:fromFRel_to_GMET}, to $\GMet$). This was enabled by our choice of allowing operations that are not necessarily nonexpansive.

\subsection{Comparison with \cite{DBLP:conf/lics/MioSV22} by Mio et al.}\label{subsec:msvcomp}
In an earlier paper \cite{DBLP:conf/lics/MioSV22}, we had investigated an extension of the MPP theory of \cite{DBLP:conf/lics/MardarePP16} (see Section \ref{subsec:MPPcomp}) along two axes:
\begin{enumerate}
    \item The carriers of quantitative algebras are not required to be objects in $\Met$. They can be $\FRel$ spaces or, more generally, $\GMet$ spaces satisfying a subset of the constraints defining metric spaces and ultrametric spaces (see \cite[\S 2.3]{DBLP:conf/lics/MioSV22}). 
    \item The operations do not have to be nonexpansive with respect to the categorical product distance, but with respect to possibly different distances on the product. %
    This is achieved by introducing the notion of  \textit{lifted signatures}. In a lifted signature $\Sigma$, each operation symbol $op \in \Sigma$ of arity $n$ has an associated lifting $L_{op}$ of the $\Set$ product endofunctor $(\_)^n: \Set\to \Set$, i.e., $L_{op}$ is an endofunctor on $\GMet$ making the following diagram commute, where $U$ is the expected forgetful functor:
\begin{equation*}
\begin{tikzcd}[row sep=small, column sep=1.4em]
	\GMet & \GMet \\
	\Set & \Set
	\arrow["U"', from=1-1, to=2-1]
	\arrow["(\_)^n"', from=2-1, to=2-2]
	\arrow["L_{op}", from=1-1, to=1-2]
	\arrow["U", from=1-2, to=2-2]
\end{tikzcd}
\end{equation*} 
For technical reasons, the paper only considers liftings that satisfy a property called \emph{preservation of isometric embeddings} \cite[Definition 3.2]{DBLP:conf/lics/MioSV22}.
    Then, the definition of quantitative algebras requires operations to be nonexpansive maps with respect to the associated lifting:
    \begin{defi}[Quantitative algebra for a lifted signature]\label{def:MSV22}\cite[Definition 3.6]{DBLP:conf/lics/MioSV22}
    Given a lifted signature $\Sigma$, a quantitative $\Sigma$-algebra is a triple $(A, d_A, \{op^A\}_{op\in \Sigma})$  such that all interpretations of operation symbols $op^A:A^n\to A$ (where $ar(op) =n$) are nonexpansive functions
    $$op^A: (A^n,L_{op}(d_A)) \rightarrow (A,d_A),$$
    where $L_{op}$ is the lifting of the $\Set$ product endofunctor associated to $\op$.%
    \end{defi}
    We thereby have a definition which follows the MPP theory in requiring operations to be nonexpansive (Definition \ref{MPP-algebra}), and yet generalises the MPP theory since nonexpansiveness can be imposed with respect to arbitrary liftings of the product, and not just with respect to the categorical product lifting computing the coordinatewise maximum.
\end{enumerate}
Since in the present paper we also present an extension of the MPP theory along these two axes, we can compare to the extension proposed in \cite{DBLP:conf/lics/MioSV22} along the same axes:
\begin{enumerate}
    \item \hl{The} present paper extends \cite{DBLP:conf/lics/MioSV22} by allowing $\GMet$ spaces to be defined by arbitrary sets of $\folang$-implications. Thanks to the formal translation defined in Section \ref{section:fromFRel_to_GMET}, we obtain the results for any $\GMet$ in a modular way starting from those for $\FRel$.
    \item \hl{We first} note that the requirement that the interpretation of an $n$-ary operation $op$ is $L_{op}$-nonexpansive (Definition \ref{def:MSV22}) can be expressed with a family of $\FRel$ quantitative equations in the framework of the present paper, analogously to \eqref{eqn:opnexpasqeq}: for every generalised metric $d_X$ on $X = \{x_1,\dots, x_n,x'_1,\dots, x'_n\}$, we take
\begin{equation*}\label{eqn:Lnexpqeqn}
	\forall(X,d_X). op(x_1,\dots, x_n) =_{\epsilon} op(x'_1,\dots, x'_n) \quad \epsilon =L_{op}(d_X)((x_1,\dots,x_n),(x'_1,\dots,x'_n)).
\end{equation*}
To see this, note that these quantitative equations correspond to the $L$--NE rules in \cite[Definition 3.11]{DBLP:conf/lics/MioSV22}, which in turn express $L_{op}$-nonexpansiveness of liftings preserving isometrics embeddings.
This means that the quantitative algebras in the present paper encompass those of \cite{DBLP:conf/lics/MioSV22}. Conversely, if we set $L_{op}$ to always be the discrete $n$-ary product functor $L^n$ (see \eqref{eq:discreteproductlifting}, or its $\GMet$ variants), then any function $op^A: (A^n, L^n(d_A)) \rightarrow (A,d_A)$ is nonexpansive. Therefore, the quantitative algebras of \cite{DBLP:conf/lics/MioSV22} are those of \autoref{defi_quantitative_algebra}, for the $\GMet$ spaces considered in \cite[\S 2.3]{DBLP:conf/lics/MioSV22}.

\end{enumerate}
Hence, the definition of quantitative algebras presented in this paper subsumes that of \cite{DBLP:conf/lics/MioSV22} while being simpler, as operations are just $\Set$ functions with no lifting or nonexpansiveness conditions attached, and extends it to a larger class of $\GMet$ spaces.

Orthogonally to these two axes, a major difference between \cite{DBLP:conf/lics/MioSV22}
and the present paper is that in \cite{DBLP:conf/lics/MioSV22} the logical judgments are quantitative inferences like in the MPP theory, i.e., implications between quantitative equations. In contrast, here we use quantitative equations, and a proof system solely based on those (see Section \ref{subsec:MPPcomp}). 

Furthermore, the results in Sections \ref{section:lifting} and \ref{section:fromFRel_to_GMET} about monad liftings constitute a novelty of the present paper.

\section{Conclusions and Directions for Future Work}
\label{conclusion_section}

We have presented an extension of the theory of quantitative algebras of Mardare, Panangaden and Plotkin \cite{DBLP:conf/lics/MardarePP16}. In our \hl{framework,}  the carriers of quantitative algebras are not restricted to be metric spaces and can be arbitrary fuzzy relations (or generalised metric spaces)\hl{,} and the interpretations of the algebraic operations are not required to be nonexpansive. We have established some key results, including the soundness and completeness of a novel proof system, the existence of free quantitative algebras, the strict monadicity of the associated Free-Forgetful adjunction, and the correspondence between monad liftings of a finitary monad \hl{on $\Set$} and quantitative extensions of an equational presentation.

A first direction for future work consists in trying to adapt and generalise  to our setting some theoretical results obtained for the framework of Mardare, Panangaden and Plotkin \cite{DBLP:conf/lics/MardarePP16}. Examples include monad composition techniques \cite{DBLP:conf/lics/BacciMPP18,DBLP:journals/corr/abs-2212-11784}, fixed-points \cite{DBLP:conf/lics/MardarePP21}, completion techniques \cite{DBLP:conf/lics/BacciMPP18}, and variety “HSP-style” theorems 
\cite{DBLP:conf/lics/MardarePP17,DBLP:conf/lics/Adamek22,Jurka2024}.

A second direction, more oriented towards applications, consists in leveraging the additional flexibility provided by our theory. For example, in \cite{dallago_et_al:LIPIcs.FSCD.2022.4} the authors investigate Curry's \emph{combinatory logic} (an algebraic counterpart of the $\lambda$-calculus) \hl{under the lens} of quantitative algebras, and they point out the need of considering operations that are not nonexpansive and carriers that are \emph{partial ultra-metrics}. 
As the latter is an example of a $\GMet$ category, in the sense of Section \ref{sec:gen:metric:spaces}, the research line of \cite{dallago_et_al:LIPIcs.FSCD.2022.4} can be carried out within the framework presented in this work. Similarly, in \cite{DBLP:conf/lics/MioSV22} the authors have investigated the \L ukaszyk–Karmowski distance on \emph{diffuse metric spaces} \cite{Hitzler2000,Castro2021} of probability distributions. This is yet another type of  $\GMet$ category that can be formalised within our framework. As a last example, in \cite{DBLP:journals/pacmpl/GavazzoF23} the authors investigate ``quantitative rewriting systems'' and need to go beyond nonexpansive operations, by admitting (in what they call ``graded rewriting systems'') Lipschitz operations with constant $\alpha>1$. As noted in Remark \ref{remark_Lipschitz}, it is possible in our theory to express, by means of quantitative equations, that operations are Lipschitz for any $\alpha>1$.

A third direction for future work consists in exploring further generalisations of our framework. For example, our choice of considering fuzzy relations $d_A:A^2\rightarrow [0,1]$ has been made, somewhat arbitrarily, as a compromise between maximal generality and the convenience of dealing with a concrete notion of numeric distance. One could alternatively work with distances $d_A: A^2\rightarrow [0,\infty]$ (valued in the extended real line) as in \cite{DBLP:conf/lics/MardarePP16} or, more generally,  $d_A: A^2\rightarrow Q$ where $Q$ is an abstract quantale \cite{DBLP:journals/logcom/PiazzaC96}. In this direction, it has been shown in the PhD thesis of the second author \cite{Sarkis2024} that all the results of this work can be extended to the setting where distances $d_A : A^2 \rightarrow L$ are valued in a complete lattice $L$. The cases $L=[0,1]$, $L=[0, \infty]$, and $L=Q$ (the complete lattice underlying a quantale) are instances of this generalization. In another direction, it could be interesting to follow the work of \cite{DBLP:conf/calco/FordMS21} and move beyond ``distances''  and towards arbitrary relational structures \hl{(see also \cite{Jurka2024})}.

A fourth direction consists in using our deductive apparatus to reason quantitatively about program distances as, e.g., suggested in the preliminary examples given in \cite[\S VI]{DBLP:conf/lics/MioSV21} in the context of process algebras. In particular, adapting the well-known framework of (equational) ``up-to techniques'' (see, e.g., \cite{DBLP:journals/acta/BonchiPPR17}) to the quantitative setting (see, e.g., \cite{bonchi_et_al:LIPIcs:2018:9555}) appears to be a promising endeavour. 

\hl{A fifth direction consists in exploring the connections between our line of work and the vast literature on \emph{fuzzy logic}}(see, e.g., \cite{BookFuzzyLogic}), \hl{where the logical apparatus of classical first order logic (quantifiers, conjunctions, implications, etc.) is replaced with a quantitative counterpart dealing with truth values in $[0,1]$ or other sets of truth values. For example, in \emph{fuzzy equational logic} and \emph{fuzzy Horn logic} } \cite[Chapters 3 and 4]{BookFuzzyEL}, \hl{equalities and implications between equalities have \textit{degrees of truth}. The recent work \cite{DBLP:journals/corr/abs-2302-01224} 
constitutes a step in this direction and is based on taking truth values in the \emph{Lawvere Quantale} $[0,\infty]$.} \hl{A related work is the recent} \cite{DBLP:conf/lics/DagninoP22} \hl{ which uses \emph{Lawvere  hyperdoctrines} to study quantitative equality in the linear logic setting.}

\hl{Finally, we conclude with a technical question which we leave here as an open problem. As discussed in \autoref{remark_inf_variables_2}, the $\FRel$ space $(A,d_A)$ in our \autoref{quantitative_equation_def} of equations ($\forall (A,d_A).s=t$) and quantitative equations ($\forall (A,d_A).s=_\epsilon t$) is not restricted in any way, and can therefore have arbitrary cardinality. Can we find a cardinal $\kappa$ such that any quantitative equational theory can be generated by a set of equations and quantitative equations such that $|A|< \kappa$ ? For example, in the case of standard Universal algebra, it is well known that one can always restrict to equations $\forall A. s= t$ (see Definition \ref{standard_equations_def}) where $|A|$ is finite, and thus $\kappa=\aleph_0$. Formally, is there a cardinal $\kappa$ such that, for any $\Phi\subseteq  \MEq(\Sigma)$ there is a set $\Psi\subseteq  \MEq(\Sigma)$ such that:}
\begin{enumerate}
    \item for all equations $\forall (A,d_A).s=t\in \Psi$, it holds that $|A|<\kappa$, 
    \item for all quantitative equations $\forall (A,d_A).s=_\epsilon t\in \Psi$, it holds that $|A|<\kappa$, and
    \item $\QModels(\Phi) = \QModels(\Psi)$?
\end{enumerate}
\hl{We mention a relevant technical fact: the category $\FRel$ is locally $\omega_1$-presentable by} \cite[Example 5.27.(3)]{AR1994} \hl{since $\FRel$ is a category of models of a Horn theory whose sentences have less than $\omega_1$ premises.}

\hl{
\emph{Acknowledgements:} The authors are grateful to the anonymous reviewers for their valuable comments and suggestions which helped to improve the final version of this article. The authors would also like to thank the participants of the 2023 Bellairs Workshop on \textit{Quantitative Logic and Reasoning}, especially the organisers Prakash Panangaden and Alexandra Silva. This research has been partially supported by the French ANR Project ANR-20-CE48-0005 ``QuaReMe''.
}

\bibliographystyle{alphaurl}
\bibliography{main}

\end{document}